\documentclass[prd,aps,nofootinbib,superscriptaddress]{revtex4}
\usepackage{amsmath,mathtools,cases,slashed}

\usepackage{color}

\allowdisplaybreaks

\newcommand{\kperpq}{\vec{k}_\perp^{\,2}}

\begin{document}

\title{Exploring twist-2 GPDs through quasi-distributions in a diquark spectator model}

\author{Shohini Bhattacharya}
\affiliation{Department of Physics, SERC,
             Temple University, Philadelphia, PA 19122, USA}

\author{Christopher Cocuzza}
\affiliation{Department of Physics, SERC,
             Temple University, Philadelphia, PA 19122, USA}

\author{Andreas Metz}
\affiliation{Department of Physics, SERC,
             Temple University, Philadelphia, PA 19122, USA}

\begin{abstract}
Quasi parton distributions (quasi-PDFs) are currently under intense investigation.
Quasi-PDFs are defined through spatial correlation functions and are thus accessible in lattice QCD.
They gradually approach their corresponding standard (light-cone) PDFs as the hadron momentum increases. 
Recently, we investigated the concept of quasi-distributions in the case of generalized parton distributions (GPDs) by calculating the twist-2 vector GPDs in the scalar diquark spectator model.
In the present work, we extend this study to the remaining six leading-twist GPDs.
For large hadron momenta, all quasi-GPDs analytically reduce to the corresponding standard GPDs.
We also study the numerical mismatch between quasi-GPDs and standard GPDs for finite hadron momenta.
Furthermore, we present results for quasi-PDFs, and explore higher-twist effects associated with the parton momentum and the longitudinal momentum transfer to the target.
We study the dependence of our results on the model parameters as well as the type of diquark.
Finally, we discuss the lowest moments of quasi distributions, and elaborate on the relation between quasi-GPDs and the total angular momentum of quarks. 
The moment analysis suggests a preferred definition of several quasi-distributions.
\end{abstract}


\date{\today}

\maketitle

\section{Introduction}

Parton distribution functions (PDFs) are important objects encoding information about the quark and gluon structure of hadrons~\cite{Collins:1981uw}.
They can be extracted from data for a large class of hard scattering processes, where the key underlying tool is factorization theorems in quantum chromodynamics (QCD) that separate the perturbatively calculable short distance part of a cross section from the long-distance part described by PDFs and other potential non-perturbative quantities~\cite{Collins:1989gx}.
On the other hand, first-principles calculations of PDFs using lattice QCD have remained challenging due to their explicit time-dependence. 
As a result, in the past almost all related studies in lattice QCD focused on moments of PDFs which are defined through time-independent local operators, while the full dependence of PDFs on the parton momentum fraction $x$ remained elusive.

The recently proposed quasi parton distributions (quasi-PDFs) offer a way to directly access the $x$-dependence of the PDFs in lattice QCD~\cite{Ji:2013dva,Ji:2014gla}. 
Quasi-PDFs are defined through spatial equal-time operators that can be computed on four-dimensional Euclidean lattices. 
They reduce to their corresponding standard (light-cone) PDFs if the hadron momentum $P^3 = |\vec{P}| \to \infty$, prior to renormalization.
However, for lattice calculations one first renormalizes, and $P^3$ is finite.
This leads to two sources of discrepancies between quasi-PDFs and standard PDFs: higher-twist corrections that are suppressed by powers of $\frac{1}{P^3}$, and a different ultraviolet (UV) behavior for these two types of PDFs.
The UV disparities can be cured order by order in perturbative QCD through a so-called matching procedure --- see for instance Refs.~\cite{Xiong:2013bka,Stewart:2017tvs,Izubuchi:2018srq}.
Other approaches for computing the $x$-dependence of PDFs and related quantities have also been suggested~\cite{Braun:1994jq, Detmold:2005gg, Braun:2007wv, Ma:2014jla, Chambers:2017dov, Hansen:2017mnd, Radyushkin:2017cyf, Orginos:2017kos, Ma:2017pxb, Radyushkin:2017lvu, Liang:2017mye, Detmold:2018kwu}.
Some of them are closely related to the concept of quasi-PDFs.

By now there has been important progress in understanding the renormalization of quasi-PDFs~\cite{Ji:2015jwa, Ishikawa:2016znu, Chen:2016fxx, Constantinou:2017sej, Alexandrou:2017huk, Chen:2017mzz, Ji:2017oey, Ishikawa:2017faj, Green:2017xeu, Spanoudes:2018zya, Zhang:2018diq, Li:2018tpe, Constantinou:2019vyb}. 
A variety of other aspects of quasi-PDFs and, generally, Euclidean correlators have also been extensively studied~\cite{Ji:2014hxa, Li:2016amo, Monahan:2016bvm, Radyushkin:2016hsy, Radyushkin:2017ffo, Carlson:2017gpk, Briceno:2017cpo, Xiong:2017jtn, Rossi:2017muf, Ji:2017rah, Wang:2017qyg, Chen:2017mie, Monahan:2017hpu, Radyushkin:2018cvn, Zhang:2018ggy, Ji:2018hvs, Xu:2018mpf, Jia:2018qee, Briceno:2018lfj, Rossi:2018zkn, Radyushkin:2018nbf, Ji:2018waw, Karpie:2018zaz, Braun:2018brg, Liu:2018tox, Ebert:2018gzl, Briceno:2018qfa, Ebert:2019okf}. 
In particular, the first lattice QCD results for quasi-PDFs and related quantities can be considered milestones in this field~\cite{Lin:2014zya, Alexandrou:2015rja, Chen:2016utp, Alexandrou:2016jqi, Zhang:2017bzy, Alexandrou:2017huk, Chen:2017mzz, Green:2017xeu, Lin:2017ani, Orginos:2017kos, Bali:2017gfr, Alexandrou:2017dzj, Chen:2017gck, Alexandrou:2018pbm, Chen:2018xof, Chen:2018fwa, Alexandrou:2018eet, Liu:2018uuj, Bali:2018spj, Lin:2018qky, Fan:2018dxu, Liu:2018hxv, Bali:2018qat, Shugert:2018pwi, Sufian:2019bol, Alexandrou:2019lfo}. 
Additionally, the convergence of quasi-PDFs to the corresponding standard PDFs has been explored in several models~\cite{Gamberg:2014zwa, Bacchetta:2016zjm, Nam:2017gzm, Broniowski:2017wbr, Hobbs:2017xtq, Broniowski:2017gfp, Xu:2018eii, Bhattacharya:2018zxi}. 
The progress in this field was recently reviewed in Refs.~\cite{Monahan:2018euv, Cichy:2018mum, Zhao:2018fyu}.

As already pointed out in Ref.~\cite{Ji:2013dva}, the concept of quasi-distributions is not limited to forward PDFs.
For example, generalized parton distributions (GPDs)~\cite{Mueller:1998fv, Ji:1996ek, Radyushkin:1996nd, Ji:1996nm, Radyushkin:1996ru} could also be addressed in this approach.
A number of compelling motivations to study GPDs exist, where we refer to review articles for more details~\cite{Goeke:2001tz, Diehl:2003ny, Belitsky:2005qn, Boffi:2007yc, Guidal:2013rya, Mueller:2014hsa, Kumericki:2016ehc}.
On the other hand, extracting GPDs from experiment is difficult.
In this situation, reliable information from lattice QCD on GPDs using quasi distributions would be very helpful.
So far, only a very limited number of papers have considered quasi-GPDs.
In Refs.~\cite{Ji:2015qla, Xiong:2015nua, Liu:2019urm} the perturbative matching for quasi-GPDs of quarks was studied. 
Recently, we explored quasi-GPDs in the Scalar Diquark-spectator Model (SDM) of the nucleon, where the focus was on the twist-2 vector GPDs $H$ and $E$~\cite{Bhattacharya:2018zxi}.
Here we extend this study by considering the remaining six leading-twist quark GPDs in the same model.
We also present some new model-independent results on moments of quasi-distributions.

Specifically, in this work we address the following points.
In Sec.~\ref{sec:GPDs} we provide definitions of all leading-twist quasi-GPDs for quarks.
For each standard GPD we consider two corresponding quasi-GPDs.
Some important kinematical relations are listed as well in that section.
In Sec. \ref{sec:analytical_results} we present, in particular, the analytical results for the standard GPDs and the quasi-GPDs in the SDM.
For $P^3 \to \infty$, all expressions for the quasi-GPDs reduce to the ones of the respective standard GPDs, where the necessary steps for this check are presented through one example.
This outcome further supports quasi-GPDs as a viable tool for studying standard GPDs.
As a byproduct, we also consider quasi-PDFs in the SDM, and elaborate for PDFs on the specific point $x = 0$ at which standard PDFs (in the SDM) are discontinuous.
It is interesting that, in the limit $P^3 \to \infty$, quasi-PDFs exactly reproduce the standard PDFs even at $x = 0$.
The numerical results are discussed in Sec.~\ref{sec:numerical_results}.
For PDFs we find considerable discrepancies between the quasi-distributions and the standard distributions around $x = 0$ and $x = 1$, and we locate the source of this feature.
For GPDs one observes the same issue around $x = 1$, as well as  in and close to the ERBL region, if that region is very narrow.
On the other hand, quasi-GPDs and standard GPDs are very similar for a considerable part of a large ERBL region. 
We also study two sources of higher-twist effects --- those related with the average longitudinal momentum fractions of the quark, and those associated with the skewness variable of GPDs.
In general, we have tried to extract robust numerical results of the SDM.
To this end we have explored the dependence of the results on variations of the model parameters.
In Sec. \ref{sec:axial-vector}, we corroborate the robustness of the SDM results by studying the impact on the unpolarized GPD $H$ of modeling the diquark as an axial-vector diquark instead of a scalar diquark.
Model-independent results on the first and second moments of quasi-distributions can be found in Sec.~\ref{sec:moments}.
They include a discussion of the relation between quasi-GPDs and the total angular momentum of quarks.
Considering moments of quasi-distributions from lattice QCD might offer a way to study systematic uncertainties.
The moment analysis also suggests a preferred definition of several quasi-distributions.
We summarize our work in Sec.~\ref{sec:summary}.

\section{Definition of GPDs}
\label{sec:GPDs}
We start by recalling the definition of twist-2 standard GPDs of quarks for a spin-$\tfrac{1}{2}$ hadron, which are specified through the Fourier transform of off-forward matrix elements of bi-local quark operators (see for instance Ref.~\cite{Diehl:2003ny})\footnote{For a generic four-vector $v$ we denote the Minkowski components by $(v^0, v^1, v^2, v^3)$ and the light-cone components by $(v^+, v^-, \vec{v}_\perp)$, with $v^+ = \frac{1}{\sqrt{2}} (v^0 + v^3)$, $v^- = \frac{1}{\sqrt{2}} (v^0 - v^3)$ and $\vec{v}_\perp = (v^1, v^2)$.}, 
\begin{equation}
F^{[\Gamma]}(x, \Delta; \lambda, \lambda')=\frac{1}{2} \int \frac{dz^{-}}{2\pi} e^{i k \cdot z} \langle p', \lambda'| \bar{\psi} (-\tfrac{z}{2})\, \Gamma \, {\cal W}(-\tfrac{z}{2},\tfrac{z}{2}) \psi (\tfrac{z}{2})|p, \lambda \rangle \bigg |_{z^{+}=0,\vec{z}_{\perp}=\vec{0}_{\perp}} \, .
\label{e:corr_standard_GPD}
\end{equation} 
In Equation~(\ref{e:corr_standard_GPD}), $\Gamma$ denotes a generic gamma matrix, and the Wilson line 
\begin{equation}
{\cal W}(-\tfrac{z}{2},\tfrac{z}{2})\bigg |_{z^{+}=0,\vec{z}_{\perp}=\vec{0}_{\perp}} = {\cal P} \, {\rm exp} \bigg ( -ig_{s} \int^{\tfrac{z^{-}}{2}}_{-\tfrac{z^{-}}{2}} \, dy^{-} A^{+}(0^{+},y^{-},\vec{0}_{\perp}) \bigg ) 
\label{e:wilson_line_standard_GPD}
\end{equation}
ensures the color gauge invariance of the operator, where ${\cal P}$ indicates path-ordering and $g_s$ the strong coupling constant.
The incoming (outgoing) hadron state in (\ref{e:corr_standard_GPD}) is characterized by the 4-momentum $p \, (p')$ and the helicity $\lambda \, (\lambda')$. 
Frequently used kinematical variables in the context of such off-forward matrix elements are
\begin{equation}
P = \frac{1}{2} (p + p') \,, \qquad
\Delta = p' - p \,, \qquad
t = \Delta^2 \,, \qquad
\xi = \frac{p^{\prime +} - p^+}{p^{\prime +} + p^+} = - \frac{\Delta^+}{2 P^+} \,.
\end{equation}
For the skewness variable one typically considers $\xi \geq 0$, because $\xi$ is non-negative for every known physical process that allows access to the GPDs.
Therefore we also limit our discussion to $\xi \geq 0$.
We work in a symmetric frame of reference where $\vec{P}_{\perp}=0$. 
Also, we take $P^{3} > 0$ and large. 
The variable $t$ is related to $\xi$ and $\vec{\Delta}_{\perp}$ through
\begin{eqnarray} 
t = - \frac{1}{1 - \xi^{2}} (4 \xi^{2} M^{2} + \vec{\Delta}^{2}_{\perp}) \, ,
\label{e:t_s}
\end{eqnarray}
where $M$ is the nucleon mass. 
Equation~(\ref{e:corr_standard_GPD}) represents a leading-twist matrix element if $\Gamma$ contains one plus-index.
The corresponding (eight) quark GPDs are then defined via~(see for instance Refs.~\cite{Diehl:2003ny, Meissner:2007rx})
\begin{eqnarray}
F^{[\gamma^{+}]}(x, \Delta; \lambda, \lambda')&=&\frac{1}{2P^{+}} \bar{u}(p', \lambda ') \bigg[ \gamma^{+} H(x, \xi, t) + \frac{i\sigma^{+\mu}\Delta_{\mu}}{2M} E(x, \xi, t) \bigg] u(p, \lambda) \, , 
\label{e:U_s} 
\\[0.1cm]
F^{[\gamma^{+}\gamma_{5}]}(x, \Delta; \lambda, \lambda')&=&\frac{1}{2P^{+}} \bar{u}(p', \lambda ') \bigg[ \gamma^{+}\gamma_5 \widetilde{H}(x, \xi, t) + \frac{\Delta^{+}\gamma_{5}}{2M} \widetilde{E}(x, \xi, t) \bigg ] u(p, \lambda) \, , 
\label{e:L_s} 
\\[0.1cm]
F^{[i\sigma^{j+}\gamma_{5}]}(x, \Delta; \lambda, \lambda')&=&-\frac{i\varepsilon^{-+ij}}{2P^{+}} \bar{u}(p', \lambda ') \bigg [i\sigma^{+i}H_{T}(x, \xi, t) + \frac{\gamma^{+}\Delta^{i}_{\perp}-\Delta^{+}\gamma^{i}_{\perp}}{2M} E_{T}(x, \xi, t) \nonumber \\[0.1cm]
&&+ \frac{P^{+}\Delta^{i}_{\perp}}{M^{2}}\widetilde{H}_{T}(x, \xi, t) - \frac{P^{+}\gamma^{i}_{\perp}}{M}\widetilde{E}_{T}(x, \xi, t) \bigg ] u(p, \lambda)\,,
\label{e:T_s}
\end{eqnarray}
where $u(p, \lambda)$ ($\bar{u}(p', \lambda ')$) is the helicity spinor for the incoming (outgoing) hadron and $\sigma^{\mu\nu}=\frac{i}{2}(\gamma^{\mu}\gamma^{\nu}-\gamma^{\nu}\gamma^{\mu})$.
We adopt the convention of $\varepsilon^{0123}=1$. 
The quarks are unpolarized in the case of the vector GPDs $H$ and $E$, longitudinally polarized for $\widetilde{H}$ and $\widetilde{E}$, and transversely polarized for $H_{T}$, $E_{T}$, $\widetilde{H}_{T}$ and $\widetilde{E}_{T}$.
In Equation~(\ref{e:T_s}), because of the relation $i\sigma^{\mu \nu}\gamma_{5}=-\frac{1}{2}\epsilon^{\mu \nu \alpha \beta}\sigma_{\alpha \beta}$, one may also work with the matrix $i\sigma^{j+}$ (instead of $i\sigma^{j+}\gamma_{5}$) to define chiral-odd quark GPDs. 
A generic GPD depends upon the average longitudinal momentum fraction $x = \frac{k^{+}}{P^{+}}$, as well as $\xi$ and $t$. 
By means of Eq.~(\ref{e:t_s}) one can consider standard GPDs as function of $x$, $\xi$ and $\vec{\Delta}_{\perp}$.
We in fact use these variables for the numerical evaluations of the GPDs later on.
We recall in passing that the support region for the standard GPDs is given by the range $-1 \leq x \leq 1$.

Quasi-GPDs, on the other hand, are defined through an equal-time spatial correlation function~\cite{Ji:2013dva},
\begin{equation}
F^{[\Gamma]}_{\rm Q}(x, \Delta;   \lambda, \lambda'; P^{3}) =\frac{1}{2} \int \frac{dz^{3}}{2\pi}  e^{ik \cdot z} \langle p',\lambda '| \bar{\psi}(-\tfrac{z}{2}) \, \Gamma \, {\cal W}_{\rm Q}(-\tfrac{z}{2}, \tfrac{z}{2}) \psi (\tfrac{z}{2})|p, \lambda \rangle \bigg |_{z^{0}=0, \vec{z}_{\perp}=\vec{0}_{\perp}} \, ,
\label{e: corr_quasi_GPD}
\end{equation}
where the Wilson line is given by,
\begin{equation}
{\cal W}_{\rm Q}(-\tfrac{z}{2},\tfrac{z}{2})\bigg |_{z^{0}=0,\vec{z}_{\perp}=\vec{0}_{\perp}} = {\cal P} \, {\rm exp} \bigg ( -ig_{s} \int^{\tfrac{z^{3}}{2}}_{-\tfrac{z^{3}}{2}} \, dy^{3} A^{3}(0, \vec{0}_{\perp}, y^{3}) \bigg ) \, .
\end{equation}
For a given standard GPD, we consider two distinct definitions of its corresponding quasi-GPD. 
The counterparts of Eqs.~(\ref{e:U_s}), (\ref{e:L_s}) and (\ref{e:T_s}) are
\begin{eqnarray}
F^{[\gamma^{0}]}(x, \Delta; \lambda, \lambda'; P^{3})&=&\frac{1}{2P^{0}}\bar{u}(p', \lambda ') \bigg [\gamma^{0} H_{\rm Q(0)}(x, \xi, t; P^{3}) + \frac{i\sigma^{0\mu}\Delta_{\mu}}{2M} E_{\rm Q(0)}(x, \xi, t; P^{3}) \bigg ] u(p, \lambda) \, ,  
\label{e:U_q}\\[0.1cm]
F^{[\gamma^{3}\gamma_5]}(x, \Delta; \lambda, \lambda'; P^{3})&=&\frac{1}{2P^{0}}\bar{u}(p', \lambda ') \bigg [\gamma^{3}\gamma_{5} \widetilde{H}_{\rm Q(3)}(x, \xi, t;P^{3}) + \frac{\Delta^{3}\gamma_{5}}{2M}\widetilde{E}_{\rm Q(3)}(x, \xi, t;P^{3}) \bigg ] u(p, \lambda) \, ,
\label{e:L_q} \\[0.1cm]
F^{[i\sigma^{j0}\gamma_{5}]}(x, \Delta; \lambda, \lambda'; P^{3})&=&-\frac{i\varepsilon^{03ij}}{2P^{0}} \bar{u}(p', \lambda ') \bigg [i\sigma^{3i}H_{T,{\rm Q(0)}}(x, \xi, t; P^{3}) + \frac{\gamma^{3}\Delta^{i}_{\perp}-\Delta^{3}\gamma^{i}_{\perp}}{2M} E_{T, {\rm Q(0)}}(x, \xi, t; P^{3}) \nonumber \\[0.1cm]
&&+ \frac{P^{3}\Delta^{i}_{\perp}}{M^{2}}\widetilde{H}_{T, {\rm Q(0)}}(x, \xi, t; P^{3}) - \frac{P^{3}\gamma^{i}_{\perp}}{M}\widetilde{E}_{T, {\rm Q(0)}}(x, \xi, t; P^{3}) \bigg ] u(p, \lambda) \, .
\label{e:T_q}
\end{eqnarray}
One can define $H_{\rm Q(3)}$ and $E_{\rm Q(3)}$ through Eq.~(\ref{e:U_q}) using the replacement $0 \rightarrow 3$ (see also Ref.~\cite{Bhattacharya:2018zxi}), while $\widetilde{H}_{\rm Q(0)}$ and $\widetilde{E}_{\rm Q(0)}$ are defined through Eq.~(\ref{e:L_q}) with $0 \leftrightarrow 3$. 
The chiral-odd quasi-GPDs $H_{T,{\rm Q(3)}}$, $E_{T,{\rm Q(3)}}$, $\widetilde{H}_{T,{\rm Q(3)}}$, and $\widetilde{E}_{T,{\rm Q(3)}}$ are defined through Eq.~(\ref{e:T_q}) with $0 \rightarrow 3$,  with the exception that $\varepsilon^{03ij}$ should be left as is. 
The factor $\frac{1}{P^{0}}$ on the r.h.s.~of ~(\ref{e:L_q}) (which appears counterintuitive due to the $\gamma^{3}\gamma_{5}$ projection) is necessary to be consistent with the definition of the corresponding helicity quasi-PDF, such that the definitions of all (sixteen) quasi-GPDs are consistent with the corresponding forward limits. 
It has been argued that the gamma matrices used in~(\ref{e:U_q}), (\ref{e:L_q}) and (\ref{e:T_q}) provide optimal behavior of the associated operators under renormalization~\cite{Constantinou:2017sej, Chen:2017mie}.
By taking the forward limit of Eqs.~(\ref{e:U_q})--(\ref{e:T_q}) one recovers the so far most frequently used definitions of the quasi-PDFs $f_{1, {\rm Q(0)}}$, $g_{1, {\rm Q(3)}}$ and $h_{1, {\rm Q(0)}}$.
In Sec.~\ref{sec:moments} below we will return to this point.

We now briefly discuss the behavior of GPDs under the replacement $\xi \rightarrow -\xi$.  
Hermiticity implies that all standard GPDs but $\widetilde{E}_{T}$ are even functions of $\xi$, while $\widetilde{E}_{T}$ is an odd function of $\xi$~\cite{Diehl:2003ny, Meissner:2007rx}.
We find the exact same (model-independent) behavior for the corresponding quasi-GPDs.
Exploiting the symmetry of quasi-GPDs under $\xi \rightarrow -\xi$ may provide more statistics for lattice calculations. 

Apart from the dependence on $\xi$ and $t$, quasi-GPDs are functions of $x = \tfrac{k^{3}}{P^{3}}$. 
The latter variable is of course different from the average plus-momentum $\tfrac{k^{+}}{P^{+}}$ that appears for standard GPDs, and it is not possible to relate these two momentum fractions in a model-independent manner.
In Sec.~\ref{sec: x&xt} we study the impact of their difference in the cut-graph approach in the diquark spectator model.
Note that the support region for the quasi-GPDs is given by $- \infty < x < \infty$.
For the calculations we also need the relation $P^{0}=\delta P^{3}$ where 
\begin{equation}
\delta = \sqrt{1+ \frac{M^{2}-t/4}{(P^{3})^{2}}} \,.
\label{e:delta}
\end{equation}
Below we frequently make use of the variable $\delta$.
Moreover, $P \cdot \Delta = 0$, from which one obtains $\Delta^{0} = - 2 \xi P^{3}$.

\section{Analytical Results in Scalar Diquark Model}
\label{sec:analytical_results}
In this section we present the analytical results in the SDM.
Details about this model can be found in Ref.~\cite{Bhattacharya:2018zxi} and references therein.

\subsection{Results for standard GPDs}
\label{sec:standard_GPDs}
We begin with the results for the standard GPDs.
To the lowest nontrivial order in the SDM, the correlator in~(\ref{e:corr_standard_GPD}) takes the form
\begin{equation}
F^{[\Gamma]}(x, \Delta; \lambda, \lambda ') = \frac{i \, g^{2}}{2 (2\pi)^4} \int dk^{-} \, d^{2}\vec{k}_{\perp} \, \frac{\bar{u}(p', \lambda') \, \Big( \slashed{k} + \frac{\slashed{\Delta}}{2} + m_{q} \Big) \, \Gamma \, \Big( \slashed{k} - \frac{\slashed{\Delta}}{2} + m_{q} \Big) \, u(p, \lambda)}{D_{\rm GPD}}  \,, 
\end{equation}
where $g$ denotes the strength of the nucleon-quark-diquark vertex, and
\begin{equation}
D_{\rm GPD} = \bigg[ \Big( k + \frac{\Delta}{2} \Big)^2 - m_q^2 + i \varepsilon \bigg] \, \bigg[ \Big( k - \frac{\Delta}{2} \Big)^2 - m_q^2 + i \varepsilon \bigg] \, \big[ (P - k)^2 - m_s^2 + i \varepsilon \big]\, . 
\label{e:deno_standard}
\end{equation}
In order to obtain the standard GPDs we have used Gordon identities and evaluated the $k^-$-integral by contour integration.
The result for the GPD $H$ can be cast in the form
\begin{eqnarray}
H(x, \xi, t)=
\begin{dcases}
0  &-1 \le x \le -\xi \, ,\\[0.1cm]
\frac{g^{2}(x+\xi)(1+\xi)(1-\xi^{2})}{4(2\pi)^{3}} \int d^{2}\vec{k}_{\perp} \frac{N_{H}}{D_{1} \, D^{-\xi \le x \le \xi}_{2}} &-\xi \le x \le \xi \, ,\\[0.1cm]
\frac{g^{2}(1-x)(1-\xi^{2})}{2(2\pi)^{3}}\int d^{2}\vec{k}_{\perp} \frac{N_{H}}{D_{1} \, D^{x \geq \xi}_{2}} & x \geq \xi \, , 
\end{dcases}
\label{e:GPD_standard}
\end{eqnarray}
and corresponding expressions hold for the other GPDs. 
The following is a compilation of the numerators of all the leading-twist standard GPDs in the SDM: 
\begin{eqnarray}
N_H & = & \vec{k}^{2}_{\perp} + (m_q + x M)^2 + (1 - x)^2 \, \frac{t}{4} - (1 - x) \xi t \, \frac{\vec{k}_\perp \cdot \vec{\Delta}_\perp}{\vec{\Delta}_\perp^2} \,,
\label{e: H_s}\\[0.1cm]
N_E & = & 2 (1 - x) M \bigg[ \, m_q + \bigg( x + 2 \xi \, \frac{\vec{k}_\perp \cdot \vec{\Delta}_\perp}{\vec{\Delta}_\perp^2} \bigg) M \bigg] \,,
\label{e: E_s}\\[0.1cm]
N_{\widetilde{H}} & = & - \, \vec{k}^{2}_{\perp} + (m_q + x M)^2 - (1 - x)^2 \, \frac{t}{4} + \xi \big[ 4 M (m_q + x M) + (1 - x) t \big] \frac{\vec{k}_\perp \cdot \vec{\Delta}_\perp}{\vec{\Delta}_\perp^2} \,,
\label{e: tH_s}\\[0.1cm]
\xi N_{\widetilde{E}} & = & 2 M \bigg[ (1 - x) \xi (m_q + M) + 2 \big[ (1 - \xi^2) m_q + (x - \xi^2) M \big] \frac{\vec{k}_\perp \cdot \vec{\Delta}_\perp}{\vec{\Delta}_\perp^2} \bigg] \,, 
\label{e: tE_s}\\[0.1cm]
N_{H_T} & = & \vec{k}^{2}_{\perp} - 2 \frac{(\vec{k}_\perp \cdot \vec{\Delta}_\perp)^2}{\vec{\Delta}_\perp^2} + (m_q + x M)^2 - (1 - x)^2 \, \frac{t}{4} + \xi \big[ 4 M (m_q + x M) + (1 - x) t \big] \frac{\vec{k}_\perp \cdot \vec{\Delta}_\perp}{\vec{\Delta}_\perp^2} \,,
\label{e: Ht_s} \\[0.1cm]
N_{E_T} & = & 2 M \bigg[ \, 4 M \, \frac{\vec{k}^{2}_{\perp} \vec{\Delta}_\perp^2 - 2 \, (\vec{k}_\perp \cdot \vec{\Delta}_\perp)^2}{ (\vec{\Delta}_\perp^2)^2} + \bigg( 1 - x - 2 \xi \, \frac{\vec{k}_\perp \cdot \vec{\Delta}_\perp}{\vec{\Delta}_\perp^2} \bigg) (m_q + M) \bigg] \,,
\label{e: Et_s} \\[0.1cm]
N_{\widetilde{H}_T} & = & - \, M^2 \bigg[ \, 4 (1 - \xi^2) \, \frac{\vec{k}^{2}_{\perp} \vec{\Delta}_\perp^2 - 2 \, (\vec{k}_\perp \cdot \vec{\Delta}_\perp)^2}{(\vec{\Delta}_\perp^2)^2} + (1 - x) \bigg( 1 - x - 4 \xi \, \frac{\vec{k}_\perp \cdot \vec{\Delta}_\perp}{\vec{\Delta}_\perp^2} \bigg) \bigg] \,,
\label{e: tHt_s}\\[0.1cm]
N_{\widetilde{E}_T} & = & 4 M \bigg[ \, 2 \xi M \, \frac{\vec{k}^{2}_{\perp} \vec{\Delta}_\perp^2 - 2 \, (\vec{k}_\perp \cdot \vec{\Delta}_\perp)^2}{(\vec{\Delta}_\perp^2)^2} - (m_q + x M) \, \frac{\vec{k}_\perp \cdot \vec{\Delta}_\perp}{\vec{\Delta}_\perp^2} \bigg] 
\label{e: tEt_s}\,.
\end{eqnarray}
The denominators in~(\ref{e:GPD_standard}) are given by
\begin{eqnarray}
D_1 & = & (1 + \xi)^2 \vec{k}^{2}_{\perp} + \frac{1}{4} (1 - x)^2 \vec{\Delta}_\perp^2 - (1 - x) (1 + \xi) \vec{k}_\perp \cdot \vec{\Delta}_\perp + (1 - x) (1 + \xi) m_q^2 + (x + \xi) (1 + \xi) m_s^2
\nonumber \\
& & - \, (1 - x) (x + \xi) M^2 \,, {\phantom{\frac{1}{4}}} 
\nonumber \\
D_2^{- \xi \le x \le \xi} & = & \xi (1 - \xi^2) \vec{k}^{2}_{\perp} + \frac{1}{4} (1 - x^2) \xi \vec{\Delta}_\perp^2 + x (1 - \xi^2) \vec{k}_\perp \cdot \vec{\Delta}_\perp + \xi (1 - \xi^2) m_q^2 - \xi (x^2 - \xi^2) M^2 \,,
\nonumber \\
D_2^{x \ge \xi} & = & (1 - \xi)^2 \vec{k}^{2}_{\perp} + \frac{1}{4} (1 - x)^2 \vec{\Delta}_\perp^2 + (1 - x) (1 - \xi) \vec{k}_\perp \cdot \vec{\Delta}_\perp + (1 - x) (1 - \xi) m_q^2 + (x - \xi) (1 - \xi) m_s^2 
\nonumber \\
& & - \, (1 - x) (x - \xi) M^2 \,. {\phantom{\frac{1}{4}}}
\end{eqnarray}
In the above equations we have used the quark mass $m_q$ and the diquark mass $m_s$.
The standard GPDs in the SDM can also be extracted from the results for the generalized transverse momentum dependent parton distributions 
listed in Ref.~\cite{Meissner:2009ww}. 
We reckoned full consistency between the results.
The standard GPDs vanish for $-1 \le x \le -\xi$ due to the absence of antiquarks to ${\cal O}(g^2)$ in the SDM.
We emphasize that the positions of the $k^-$-poles in~(\ref{e:deno_standard}) depend on $x$. 
This leads to different analytical expressions for the standard GPDs in the ERBL and DGLAP regions. 
The GPDs remain continuous at the boundaries $x = \pm \, \xi$ between these regions (see also Ref.~\cite{Bhattacharya:2018zxi}), though their derivatives are discontinuous.
Note also that spectator models typically lead to discontinuous higher-twist GPDs~\cite{Aslan:2018zzk, Aslan:2018tff}.

The GPD $\widetilde{E}$ exhibits a singularity as $\xi \rightarrow 0$ which is why we show $\xi \widetilde{E}$ in Eq.~(\ref{e: tE_s}) and later on for the numerics.
For the chiral-odd GPDs, one has integrals of the type $\int d^{2}\vec{k}_{\perp}k^{i}_{\perp}k^{j}_{\perp} \ldots =A\, \delta^{ij}_{\perp} + B \, \Delta^{i}_{\perp}\Delta^{j}_{\perp}$. 
Such integrals give rise to terms like $\tfrac{\vec{k}^{2}_{\perp}\vec{\Delta}^{2}_{\perp}-2(\vec{k}_{\perp} \cdot \vec{\Delta}_{\perp})^{2}}{(\vec{\Delta}_\perp^2)^2}$ as can be seen in Eqs.~(\ref{e: Ht_s})--(\ref{e: tEt_s}).

Our model results must satisfy the symmetry behavior under the replacement $\xi \rightarrow -\xi$ discussed in Sec.~\ref{sec:GPDs} above.
In order to verify that the results pass this test, it is necessary to replace the integration variable $\vec{k}_{\perp}$ with $- \vec{k}_{\perp}$.
One then finds that the numerators in Eqs.~(\ref{e: H_s})--(\ref{e: tEt_s}) are indeed even under $\xi \rightarrow -\xi$ except the one for $\widetilde{E}_{T}$, which is odd under this transformation.
The analysis of the denominators requires more care.
In order to locate the position of the poles in the complex $k^-$-plane, and hence to arrive at the above expressions of the standard GPDs, we have considered $\xi > 0$. 
Keeping this in mind, one can verify that $\xi \rightarrow -\xi$ switches the position of the poles of the quark propagators only such that the denominators in the ERBL and DGLAP regions are even in $\xi$. 
We also note that our analytical results for the quasi-GPDs below show the exact same behavior under $\xi \rightarrow -\xi$ as the respective standard GPDs.

In the SDM to ${\cal O}(g^2)$, all the leading-twist standard GPDs are UV-finite, except $H$ and $\widetilde{H}$.
(We consider the fact that the chiral-odd GPD $H_{T}$ is UV-finite to be an artifact of the SDM.
In the quark-target model in perturbative QCD this function shows the well-known UV-divergence~\cite{Meissner:2007rx, Xiong:2015nua}.)
For the numerics we impose a cut-off on the transverse quark momenta on all the standard GPDs as well as the (UV-finite) quasi-GPDs.

\subsection{Results for quasi-GPDs}
The quasi-GPD correlator in Eq.~(\ref{e: corr_quasi_GPD}) in the SDM reads
\begin{equation}
F_{\rm Q}^{[\Gamma]}(x, \Delta; \lambda, \lambda'; P^3) = \frac{i \, g^2}{2 (2\pi)^4} \int dk^0 \, d^2\vec{k}_\perp \, \frac{\bar{u}(p', \lambda') \, \Big( \slashed{k} + \frac{\slashed{\Delta}}{2} + m_q \Big) \, \Gamma \, \Big( \slashed{k} - \frac{\slashed{\Delta}}{2} + m_q \Big) \, u(p, \lambda)}{D_{\rm GPD}} \,.
\end{equation}
We again have used Gordon identities to obtain the quasi-GPDs.
Before carrying out the $k^0$-integral one has 
\begin{equation}
H_{{\rm Q}(0/3)}(x, \xi, t; P^3) = \frac{i \, g^2 P^3}{(2\pi)^4} \int dk^0 \, d^2\vec{k}_\perp \, \frac{N_{H(0/3)}}{D_{\rm GPD}} \,,
\end{equation}
and corresponding expressions for the other quasi-GPDs. 
For completeness we first quote the numerators for the unpolarized quasi-GPDs, $H_{\rm Q(0/3)}$ and $E_{\rm Q(0/3)}$, from our previous work~\cite{Bhattacharya:2018zxi}:
\begin{eqnarray}
N_{H(0)} & = & \delta (k^0)^2 - \frac{2}{P^3} \bigg[ \, x (P^3)^2 - m_q M - x \, \frac{t}{4} - \frac{1}{2} \, \delta \xi t \, \frac{\vec{k}_\perp \cdot \vec{\Delta}_\perp}{\vec{\Delta}_\perp^2} \bigg] k^0 
\nonumber \\[0.1cm]
&& + \, \delta \bigg[ \, x^2 (P^3)^2 + \kperpq + m_q^2 + (1 - 2x) \, \frac{t}{4} - \delta \xi t \, \frac{\vec{k}_\perp \cdot \vec{\Delta}_\perp}{\vec{\Delta}_\perp^2} \bigg] \,,
\\[0.1cm]
N_{H(3)} & = & - \, (k^0)^2 + \frac{2}{\delta P^3} \bigg[ \, x \big( (P^3)^2 + M^2 \big) - \frac{t}{4} \bigg] k^0 - x^2 (P^3)^2 + \kperpq + m_q \big( m_q + 2 x M \big) + \frac{t}{4} - (1 - x) \, \frac{\xi t}{\delta} \, \frac{\vec{k}_\perp \cdot \vec{\Delta}_\perp}{\vec{\Delta}_\perp^2} \,, \phantom{aa}
\\[0.1cm]
N_{E(0)} & = & - 2 M \delta \bigg( m_q + x M + 2 M \delta \xi \, \frac{\vec{k}_\perp \cdot \vec{\Delta}_\perp}{\vec{\Delta}_\perp^2}
\bigg) \bigg( \frac{k^0}{\delta P^3} - 1 \bigg) \,,
\\[0.1cm]
N_{E(3)} & = & 2 (1 - x) M \bigg( \frac{M}{\delta P^3} \, k^0 + m_q + 2 \, \frac{M \xi}{\delta} \, \frac{\vec{k}_\perp \cdot \vec{\Delta}_\perp}{\vec{\Delta}_\perp^2} \bigg) \,.
\end{eqnarray}
We now turn to the new results by first considering the case of longitudinal quark polarization, that is, the quasi-GPDs $\widetilde{H}_{\rm Q(0/3)}$ and $\widetilde{E}_{\rm Q(0/3)}$. 
They read
\begin{eqnarray}
N_{\widetilde{H}(0)} &=&-(k^{0})^{2}+2k^{0} \bigg[ x\delta P^{3}-2\xi\frac{\vec{k}_{\perp}\cdot\vec{\Delta}_{\perp}}{\vec{\Delta}^{2}_{\perp}}(1-\delta^{2})P^{3} \bigg]-\vec{k}^{2}_{\perp}-\frac{t}{4}-x^{2}(P^{3})^{2}+2x(1-\delta^{2})(P^{3})^{2}\nonumber \\[0.1cm]
&&+2xM(m_{q}+M)+m^{2}_{q}+4\delta \xi\frac{\vec{k}_{\perp}\cdot\vec{\Delta}_{\perp}}{\vec{\Delta}^{2}_{\perp}} \bigg[(1-\delta^{2})(P^{3})^{2}+M(m_{q}+M)\bigg] \,, 
\\[0.1cm]
N_{\widetilde{H}(3)} &=&\delta (k^{0})^{2}+2\frac{k^{0}}{P^{3}} \bigg[(1-x-\delta^{2})(P^{3})^{2}+M(m_{q}+M) \bigg] + \delta \bigg[-\vec{k}^{2}_{\perp}-\frac{t}{4}+x^{2}(P^{3})^{2}+m^{2}_{q} \bigg]
\nonumber \\[0.1cm]
&&+4\xi\frac{\vec{k}_{\perp}\cdot\vec{\Delta}_{\perp}}{\vec{\Delta}^{2}_{\perp}} \bigg[(1-x)(1-\delta^{2})(P^{3})^{2}+M(m_{q}+M) \bigg] \, , 
\\[0.1cm]
\xi N_{\widetilde{E}(0)} &=&4k^{0}M^{2}\frac{1}{P^{3}}\frac{\vec{k}_{\perp}\cdot\vec{\Delta}_{\perp}}{\vec{\Delta}^{2}_{\perp}}+2M\xi(1-x)(m_{q}+M)-4\delta M\frac{\vec{k}_{\perp}\cdot\vec{\Delta}_{\perp}}{\vec{\Delta}^{2}_{\perp}} \bigg[ M {\xi}^{2}-m_{q}(1 - {\xi}^{2}) \bigg] \,, 
\\[0.1cm]
\xi N_{\widetilde{E}(3)} &=&-2k^{0}\frac{\xi}{P^{3}}M(m_{q}+M)+2\delta \xi M(m_{q}+M)-4 M \frac{\vec{k}_{\perp}\cdot\vec{\Delta}_{\perp}}{\vec{\Delta}^{2}_{\perp}} \bigg[ M({\xi}^{2}-x)-m_{q}(1-\xi^{2}) \bigg] \,.
\end{eqnarray}
Note that the quasi-GPDs $\widetilde{E}_{\rm Q(0/3)}$ have a pole at $\xi=0$, just like their light-cone counterpart. 
We next list the numerators of the quasi-GPDs that appear for transverse quark polarization:
\begin{eqnarray}
N_{H_{T}(0)} &=& \delta (k^{0})^{2} -\frac{k^{0}}{P^{3}} \bigg [ \frac{\vec{\Delta}^{2}_{\perp}}{2} -2m_{q} M + \bigg ( 2x-2\xi^{2}(1-\delta^{2}) \bigg ) (P^{3})^{2} \bigg ] + \delta \bigg [ \vec{k}^{2}_{\perp} -2 \frac{(\vec{k}_{\perp}\cdot \vec{\Delta}_{\perp})^{2}}{\vec{\Delta}^{2}_{\perp}} +\frac{\vec{\Delta}^{2}_{\perp}}{4} +m^{2}_{q} \bigg ] 
\nonumber \\[0.1cm]
&& + \delta \bigg [ x^{2} - \xi^{2} (1-\delta^{2})\bigg ] (P^{3})^{2} -4 \xi \frac{\vec{k}_{\perp}\cdot \vec{\Delta}_{\perp}}{\vec{\Delta}^{2}_{\perp}} \bigg [(x-\xi^{2})(1-\delta^{2})(P^{3})^{2} -m_{q}M+\frac{\vec{\Delta}^{2}_{\perp}}{4} \bigg ] \,, 
\\[0.1cm]
N_{H_{T}(3)} &=&-(k^{0})^{2}-2k^{0} \bigg[2\xi(1-\delta^{2})P^{3}\,\frac{\vec{k}_{\perp}\cdot\vec{\Delta}_{\perp}}{\Delta^{2}_{\perp}}-x\delta P^{3}\bigg]\nonumber \\[0.1cm]
&&+\vec{k}^{2}_{\perp}-2\frac{( \vec{k}_{\perp}\cdot\vec{\Delta}_{\perp})^{2}}{\vec{\Delta}^{2}_{\perp}}+(1-2x)\frac{\vec{\Delta}^{2}_{\perp}}{4}+m^{2}_{q}+2xm_{q}M-(P^{3})^{2} \bigg[x^{2}+(1-2x)\xi^{2}(1-\delta^{2}) \bigg]
\nonumber \\[0.1cm]
&&-4\delta \xi\frac{\vec{k}_{\perp}\cdot\vec{\Delta}_{\perp}}{\Delta^{2}_{\perp}} \bigg[\frac{\vec{\Delta}^{2}_{\perp}}{4}-m_{q}M-\xi^{2}(1-\delta^{2})(P^{3})^{2} \bigg] \,, 
\\[0.1cm]
N_{\widetilde{H}_{T}(0)}&=&\delta N_{\widetilde{H}_{T}(3)} \,,
\label{e: tHt_related_form}
\\[0.1cm]
N_{\widetilde{H}_{T}(3)}& = &- 4M^{2}(1-\xi^{2}) \frac{\vec{k}^{2}_{\perp} \vec{\Delta}_\perp^2 - 2 \, (\vec{k}_\perp \cdot \vec{\Delta}_\perp)^2}{ (\vec{\Delta}_\perp^2)^2} - \frac{k^{0}}{\delta P^{3}} \bigg[ 2\delta \xi \frac{\vec{k}_{\perp}\cdot\vec{\Delta}_{\perp}}{\vec{\Delta}^{2}_{\perp}}-(1-x)\bigg]M^{2} 
\nonumber \\[0.1cm]
&& +2\frac{\xi}{\delta}\frac{\vec{k}_{\perp}\cdot\vec{\Delta}_{\perp}}{\vec{\Delta}^{2}_{\perp}} \bigg[(1-x)+\delta^{2} \bigg]M^{2}-(1-x)M^{2} \,,
\\[0.1cm]
N_{E_{T}(0)} &=& 8\delta M^{2} \frac{\vec{k}^{2}_{\perp} \vec{\Delta}_\perp^2 - 2 \, (\vec{k}_\perp \cdot \vec{\Delta}_\perp)^2}{ (\vec{\Delta}_\perp^2)^2} - 2\delta M(m_{q}+M) \bigg[\frac{k^{0}}{\delta P^{3}}-1 \bigg]-4\xi M(m_{q}+M)\frac{\vec{k}_{\perp}\cdot\vec{\Delta}_{\perp}}{\vec{\Delta}^{2}_{\perp}}\, , \\[0.1cm]
N_{E_{T}(3)} &=& 8 M^{2} \frac{\vec{k}^{2}_{\perp} \vec{\Delta}_\perp^2 - 2 \, (\vec{k}_\perp \cdot \vec{\Delta}_\perp)^2}{ (\vec{\Delta}_\perp^2)^2}
+2(1-x)M(m_{q}+M)-4\delta \xi M(m_{q}+M)\frac{\vec{k}_{\perp}\cdot\vec{\Delta}_{\perp}}{\vec{\Delta}^{2}_{\perp}}\, ,
\\[0.1cm]
N_{\widetilde{E}_{T}(0)}&=& 8\delta^{2} \xi M^{2} \frac{\vec{k}^{2}_{\perp} \vec{\Delta}_\perp^2 - 2 \, (\vec{k}_\perp \cdot \vec{\Delta}_\perp)^2}{ (\vec{\Delta}_\perp^2)^2} - 4\delta M(m_{q}+x M)\frac{\vec{k}_{\perp}\cdot\vec{\Delta}_{\perp}}{\vec{\Delta}^{2}_{\perp}}\, , \\[0.1cm]
N_{\widetilde{E}_{T}(3)}&=& 8\frac{\xi}{\delta} M^{2} \frac{\vec{k}^{2}_{\perp} \vec{\Delta}_\perp^2 - 2 \, (\vec{k}_\perp \cdot \vec{\Delta}_\perp)^2}{ (\vec{\Delta}_\perp^2)^2} -4 M(m_{q}+\frac{k^{0}}{\delta P^{3}}M)\frac{\vec{k}_{\perp}\cdot\vec{\Delta}_{\perp}}{\vec{\Delta}^{2}_{\perp}} \,.
\end{eqnarray}
The quasi-GPDs $\widetilde{H}_{T, {\rm Q(0)}}$ and $\widetilde{H}_{T, {\rm Q(3)}}$ corresponding to two different Dirac structures are related through Eq.~(\ref{e: tHt_related_form}). 
This is the only quasi-GPD whose two different projections have such a simple relation. 
We repeat that all quasi-GPDs have support in the range $-\infty < x < \infty$. 
However, for large $P^3$ they are all power-suppressed outside the region $- \xi \leq x \leq 1$.
We also observe that the numerators of the quasi-GPDs $E_{T, {\rm Q(3)}}$ and $\widetilde{E}_{T, {\rm Q(0)}}$ are the only ones that do not depend on $k^{0}$. 

The denominator $D_{\rm GPD}$ can be written as
\begin{eqnarray}
D_{\rm GPD} = (k^{0}-k^{0}_{1+})(k^{0}-k^{0}_{1-})(k^{0}-k^{0}_{2+})(k^{0}-k^{0}_{2-})(k^{0}-k^{0}_{3+})(k^{0}-k^{0}_{3-}) \,,
\end{eqnarray}
where the poles from the quark propagators, with 4-momenta $(k-\tfrac{\Delta}{2})$ and $(k+\tfrac{\Delta}{2})$, and from the spectator propagator are given by
\begin{eqnarray}
k^{0}_{1\pm} &=& -\xi P^{3} \pm \sqrt{(x+\delta \xi)^{2}(P^{3})^{2}+ \bigg (\vec{k}_{\perp}-\dfrac{\vec{\Delta}_{\perp}}{2} \bigg )^{2} + m^{2}_{q} -i\varepsilon} \,, 
\\[0.1cm]
k^{0}_{2\pm} &=& \xi P^{3} \pm \sqrt{(x-\delta \xi)^{2}(P^{3})^{2}+ \bigg (\vec{k}_{\perp}+\dfrac{\vec{\Delta}_{\perp}}{2} \bigg )^{2} + m^{2}_{q} -i\varepsilon} \,, 
\\[0.1cm]
k^{0}_{3\pm} &=& \delta P^{3} \pm \sqrt{(1-x)^{2}(P^{3})^{2}+\vec{k}^{2}_{\perp}+m^{2}_{s}-i\varepsilon} \,.
\end{eqnarray}
It is important to realize that, while the position of the poles depends on $x$, they never switch half planes.   
Specifically, $k^{0}_{1-}$, $k^{0}_{2-}$ and $k^{0}_{3-}$ always lie in the upper half plane, while the other three poles lie in the lower half plane. 
After performing the $k^0$-integral, one therefore has the same functional form for the quasi-GPDs for any $x$, which implies that all quasi-GPDs  are continuous as a function of $x$ --- in this context, see also Ref.~\cite{Bhattacharya:2018zxi}.

\subsection{Recovering standard GPDs from quasi-GPDs}
We have checked that for $P^{3} \to \infty$ the analytical results of all quasi-GPDs reduce to the ones of the respective standard GPDs.
Here we provide the most important steps involved in this test.
We start with the poles of the propagators, which can be expanded as
\begin{eqnarray}
& k^{0}_{1+}  = 
\begin{dcases}
xP^{3}+\frac{1}{2(x+\xi)P^{3}} \bigg [ \vec{k}^{2}_{\perp}-\vec{k}_{\perp} \cdot \vec{\Delta}_{\perp} -\frac{t}{4}(x \, \xi +1) + m^{2}_{q}+ x \, \xi M^{2} -i\varepsilon \bigg ] + \mathcal{O}\bigg (\frac{1}{(P^{3})^{2}} \bigg ) & x \geq -\xi \, , \\[0.1cm]
-(x+2 \xi)P^{3}-\frac{1}{2(x+\xi)P^{3}} \bigg [ \vec{k}^{2}_{\perp}-\vec{k}_{\perp} \cdot \vec{\Delta}_{\perp}-\frac{t}{4}(x \, \xi +1) +m^{2}_{q}+ x \, \xi M^{2} -i\varepsilon \bigg ]+ \mathcal{O}\bigg (\frac{1}{(P^{3})^{2}} \bigg ) & x \leq -\xi \, ,
\end{dcases}
\\[0.4cm]
& k^{0}_{1-} =
\begin{dcases}
-(x+2 \xi)P^{3}-\frac{1}{2(x+\xi)P^{3}} \bigg [ \vec{k}^{2}_{\perp}-\vec{k}_{\perp} \cdot \vec{\Delta}_{\perp}-\frac{t}{4}(x \, \xi +1) +m^{2}_{q}+ x \, \xi M^{2} -i\varepsilon \bigg ]+ \mathcal{O}\bigg (\frac{1}{(P^{3})^{2}} \bigg ) & x \geq -\xi \, , \\[0.1cm]
xP^{3}+\frac{1}{2(x+\xi)P^{3}} \bigg [ \vec{k}^{2}_{\perp}-\vec{k}_{\perp} \cdot \vec{\Delta}_{\perp} -\frac{t}{4}(x \, \xi +1) + m^{2}_{q}+ x \, \xi M^{2} -i\varepsilon \bigg ] + \mathcal{O}\bigg (\frac{1}{(P^{3})^{2}} \bigg ) & x \leq -\xi \, , 
\end{dcases}
\\[0.4cm]
& k^{0}_{2+} =
\begin{dcases}
xP^{3}+\frac{1}{2(x-\xi)P^{3}} \bigg [ \vec{k}^{2}_{\perp}+\vec{k}_{\perp} \cdot \vec{\Delta}_{\perp} +\frac{t}{4}(x \, \xi -1) + m^{2}_{q}- x \, \xi M^{2} -i\varepsilon \bigg ] + \mathcal{O}\bigg (\frac{1}{(P^{3})^{2}} \bigg ) & x \geq \phantom{+} \xi \, , \\[0.1cm]
-(x-2 \xi)P^{3}+\frac{1}{2(\xi - x)P^{3}} \bigg [ \vec{k}^{2}_{\perp}+\vec{k}_{\perp} \cdot \vec{\Delta}_{\perp}+\frac{t}{4}(x \, \xi -1) +m^{2}_{q}- x \, \xi M^{2} -i\varepsilon \bigg ]+ \mathcal{O}\bigg (\frac{1}{(P^{3})^{2}} \bigg ) & x \leq \phantom{+} \xi \, ,
\end{dcases}
\\[0.4cm]
& k^{0}_{2-} =
\begin{dcases}
-(x-2 \xi)P^{3}-\frac{1}{2(x - \xi)P^{3}} \bigg [ \vec{k}^{2}_{\perp}+\vec{k}_{\perp} \cdot \vec{\Delta}_{\perp}+\frac{t}{4}(x \, \xi -1) +m^{2}_{q}- x \, \xi M^{2} -i\varepsilon \bigg ]+ \mathcal{O}\bigg (\frac{1}{(P^{3})^{2}} \bigg ) & x \geq \phantom{+} \xi \, , \\[0.1cm]
xP^{3}-\frac{1}{2(\xi -x )P^{3}} \bigg [ \vec{k}^{2}_{\perp}+\vec{k}_{\perp} \cdot \vec{\Delta}_{\perp} +\frac{t}{4}(x \, \xi -1) + m^{2}_{q}- x \, \xi M^{2} -i\varepsilon \bigg ] + \mathcal{O}\bigg (\frac{1}{(P^{3})^{2}} \bigg ) & x \leq \phantom{+} \xi \, , 
\end{dcases}
\\[0.4cm]
& k^{0}_{3+} =
\begin{dcases}
xP^{3} +\frac{1}{2(x-1)P^{3}} \bigg [ \vec{k}^{2}_{\perp}+\frac{t}{4}(1-x)-(1-x)M^{2}+m^{2}_{s}-i\varepsilon \bigg ] + \mathcal{O}\bigg (\frac{1}{(P^{3})^{2}} \bigg ) \phantom{..............} & x \geq \phantom{+} 1 \, , \\[0.1cm]
-(x-2)P^{3}+\frac{1}{2(1-x)P^{3}} \bigg [ \vec{k}^{2}_{\perp} -\frac{t}{4}(1-x)+(1-x)M^{2}+m^{2}_{s}-i\varepsilon \bigg ] + \mathcal{O}\bigg (\frac{1}{(P^{3})^{2}} \bigg ) \phantom{..............} & x \leq \phantom{+} 1 \, ,
\end{dcases} 
\\[0.4cm]
& k^{0}_{3-} =
\begin{dcases}
-(x-2)P^{3}+\frac{1}{2(x-1)P^{3}} \bigg [ -\vec{k}^{2}_{\perp} +\frac{t}{4}(1-x)-(1-x)M^{2}-m^{2}_{s}+i\varepsilon \bigg ] + \mathcal{O}\bigg (\frac{1}{(P^{3})^{2}} \bigg ) \phantom{.........} & x \geq \phantom{+} 1 \, \\[0.1cm]
xP^{3} +\frac{1}{2(1-x)P^{3}} \bigg [ -\vec{k}^{2}_{\perp}-\frac{t}{4}(1-x)+(1-x)M^{2}-m^{2}_{s}+i\varepsilon \bigg ] + \mathcal{O}\bigg (\frac{1}{(P^{3})^{2}} \bigg ) \phantom{...........} & x \leq \phantom{+} 1 \, ,
\end{dcases}
\end{eqnarray}
It is evident from these equations that the analytical expressions of the expansions of the poles depend on $x$, but the poles always lie in the same half plane, as already discussed above.

In the following we focus on the quasi-GPD $H_{\rm Q(0)}$. 
We first note that the dominant contribution is from those residues for which the leading order term is $x \, P^{3}$. 
Specifically, for the other residues the numerator of $H_{\rm Q(0)}$ has a leading contribution of order $(P^{3})^{3}$, while the leading contribution of the denominator is of order $(P^{3})^{5}$, resulting in an overall suppression like $1/(P^{3})^{2}$.
For $x \leq -\xi$ we close the integration contour in the lower half plane.
Then none of the poles have $x \, P^{3}$ as leading term, which then leads to a power-suppressed contribution.
A corresponding discussion applies for $x \ge 1$ if one closes the integration contour in the upper half plane.

For the DGLAP region ($x \geq \xi$), we close the integration contour in the upper half plane.
Then the dominant contribution comes from the residue at the pole $k^{0}_{3-}$.  
Therefore in that region
\begin{equation}
\lim_{P^{3} \rightarrow \infty} H_{\rm Q (0)} =  - \lim_{P^{3} \rightarrow \infty} \frac{g^{2}\, P^{3}}{(2\pi)^{3}} \int d^{2}\vec{k}_{\perp} \frac{N_{H(0)}(k^{0}_{3-})}{(k^{0}_{3-}-k^{0}_{1+})(k^{0}_{3-}-k^{0}_{1-})(k^{0}_{3-}-k^{0}_{2+})(k^{0}_{3-}-k^{0}_{2-})(k^{0}_{3-}-k^{0}_{3+})} \, .
\label{e: matching_HQ0_DGLAP}
\end{equation}
We first determine the leading term of the numerator in (\ref{e: matching_HQ0_DGLAP}) which is given by
\begin{eqnarray}
N_{H(0)}(k^{0}_{3-}) & = & \delta \, (k^{0}_{3-})^{2} -2\frac{k^{0}_{3-}}{P^{3}} \bigg [ x(P^{3})^{2} -m_{q} M -x \frac{t}{4}-\frac{\delta \xi t}{2}\frac{\vec{k}_{\perp} \cdot \vec{\Delta}_{\perp}}{\vec{\Delta}^{2}_{\perp}} \bigg ] \nonumber \\[0.1cm]
&& + \, \delta \bigg [ x^{2} (P^{3})^{2} + \vec{k}^{2}_{\perp}+m^{2}_{q}+(1-2x) \frac{t}{4}-\delta \xi t \frac{\vec{k}_{\perp} \cdot \vec{\Delta}_{\perp}}{\vec{\Delta}^{2}_{\perp}} \bigg ] \,. 
\label{e: matching_HQ0_numerator_DGLAP}
\end{eqnarray}
Then using  
\begin{eqnarray}
\delta (k^{0}_{3-})^{2} & = &  \Bigg ( 1+ \dfrac{M^{2} -\dfrac{t}{4}}{2 (P^{3})^{2}} \Bigg ) \times x^{2} (P^{3})^{2} \Bigg ( 1-\dfrac{\vec{k}^{2}_{\perp}+\dfrac{t}{4}(1-x) -(1-x)M^{2}+m^{2}_{s}\,}{2 x (1-x) (P^{3})^{2}} \Bigg )^{2} + \ldots
\nonumber \\[0.1cm]
& = & x^{2}(P^{3})^{2} +\frac{1}{2} x^{2} \bigg (M^{2} -\frac{t}{4} \bigg ) -\frac{x}{(1-x)}\bigg ( \vec{k}^{2}_{\perp}+\frac{t}{4}(1-x) -(1-x)M^{2}+m^{2}_{s} \bigg ) + \ldots \, ,
\end{eqnarray}
where $\ldots$ indicates suppressed terms, and 
\begin{eqnarray}
2\frac{k^{0}_{3-}}{P^{3}}  \approx  2x + \frac{1}{(1-x)(P^{3})^{2}}\bigg ( \vec{k}^{2}_{\perp}+\frac{t}{4}(1-x) -(1-x)M^{2}+m^{2}_{s} \bigg ) + \ldots \, ,
\end{eqnarray}
provides
\begin{eqnarray}
N_{H(0)}(k^{0}_{3-}) & = & x^{2}(P^{3})^{2} +\frac{1}{2} x^{2} \bigg (M^{2} -\frac{t}{4} \bigg ) -\frac{x}{(1-x)}\bigg ( \vec{k}^{2}_{\perp}+\frac{t}{4}(1-x) -(1-x)M^{2}+m^{2}_{s} \bigg )  
\nonumber \\[0.1cm]
&& -2x^{2}(P^{3})^{2} + \frac{x}{(1-x)}\bigg (\vec{k}^{2}_{\perp}+\dfrac{t}{4}(1-x) -(1-x)M^{2}+m^{2}_{s}\bigg ) +2x \big (m_{q}M +x \frac{t}{4}) + x \xi t \frac{\vec{k}_{\perp}\cdot \vec{\Delta}_{\perp}}{\vec{\Delta}^{2}_{\perp}} 
\nonumber \\[0.1cm]
&& + x^{2}(P^{3})^{2} + \frac{1}{2} x^{2} \bigg (M^{2} -\frac{t}{4} \bigg ) + \vec{k}^{2}_{\perp} +m^{2}_{q} +(1-2x)\frac{t}{4} -\xi t \frac{\vec{k}_{\perp}\cdot \vec{\Delta}_{\perp}}{\vec{\Delta}^{2}_{\perp}} + \ldots 
\nonumber \\[0.1cm]
&& = N_{H} + \ldots \,.
\label{e: matching_HQ0_numerator_result_DGLAP}
\end{eqnarray}
On the other hand, the denominator in (\ref{e: matching_HQ0_DGLAP}) simplifies as
\begin{align}
& (k^{0}_{3-}-k^{0}_{1+})(k^{0}_{3-}-k^{0}_{1-})(k^{0}_{3-}-k^{0}_{2+})(k^{0}_{3-}-k^{0}_{2-})(k^{0}_{3-}-k^{0}_{3+}) 
\nonumber \\[0.1cm]
& =  -8(P^{3})^{3} (x^{2}-\xi^{2})(1-x)(k^{0}_{3-}-k^{0}_{1+})(k^{0}_{3-}-k^{0}_{2+}) + \ldots 
\nonumber \\[0.1cm]
& = -\dfrac{2P^{3}}{(1-x)(1-\xi^{2})} D_{1}D^{x \geq \xi}_{2} + \dots \, .
\label{e: matching_HQ0_denominator_result_DGLAP}
\end{align}
Using Eqs.~(\ref{e: matching_HQ0_numerator_result_DGLAP}) and (\ref{e: matching_HQ0_denominator_result_DGLAP}) in Eq.~(\ref{e: matching_HQ0_DGLAP}), one readily confirms
\begin{equation}
\lim_{P^{3} \rightarrow \infty} H_{\rm Q(0)} = \dfrac{g^{2}(1-x)(1-\xi^{2})}{2(2\pi)^{3}}\int d^{2}\vec{k}_{\perp} \dfrac{N_{H}}{D_{1} \, D^{x \geq \xi}_{2}} = H \, .
\end{equation}
The overall logic to analytically recover $H$ in the ERBL region ($-\xi \leq x \leq \xi$)\, remains the same as discussed above. 
In this case it is convenient to close the integration contour in the lower half plane, so that the dominant contribution comes from the residue at $k^{0}_{1+}$ only. 
With a very similar analysis we have shown that all the quasi-GPDs reduce to the corresponding standard GPDs in the large-$P^{3}$ limit.

\subsection{Results for quasi-PDFs}
Starting from the expressions of the standard GPDs and taking $\Delta=0$ (which implies $\xi = t = 0$), one obtains the following expressions for the standard PDFs: 
\begin{eqnarray}
f_{1}(x) = H(x,0,0) &=&\frac{g^{2}(1-x)}{2(2\pi)^{3}}\int d^{2}\vec{k}_{\perp}\frac{\vec{k}^{2}_{\perp}+(m_{q}+xM)^{2}}{[\vec{k}^{2}_{\perp}+xm^{2}_{s}+(1-x)m^{2}_{q}-x(1-x)M^{2}]^{2}}\, , 
\label{e:f1}\\ [0.1cm]
g_{1}(x) = \tilde{H}(x,0,0) &=&\frac{g^{2}(1-x)}{2(2\pi)^{3}}\int d^{2}\vec{k}_{\perp}\frac{-\vec{k}^{2}_{\perp}+(m_{q}+xM)^{2}}{[\vec{k}^{2}_{\perp}+xm^{2}_{s}+(1-x)m^{2}_{q}-x(1-x)M^{2}]^{2}}\, ,\label{e:g1}\\ [0.1cm]
h_{1}(x) = H_{T}(x,0,0) &=&\frac{g^{2}(1-x)}{2(2\pi)^{3}}\int d^{2}\vec{k}_{\perp}\frac{(m_{q}+xM)^{2}}{[\vec{k}^{2}_{\perp}+xm^{2}_{s}+(1-x)m^{2}_{q}-x(1-x)M^{2}]^{2}} \label{e:h1}\, .
\end{eqnarray}
Only three GPDs survive in this limit --- $E$, $\widetilde{E}$, $E_{T}$ and $\widetilde{H}_{T}$ vanish because $\Delta$ appears in their prefactor in the parameterizations in~(\ref{e:U_s}), (\ref{e:L_s}) and (\ref{e:T_s}), while $\widetilde{E}_{T}$ drops out since $\bar{u}'\gamma^{i}_{T}u$ vanishes in the forward limit. 
The GPD $H$ reduces to the unpolarized PDF $f_{1}$, whereas $\widetilde{H}$ reduces to the helicity PDF $g_{1}$, and $H_{T}$ to the transversity PDF $h_{1}$.
Our results for the forward PDFs agree with the ones published in Ref.~\cite{Meissner:2007rx}. 
In general, like for standard GPDs, the region of support for PDFs is $-1 \le x \le 1$.
In the SDM to $\mathcal{O}(g^{2})$, they also vanish for $-1 \le x < 0$. 
Below we give a separate discussion for the point $x = 0$, where the forward PDFs in the SDM are discontinuous.
  
For the quasi-PDFs one has
\begin{equation}
f_{1,\rm Q(0/3)}(x; P^3) = \frac{i g^{2} P^{3}}{(2\pi)^{4}} \int dk^{0} \, d^{2}\vec{k}_\perp \, 
\frac{N_{f1(0/3)}}{D_{\rm PDF}} \,, 
\label{e:f1Q_SDM}
\end{equation}
and corresponding expressions for the other quasi-PDFs. 
The numerators are given by
\begin{eqnarray}
N_{f1(0)}&=&\delta_{0}(k^{0})^{2} - \frac{2k^{0}}{P^{3}} \Big( x(P^{3})^{2} - m_q M \Big) + \delta_{0} \Big(\vec{k}^{2}_{\perp}+x^{2}(P^{3})^{2} + m^{2}_{q} \Big)\, , \label{e:f1_q} \\[0.1cm] 
N_{f1(3)}&=&-(k^0)^2 + k^{0}\Big(2x\delta_{0}P^{3}\Big) + \vec{k}^{2}_{\perp}- x^{2}(P^{3})^{2}+ m_q \Big( m_q + 2xM \Big)\, ,
\\[0.1cm]
N_{g1(0)}&=&-(k^0)^2 + k^{0}\Big(2x\delta_{0}P^{3}\Big) - \vec{k}^{2}_{\perp}- x^{2}(P^{3})^{2}+ m_q \Big( m_q + 2xM \Big)\, , \\ [0.1cm]
N_{g1(3)}&=&\delta_{0}(k^{0})^{2} - \frac{2k^{0}}{P^{3}} \Big( x(P^{3})^{2} - m_q M \Big) + \delta_{0} \Big(-\vec{k}^{2}_{\perp}+x^{2}(P^{3})^{2} + m^{2}_{q} \Big)\, ,
\\[0.1cm]
N_{h1(0)}&=&\delta_{0}(k^{0})^{2}-\frac{2k^{0}}{P^{3}}\Big(x(P^{3})^{2}-m_{q}M\Big)+\delta_{0}\Big(x^{2}(P^{3})^{2}+m^{2}_{q}\Big)\, , \\ [0.1cm]
N_{h1(3)}&=&-(k^{0})^{2}+k^{0}\Big(2x\delta_{0}P^{3}\Big)-x^{2}(P^{3})^{2}+m_{q}\Big(m_{q}+2xM\Big)\, ,
\label{e:h1_q}
\end{eqnarray}
and the denominator reads
\begin{equation}
D_{\rm PDF} = \big[k^{2} - m^{2}_{q} + i \varepsilon \big]^{2} \, \big[ (P - k)^{2} - m^{2}_{s} + i \varepsilon \big] \,.
\end{equation}
The results for the quasi-PDFs follow directly from the ones for the quasi-GPDs.
In Eqs.~(\ref{e:f1_q})--(\ref{e:h1_q}) we have used $\delta_{0} = \delta (t=0)$. 
Like for quasi-GPDs, the support range of quasi-PDFs is $-\infty < x < \infty$. 
The process of analytically recovering standard PDFs from the corresponding quasi-PDFs has been discussed in Ref.~\cite{Bhattacharya:2018zxi}. 
Results for the quasi-PDFs associated with the gamma matrices $\gamma^{3}/\gamma^{3}\gamma_{5}/i\sigma^{j3}\gamma_{5}$ were already presented in~\cite{Gamberg:2014zwa}, but in the so-called cut-graph approximation. 
In Ref.~\cite{Bhattacharya:2018zxi}, we have discussed the differences of that approach compared to a full calculation that includes all contributions.
Note that we have calculated all the forward distributions independently using a trace technique, and have found complete agreement with the results obtained from the quasi-GPDs.

\subsection{The point $x=0$ for standard PDFs and quasi-PDFs}
In the SDM all three standard PDFs are discontinuous at $x = 0$.
(We have argued in Ref.\cite{Bhattacharya:2018zxi} that for $f_1$ this discontinuity may not be an artifact of the model.)
Specifically, in the case of $f_1$ one has
\begin{equation}
\lim_{\varepsilon \rightarrow 0} f_1(-\varepsilon) = 0 \,, \qquad
\lim_{\varepsilon \rightarrow 0} f_1(\varepsilon) = \frac{g^{2}}{2 (2\pi)^{3}} \int d^{2} \vec{k}_{\perp} \dfrac{1}{(\vec{k}^{2}_{\perp}+m^{2}_{q})} \,.
\end{equation}
Also, for $x = 0$ the contour integration that can be used for any other value of $x$ does not work.
Two questions arise at this point.
Can one still assign an unambiguous value to the standard PDFs for $x = 0\,$?
And, if so, does the corresponding quasi-PDF reproduce that value for $P^3 \to \infty\,$?
One readily verifies that using the well-known identity
\begin{equation}
\int dk^- \frac{1}{k^- - k_{\rm pole}^- + i \varepsilon} = {\rm PV} \int dk^- \frac{1}{k^- - k_{\rm pole}^-} - i \pi \delta(k^- - k_{\rm pole}^-)
\end{equation}
allows one to compute the standard PDFs for $x = 0$.
In the case of $f_1$ one finds
\begin{equation}
f_{1}(x=0) = \frac{g^{2}}{4 (2\pi)^{3}} \int d^{2} \vec{k}_{\perp} \dfrac{1}{(\vec{k}^{2}_{\perp}+m^{2}_{q})} 
= \frac{1}{2} \lim_{\varepsilon \rightarrow 0} \Big( f_1(- \varepsilon) + f_1(\varepsilon) \Big) \,.
\label{e:f1_x0}
\end{equation}
Corresponding equations hold for $g_1$ and $h_1$.
We next investigate if the quasi-PDFs $f_{\rm Q(0/3)}$ analytically reproduces the result in Eq.~(\ref{e:f1_x0}).
It turns out that this is indeed true, that is,
\begin{eqnarray}
\lim\limits_{P^{3} \rightarrow \infty} f_{1 \rm Q(0/3)}(x=0) = f_1(x = 0) \,.
\end{eqnarray}
The very same conclusion applies to $g_1$ and $h_1$.
It is interesting that the quasi-PDFs reproduce exactly all the features of the corresponding standard PDFs around the point where the latter are discontinuous.

Before closing this section, we take up the impact of including a form factor (rather than a $\vec{k}_{\perp}$ cut-off) on the discontinuity (continuity) feature exhibited by the standard PDFs (quasi-PDFs). As an example, we study the impact of the form factor $\mathcal{I}(k) = k^{2}-m^{2}_{q}/(k^{2}-\Lambda^{2})^{2}$ on the PDF $f_{1}$, as was done in Ref.~\cite{Bacchetta:2008af}. By performing the contour integration and then setting $x=0$, we find
\begin{equation}
f_{1}(0^{+}) = \dfrac{g^{2}}{24(2\pi)^{2}} \dfrac{1}{\Lambda^{4}}\bigg ( 1+ 2\dfrac{m^{2}_{q}}{\Lambda^{2}} \bigg ) \, .
\end{equation}
Note that reversing the order of operations -- setting $x=0$ and then performing the contour integration -- provides half of this result. Regardless, the discontinuity at $x=0$ persists but decreases as $\Lambda$ gets larger. Since the continuity property of the quasi-PDFs is unaffected by the inclusion of such a multiplicative form factor in our analytical expressions, discrepancies between standard and quasi-PDFs will still be prominent in the small $x$ region. Furthermore, in Ref.~\cite{Gamberg:2014zwa}, a cut-graph model with the same form factor was used to study the up and down quark distributions in the proton considering contributions from both scalar and axial-vector diquarks. Substantial discrepancies were observed between the quasi and standard distributions at large $x$ values. Since for not too small $x$ values, calculation of quasi-PDFs with or without on-shell diquarks does not lead to huge qualitative differences (see Fig. 5 in Ref.~\cite{Bhattacharya:2018zxi}), we emphasize that considerable differences will persist at large $x$ in our case as well (where the calculation is performed with the spectator off-shell). We therefore expect similar conclusions for the GPDs as well, although one cannot simply resort to doing calculations for standard and quasi-GPDs with spectator on-shell (since one needs to pick up a quark pole to get the ERBL region). To this end, we emphasize that there is no clear prescription for using form factors to regulate the UV divergences as the model no longer follows from a Lagrange density.

\section{Numerical Results in Scalar Diquark Model}
\label{sec:numerical_results}
For the numerical analysis we proceed along the lines of our previous work~\cite{Bhattacharya:2018zxi}. 
For completeness we first repeat the numerical values of the parameters.
We use $g = 1$ for the strength of the nucleon-quark-diquark coupling.
None of the general conclusions depend on the precise value of $g$. 
Our ``standard values" for the mass parameters are $M = 0.939 \, \textrm{GeV}$, $m_s = 0.7 \, \textrm{GeV}$ and $m_q = 0.35 \, \textrm{GeV}$. 
Elaborating on our choice of parameters, we mention that our starting point comes from the Ref.~\cite{Bacchetta:2008af} where the value $m_q = 0.3 \, \rm GeV$ was chosen and the value $m_s = 0.822 \, \rm GeV$ was found from the phenomenological fits of Refs.~\cite{Chekanov:2002pv,Gluck:2000dy}. We have therefore chosen values similar to these, but have adjusted so that the convergence of the quasi distributions to the standard distributions is maximal. From Fig. 4 in~\cite{Bhattacharya:2018zxi}, one can see that the relative difference between the quasi and standard
$f_1$ is quite sensitive to adjustments in $m_s$. Thus it was important to reduce the value of $m_s$ to $m_s = 0.7 \, \rm GeV$ to achieve the best convergence. One can also see that the convergence is barely affected by changes in $m_q$, so we increase the value of $m_q$ to
$m_q = 0.35 \, \rm GeV$ to satisfy $m_s + m_q > M$. In short, after exploring the sensitivity of our results to variations in $m_s$ and $m_q$, we maintain that such a choice of the parameters, as discussed at length in Ref.~\cite{Bhattacharya:2018zxi}, is ``optimal" with regard to the question of convergence of the quasi-distributions to their light-cone counterparts.
For most of our plots, the cut-off for the $|\vec{k}_{\perp}|$ integration is $\Lambda = 1 \, \textrm{GeV}$, and the transverse momentum transfer is $|\vec{\Delta}_\perp| = 0$.
We also shall show some plots and comment extensively on the dependence of the various distributions on $\Lambda$ and $|\vec{\Delta}_{\perp}|$.
We begin with discussing the PDFs.

\subsection{Results for quasi-PDFs}
\begin{figure}[t]
\includegraphics[width=6.5cm]{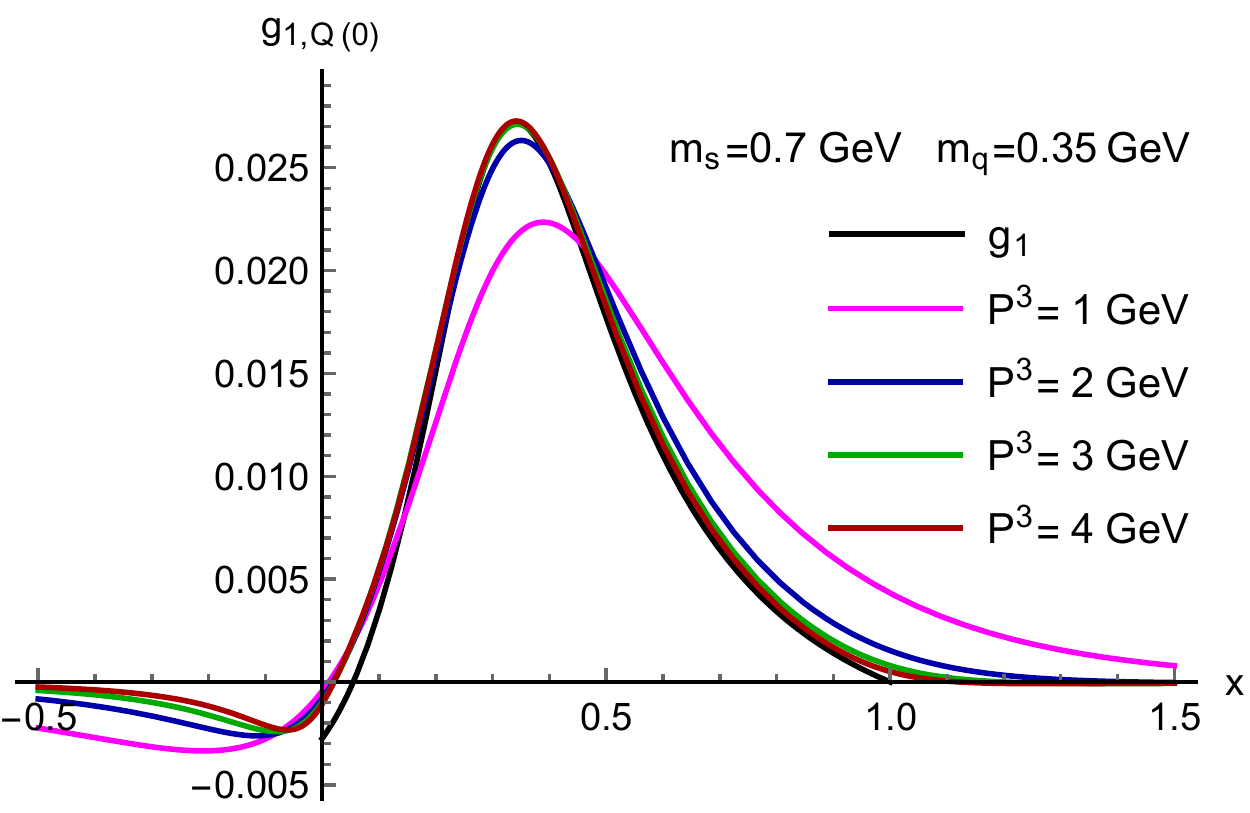}
\hspace{1.5cm}
\includegraphics[width=6.5cm]{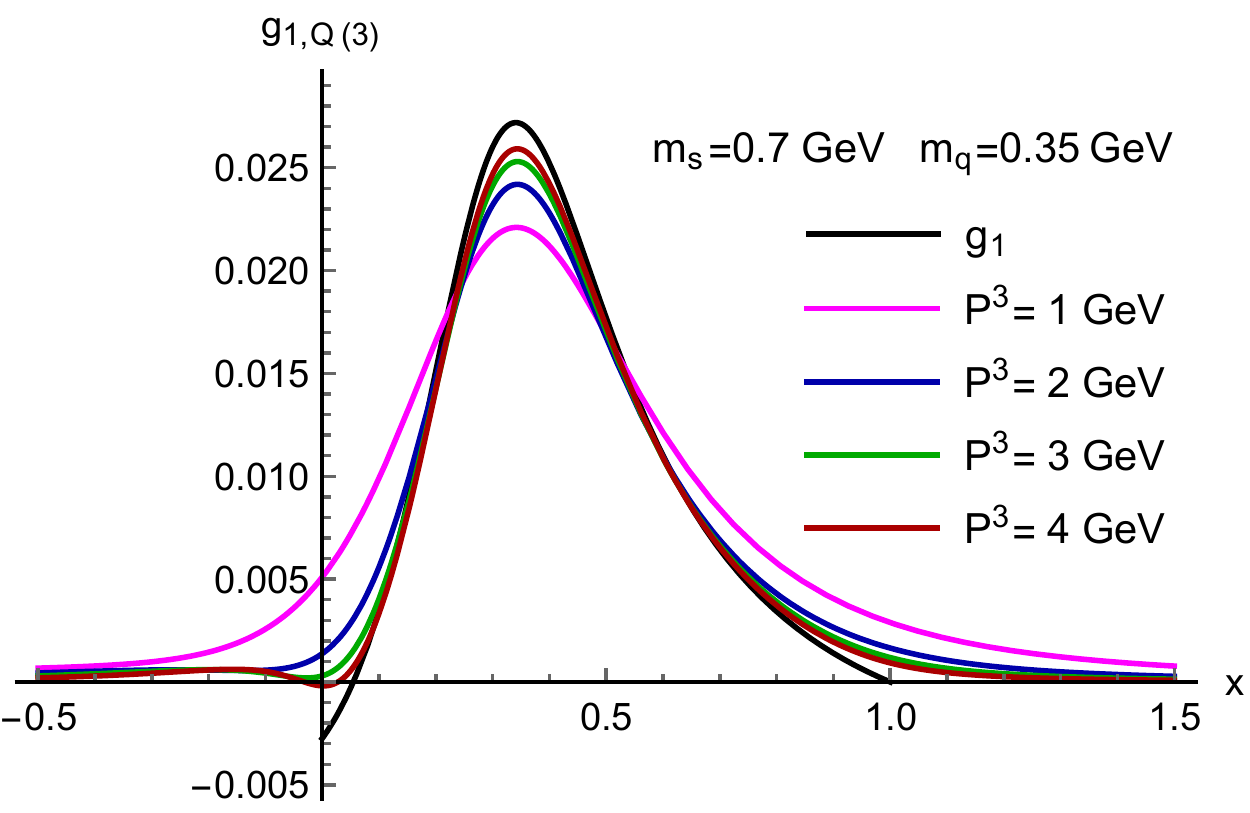}
\caption{Quasi-PDF $g_{1,{\rm Q}}$ as a function of $x$ for different values of $P^3$. 
Left panel: results for $g_{1,{\rm Q}(0)}$.
Right panel: results for $g_{1,{\rm Q}(3)}$.
The standard PDF $g_1$ is shown for comparison.}
\label{f:g1Q}
\end{figure}

\begin{figure}[t]
\includegraphics[width=6.5cm]{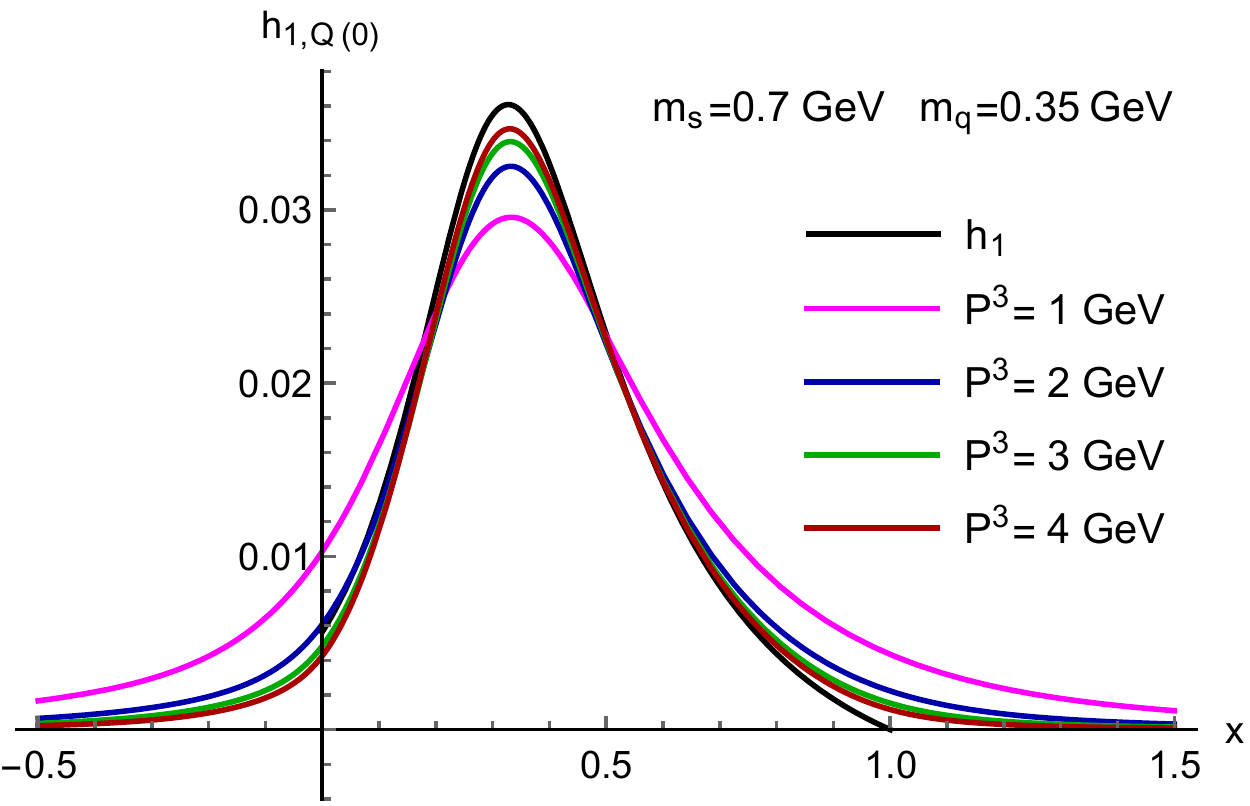}
\hspace{1.5cm}
\includegraphics[width=6.5cm]{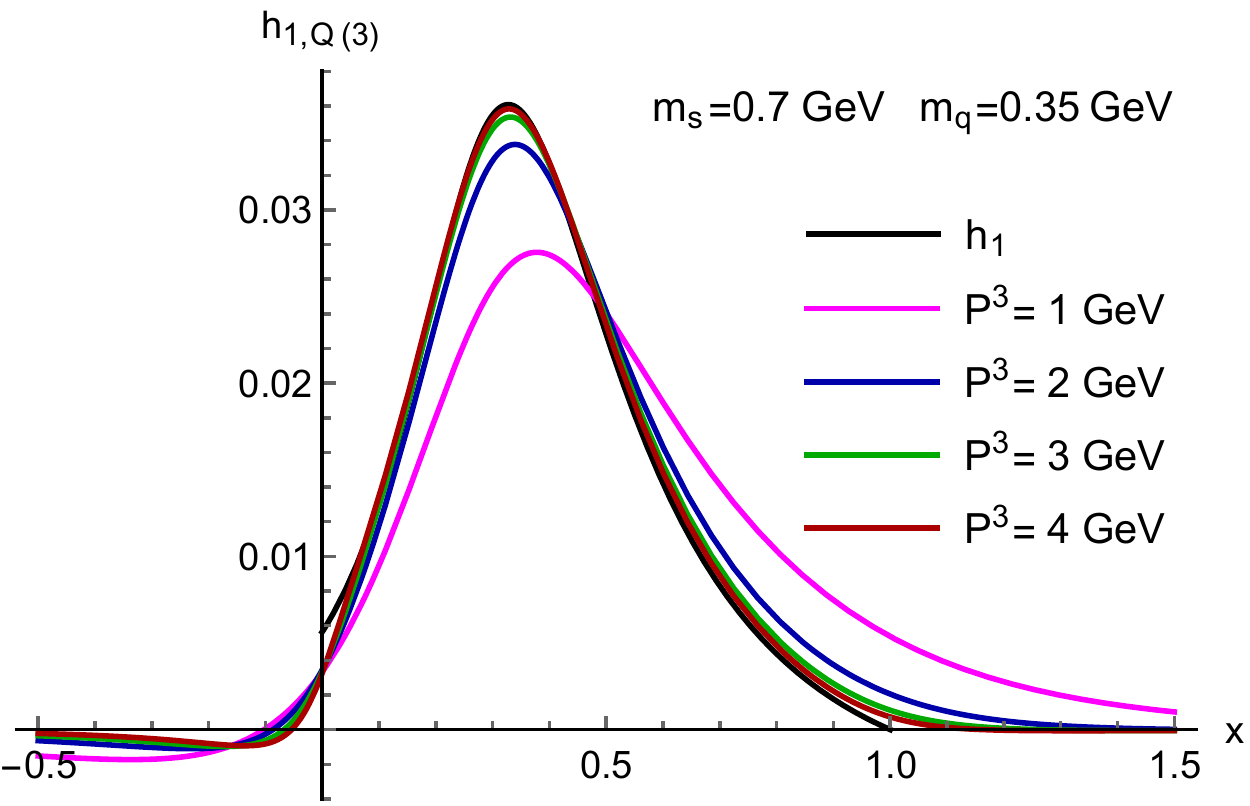}
\caption{Quasi-PDF $h_{1,{\rm Q}}$ as a function of $x$ for different values of $P^3$. 
Left panel: results for $h_{1,{\rm Q}(0)}$.
Right panel: results for $h_{1,{\rm Q}(3)}$.
The standard PDF $h_1$ is shown for comparison.}
\label{f:h1Q}
\end{figure}

\begin{figure}[!]
\includegraphics[width=6.5cm]{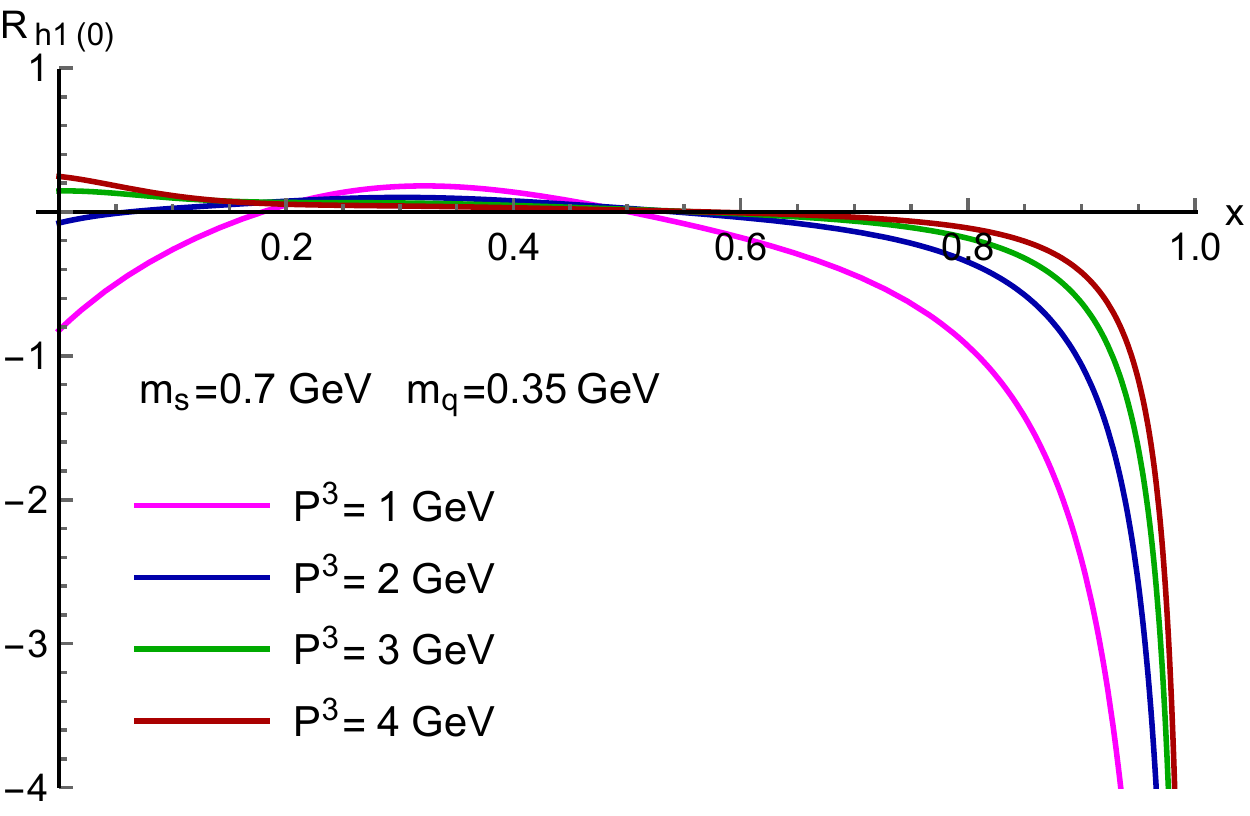}
\hspace{1.5cm}
\includegraphics[width=6.5cm]{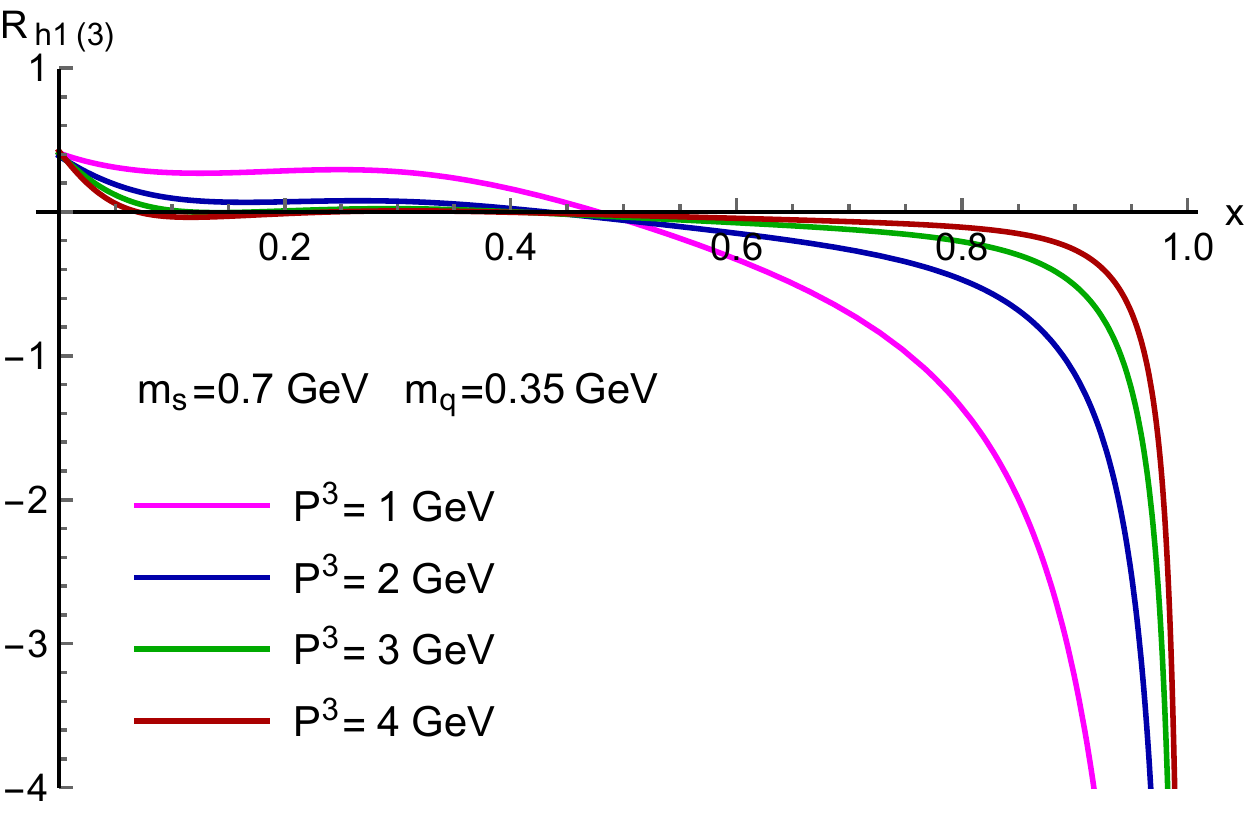}
\caption{Relative difference between quasi-PDFs and $h_1$ as a function of $x$ for different values of $P^3$.
Left panel: results for $R_{h1(0)}$.
Right panel: results for $R_{h1(3)}$.
The maximum values of $x$ for the curves are chosen such that $|R_{h1}| \le 4$.}
\label{f:h1Q_rel}
\end{figure}

\begin{figure}[!]
\includegraphics[width=6.5cm]{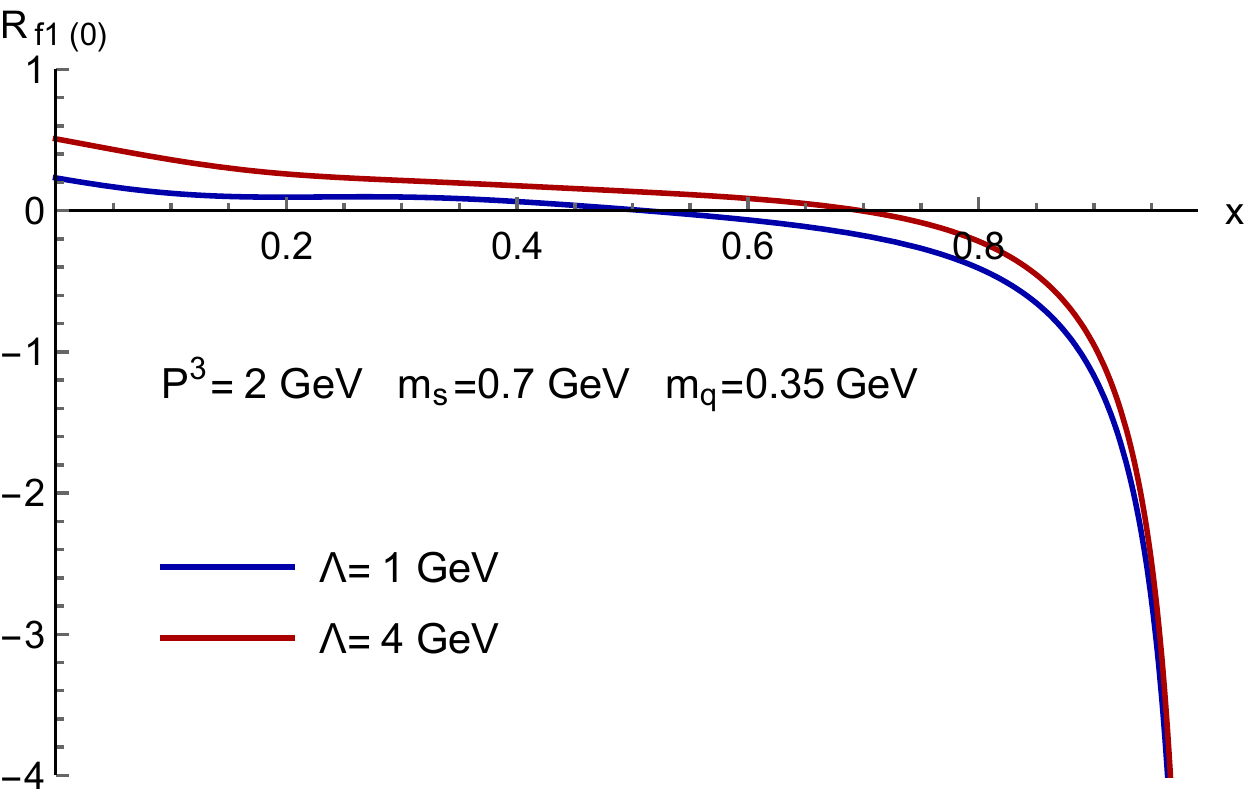}
\hspace{1.5cm}
\includegraphics[width=6.5cm]{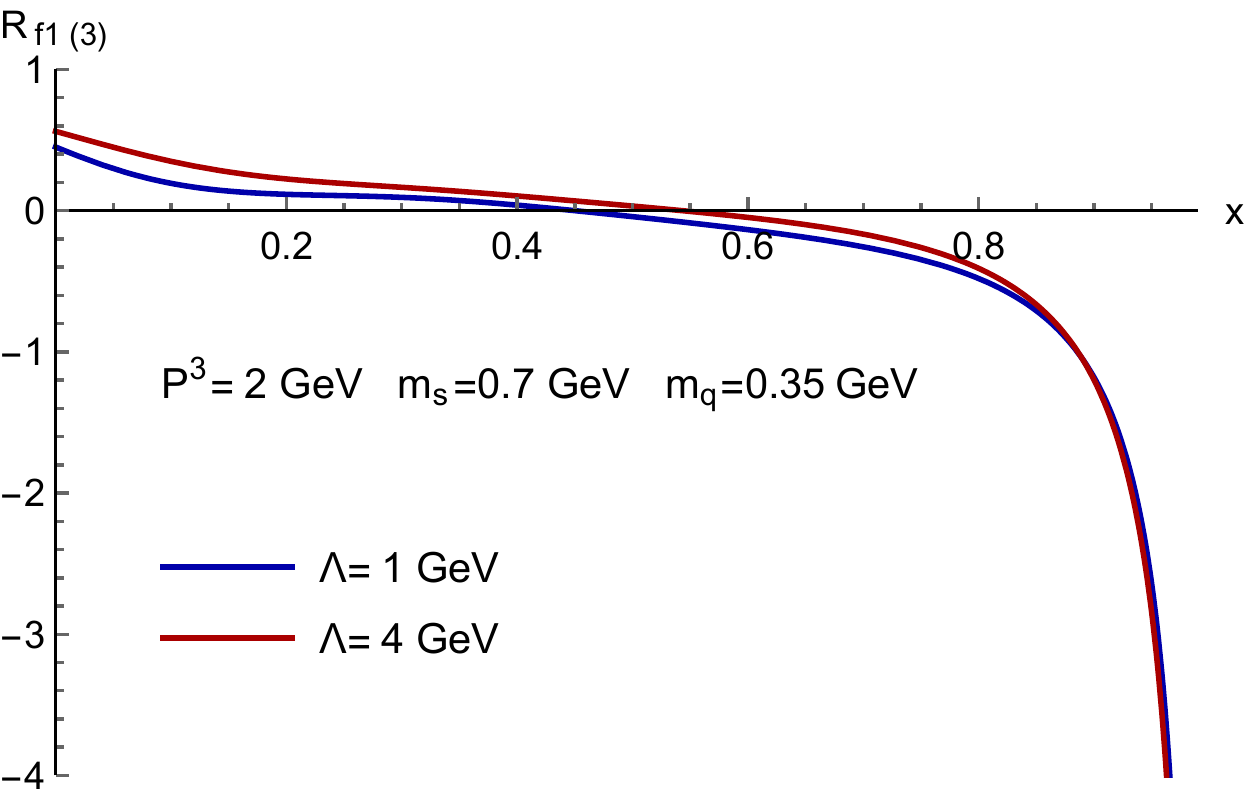}
\caption{Relative difference between quasi-PDFs and $f_1$ as a function of $x$ for different values of the cut-off $\Lambda$ for the $k_\perp$-integration.
Left panel: results for $R_{f1(0)}$.
Right panel: results for $R_{f1(3)}$.}
\label{f:lambda}
\end{figure}

Results for the quasi-PDFs $g_{1, \rm Q(0/3)}$ and $h_{1, \rm Q(0/3)}$ are shown in Fig.~\ref{f:g1Q} and Fig.~\ref{f:h1Q}, respectively. 
Comparing these results with the corresponding plots for $f_{1, \rm Q(0/3)}$ in Ref.~\cite{Bhattacharya:2018zxi}, one qualitatively observes the same features. 
First, for $P^{3} = 2 \, \textrm{GeV}$ and above, there is not much difference between $g_{1, \rm Q(0)}$ and $g_{1, \rm Q(3)}$.
The same holds in the case of the quasi-transversities.
Second, considerable differences appear between quasi-PDFs and standard PDFs as $x \rightarrow 0$ and $x \rightarrow 1$. 
As discussed in detail in~\cite{Bhattacharya:2018zxi}, the discrepancy at small $x$ can be expected since the standard PDFs are discontinuous at $x=0$. 
The quasi-PDFs are continuous, but for large $P^3$ must approach the corresponding standard PDF, which automatically results in large deviations in the region around $x = 0$.
To better illustrate the discrepancy at large $x$ we consider the relative difference, which in the case of $f_1$ we define as~\cite{Bhattacharya:2018zxi}
\begin{equation}
R_{f1(0/3)}(x; P^3) = \frac{f_1(x) - f_{1,{\rm Q}(0/3)}(x; P^3)}{f_1(x)} \,.
\label{e:rel_difference}
\end{equation}
In Fig.~\ref{f:h1Q_rel}, this quantity is shown for the transversity distributions.
Like for $f_1$, at $P^{3} = 2 \, \textrm{GeV}$ one can hardly go above $x=0.8$ for the relative difference to stay below $50 \%$. 
This statement holds true for the helicity distributions as well. 
We shed some more light on the large-$x$ discrepancy in Sec.~\ref{sec: x&xt}.

We forgo showing plots for the dependence of the PDFs (and the GPDs) on the mass parameters $m_s$ and $m_q$.
Our findings in this context can be summarized as follows.
The impact of changing $m_s$ is typically larger. 
Specifically, discrepancies get somewhat larger when increasing $m_s$, especially in the large-$x$ region.
This feature is partly related to the increasing (with $m_s$) difference between the momentum fractions that enter the standard PDFs and the quasi-PDFs.
We refer to Sec.~\ref{sec: x&xt} for further discussion of this point. 
Within the range [0.01, 0.35]$\,$GeV which we have explored, we find only a mild dependence on $m_q$. Analytically, this is caused by the fact that $m_q$ is small compared to the other scales in the problem such as $M$, $m_s$, $P^3$, cutoff for $k_\perp$. 
Transversity is the only exception with regard to the $m_{q}$ dependence especially in the small-$x$ region. 
This can be understood from the analytical result in Eq.~(\ref{e:h1}). 
For small $x$, the quark mass term in the numerator dominates resulting in a larger sensitivity to $m_q$ of this distribution compared to $f_1$ and $g_1$.
The latter distributions have a $\kperpq$ in the numerator --- in addition to the $(m_q + x M)^2$ term --- which gives rise to the (standard) logarithmic UV-divergence and, in particular, a very mild dependence on $m_q$.
As already discussed above, the absence of the UV divergence for the transversity is an artifact of the model, and therefore so is the stronger dependence of $h_1$ on $m_q$ at small $x$.     
For the GPDs we find a very similar overall pattern upon variation of $m_s$ and $m_q$.
In the ERBL region there can be some deviations from this pattern.
But the effects are not very significant, and we therefore refrain from further elaborating on them. 

In Fig.~\ref{f:lambda}, we show the relative difference for $f_1$ for two values ($1 \, \textrm{GeV}$ and $4 \, \textrm{GeV}$) of the cut-off $\Lambda$ for the $k_\perp$-integration.
For $x \lesssim 0.5$ the relative difference increases with an increase of $\Lambda$. 
But at least for $ f_{1,{\rm Q}(3)}$ this effect is mild, given that the two values of $\Lambda$ are very different. 
We find very similar results for the transversity distribution.
On the other hand, for $g_1$ the impact (on the relative difference) of changing $\Lambda$ is larger.
This applies in particular in the region around the point at which $g_1$ changes sign --- see Fig.~\ref{f:g1Q}.
It is obvious from the definition in Eq.~(\ref{e:rel_difference}) that in such a case the relative difference is not a very good measure.
A very similar situation occurs for GPDs if they switch sign. 
Overall, our choice $\Lambda = 1 \, \textrm{GeV}$ typically minimizes the difference between the quasi distributions and the standard distributions. 
Also, the fact that some of the standard distributions have a logarithmic divergence does not necessarily lead to a much poorer convergence as 
$\Lambda$ increases, unless one considers cut-off values much larger than $4 \, \textrm{GeV}$. 
Thus our model can give a faithful description of these distributions with $1 \, \rm GeV <\Lambda <4 \, \rm GeV$.

\subsection{A particular higher-twist contribution in the cut-diagram approximation}
\label{sec: x&xt}
\begin{figure}[t]
\includegraphics[width=6.5cm]{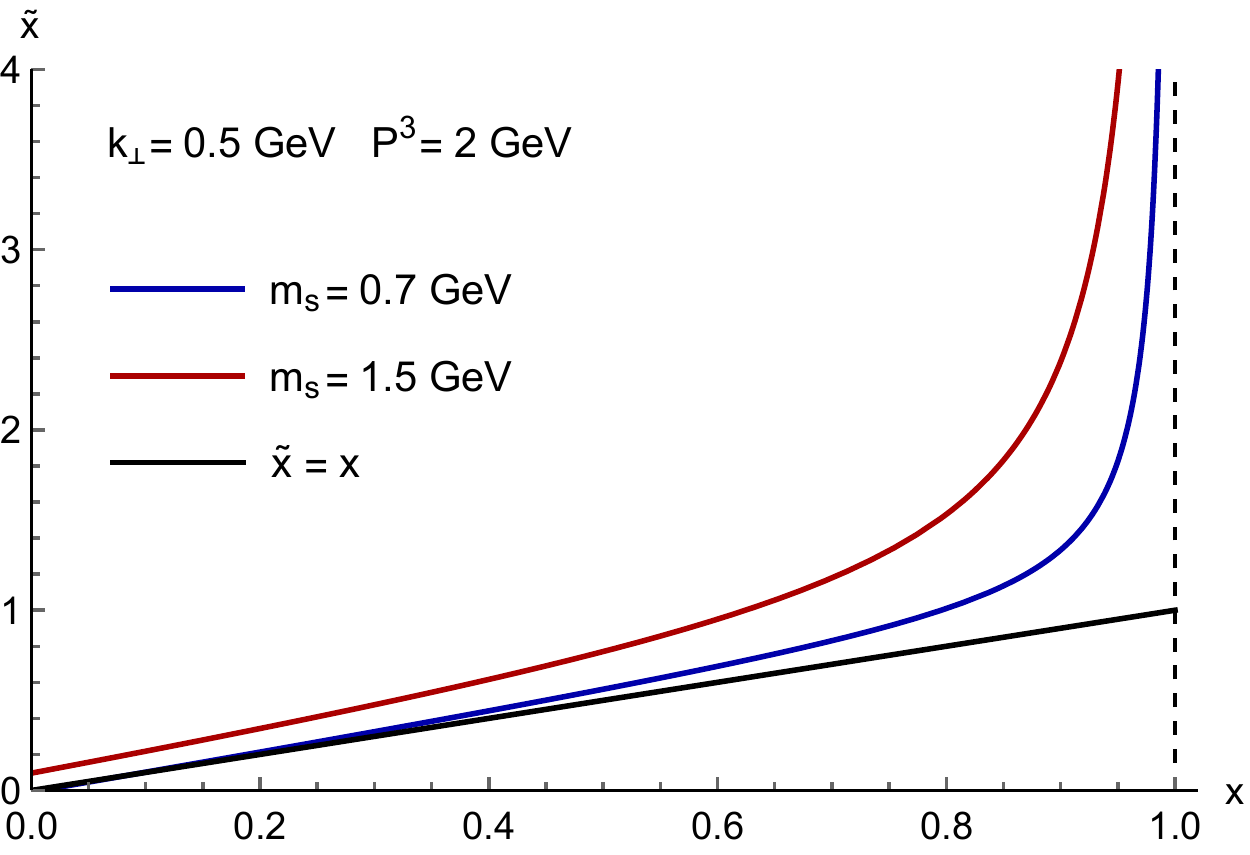}
\caption{Momentum fraction $\tilde{x}$ as a function of $x$ as given in Eq.~(\ref{e: x&xt}) in cut-graph approach, for different values of $m_s$.}
\label{f:xtilde}
\end{figure}

\begin{figure}[t]
\includegraphics[width=6.5cm]{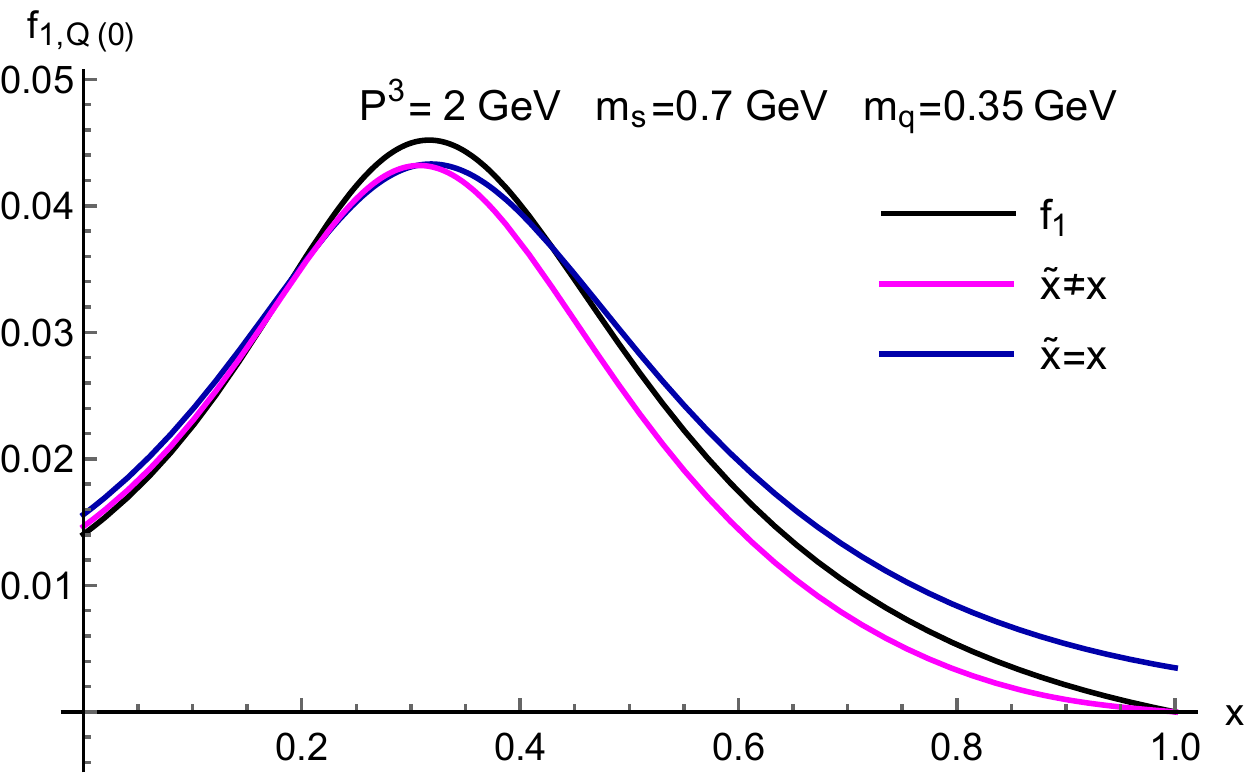}
\hspace{1.5cm}
\includegraphics[width=6.5cm]{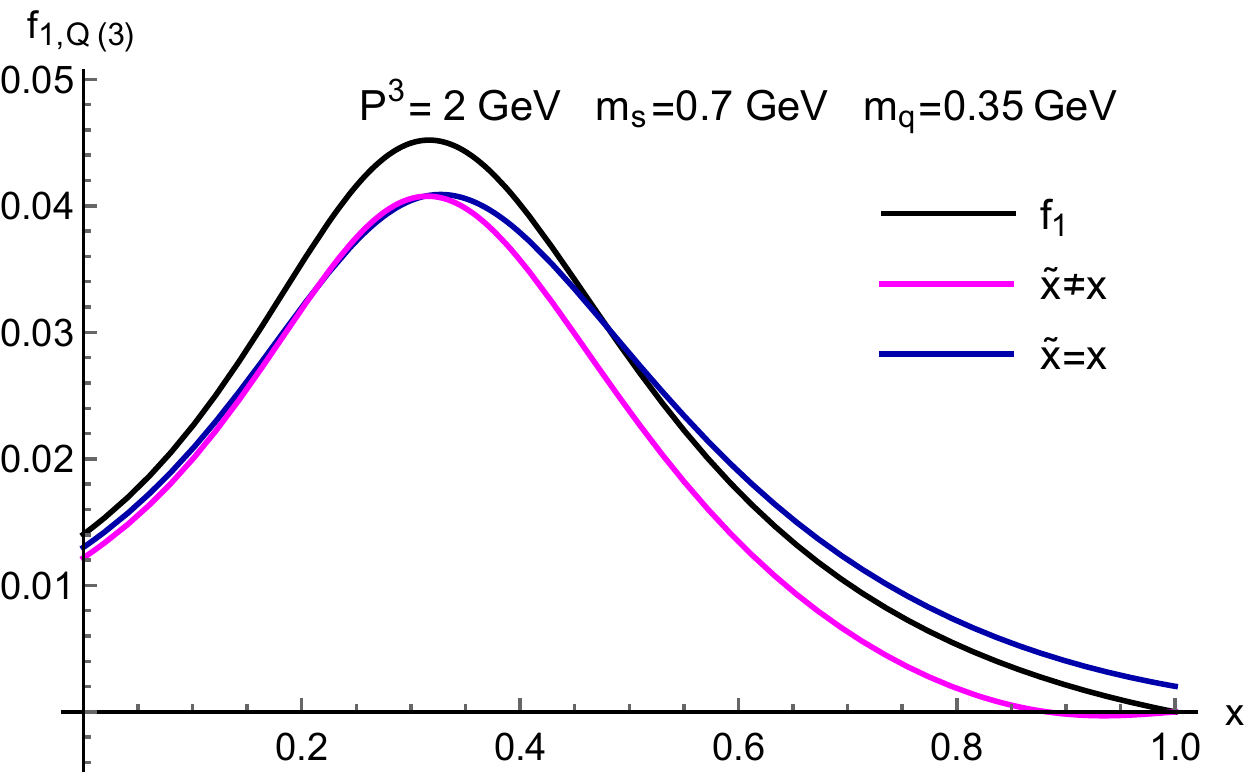}
\caption{Impact of difference between $x$ and $\tilde{x}$ as given in Eq.~(\ref{e: x&xt}) in cut-graph approach for $f_1$.
Only the contribution from the spectator pole is included.
Left panel: results for $f_{1,{\rm Q}(0)}$.
Right panel: results for $f_{1,{\rm Q}(3)}$.
Note that the curves for $\tilde{x} \neq x$ go to 0 for $x \to 1$, like the standard distributions do.}
\label{f:f1_xvsxtilde}
\end{figure}

We repeat that the two momentum fractions $\tfrac{k^{+}}{P^{+}}$ and $\tfrac{k^{3}}{P^{3}}$ are different and that they cannot be related in a model-independent way.
In this section we denote the latter by $\tilde{x}$, and study the impact of the difference between $x$ and $\tilde{x}$ in the (model-dependent) cut-graph approach in the SDM.
In Ref.~\cite{Bhattacharya:2018zxi} we found for $f_1$ that numerically, for $P^3 \geq 2 \, \textrm{GeV}$ and the range $0 \leq x \leq 1$, the difference between the cut-graph approximation and the full calculation in the SDM is rather small, except in the small-$x$ region.
For more discussion of this approach we refer to~\cite{Gamberg:2014zwa, Bhattacharya:2018zxi}.
In the cut-graph model one puts the di-quark spectator on-shell, that is, $(P - k)^2 = m_s^2$ (see Eq.~(\ref{e:deno_standard})).
One can then derive the relation
\begin{equation}
\tilde{x} = \frac{x}{2}(1+\delta_{0}) + \frac{\vec{k}^{2}_{\perp}+m^{2}_{s}-(1-x)M^{2}}{2(1-x)(1+\delta_{0})(P^{3})^{2}} 
= x + \frac{1}{4 (P^3)^2} \bigg( \frac{\kperpq + m_s^2}{1 - x} - (1 - x) M^2 \bigg) + {\cal O} \bigg( \frac{1}{(P^3)^4} \bigg) \,.
\label{e: x&xt}
\end{equation}
Obviously, the difference between $\tilde{x}$ and $x$ is of order ${\cal O}(1/ (P^3)^2)$ and therefore power-suppressed.
A numerical comparison of the two variables can be found in Fig.~\ref{f:xtilde}.
Their difference gets larger as $m_s$ increases, as can also be expected based on Eq.~(\ref{e: x&xt}).
Most importantly, due to the $1/(1-x)$ factor, one finds $\tilde{x} \to \infty$ as $x \to 1$, which implies very large differences between the two momentum fractions at large $x$ --- see also Ref.~\cite{Gamberg:2014zwa}.
One can therefore speculate that the considerable discrepancies between the quasi-distributions and the corresponding standard distributions at large $x$ are mostly caused by the (huge) discrepancy between $\tilde{x}$ and $x$.
In Fig.~\ref{f:f1_xvsxtilde} we explore this point for $f_1$.
The quasi-PDF $f_{1,\rm Q(0)}$ indeed provides, at large $x$, a better agreement with the standard PDF, while this is not true for $f_{1,\rm Q(3)}$, unless one goes to extremely large $x$.
In the case of $g_1$ and $h_1$ (not shown) we find that the ``recipe" of distinguishing between $\tilde{x}$ and $x$ works better for $g_{1,\rm Q(3)}$ and $h_{1,\rm Q(0)}$, respectively.
The fact that, overall, this ``recipe'' does not lead to a much better agreement between quasi-PDFs and standard PDFs (at large $x$) can be traced back to other higher-twist contributions in the cut-graph approach that also diverge for $x \to 1$.

\subsection{Results for quasi-GPDs}
\begin{figure}[t]
\includegraphics[width=6.5cm]{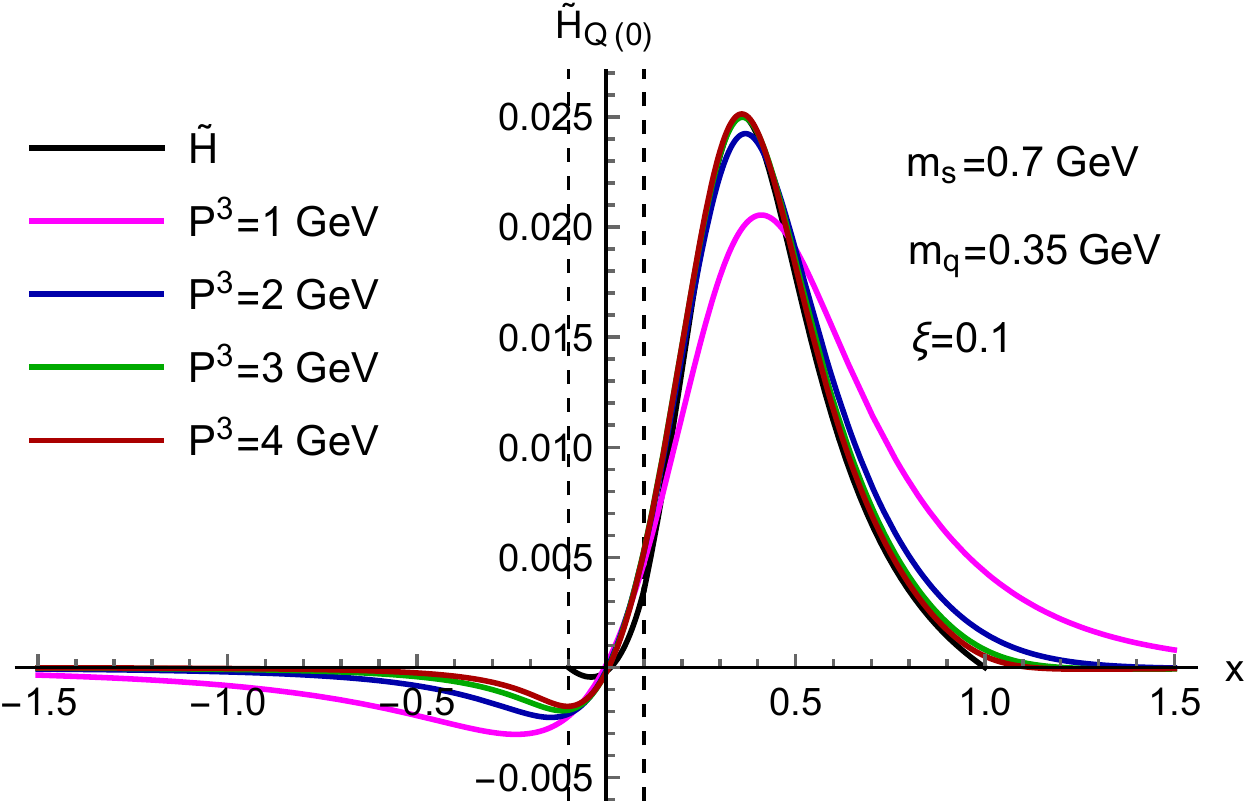}
\hspace{1.5cm}
\includegraphics[width=6.5cm]{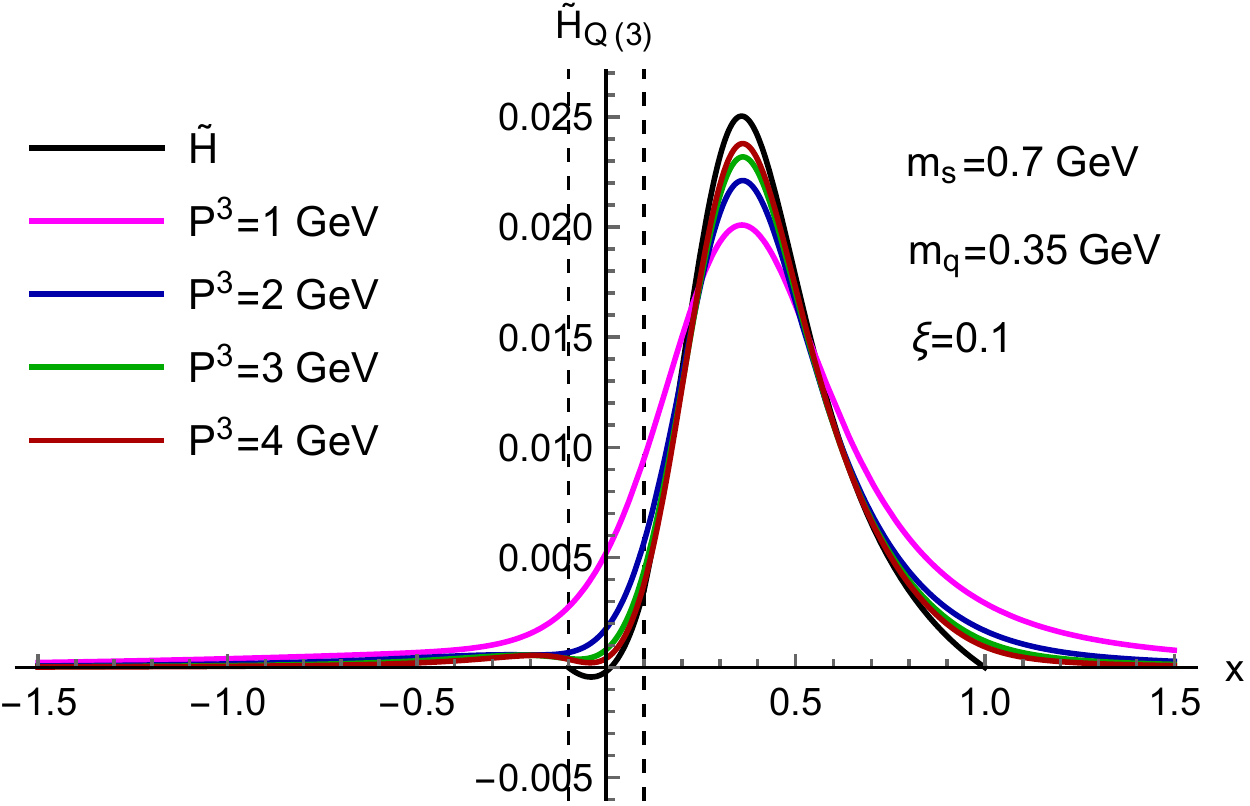}
\caption{Quasi-GPDs $\widetilde{H}_{\rm Q(0)}$ and $\widetilde{H}_{\rm Q(3)}$ as a function of $x$ for $\xi=0.1$ and different values of $P^3$.
The standard GPD $\widetilde{H}$ is shown for comparison.
The limits of the ERBL region are indicated by vertical dashed lines.}
\label{f:H_tilde}
\end{figure}

\begin{figure}[t]
\includegraphics[width=6.5cm]{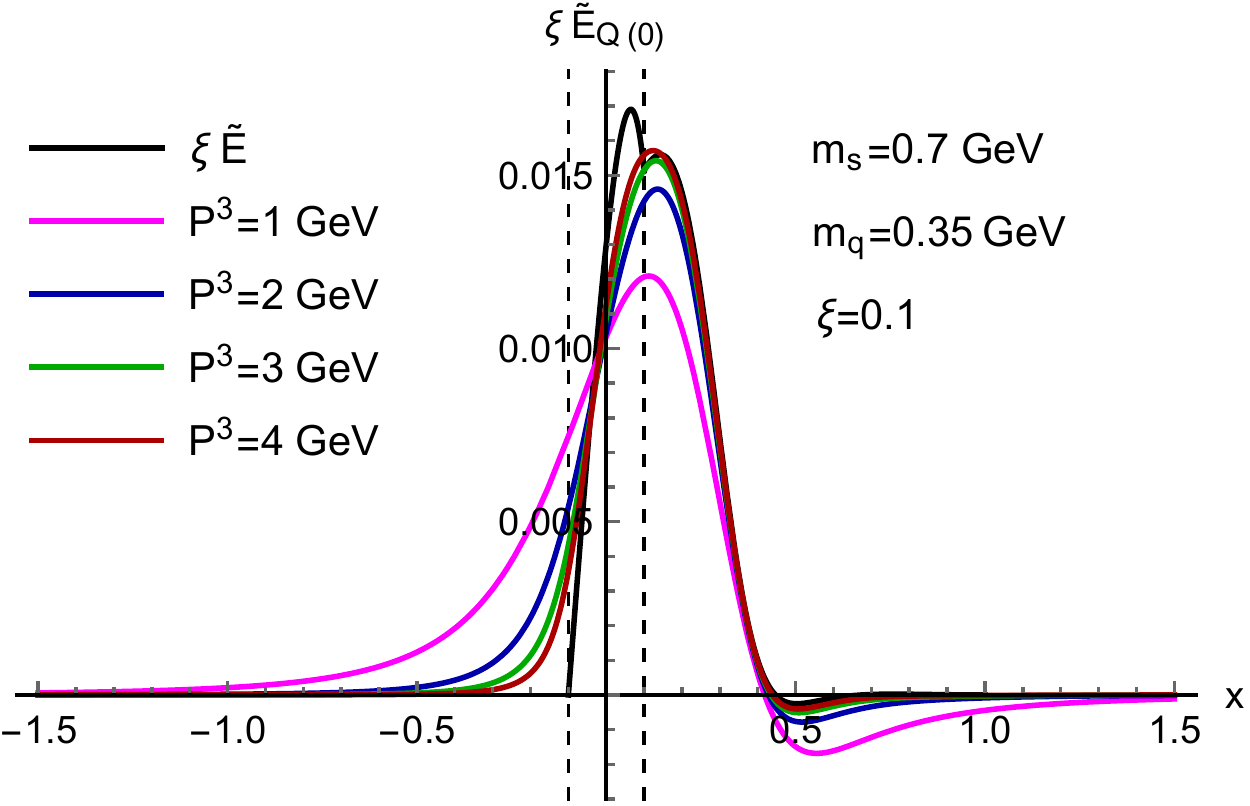}
\hspace{1.5cm}
\includegraphics[width=6.5cm]{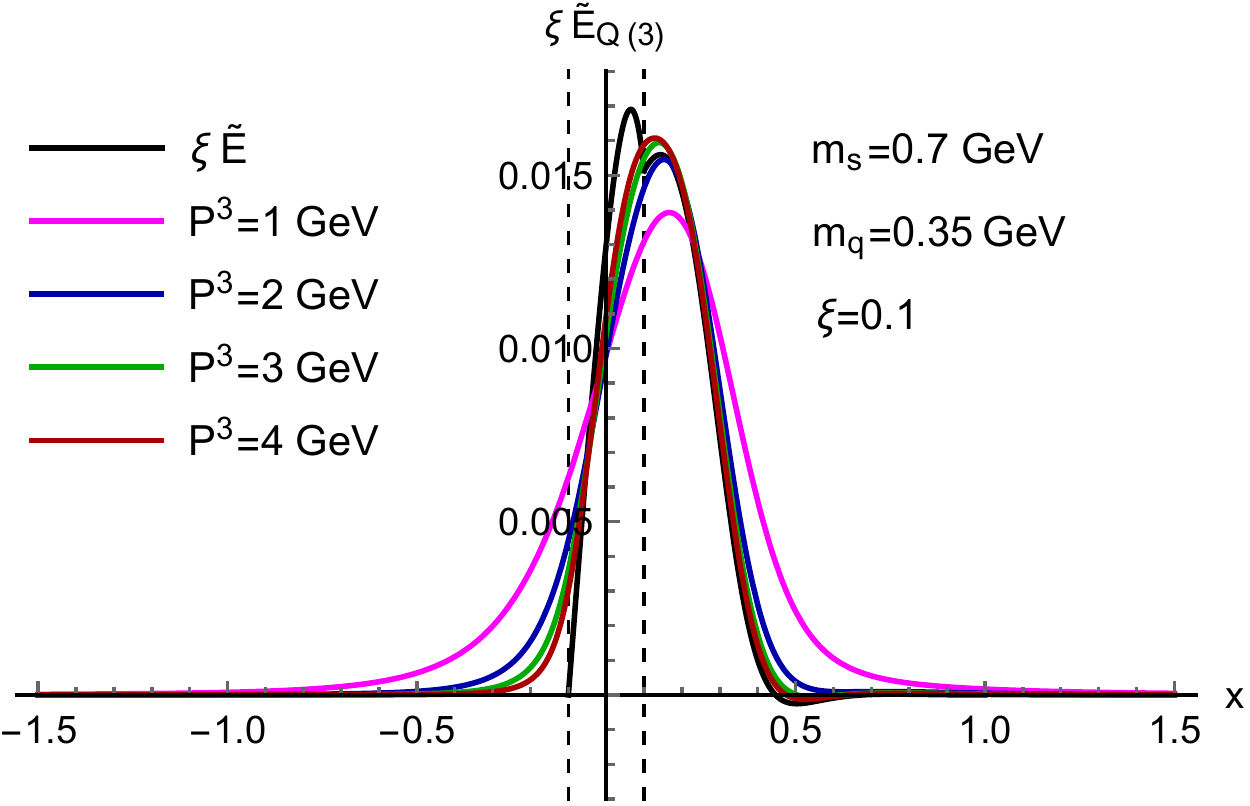}
\caption{Quasi-GPDs $\xi \widetilde{E}_{\rm Q(0)}$ and $\xi \widetilde{E}_{\rm Q(3)}$ as a function of $x$ for $\xi=0.1$ and different values of $P^3$.
The standard GPD $\xi \widetilde{E}$ is shown for comparison.
The limits of the ERBL region are indicated by vertical dashed lines.}
\label{f:E_tilde}
\end{figure}

\begin{figure}[!]
\includegraphics[width=6.5cm]{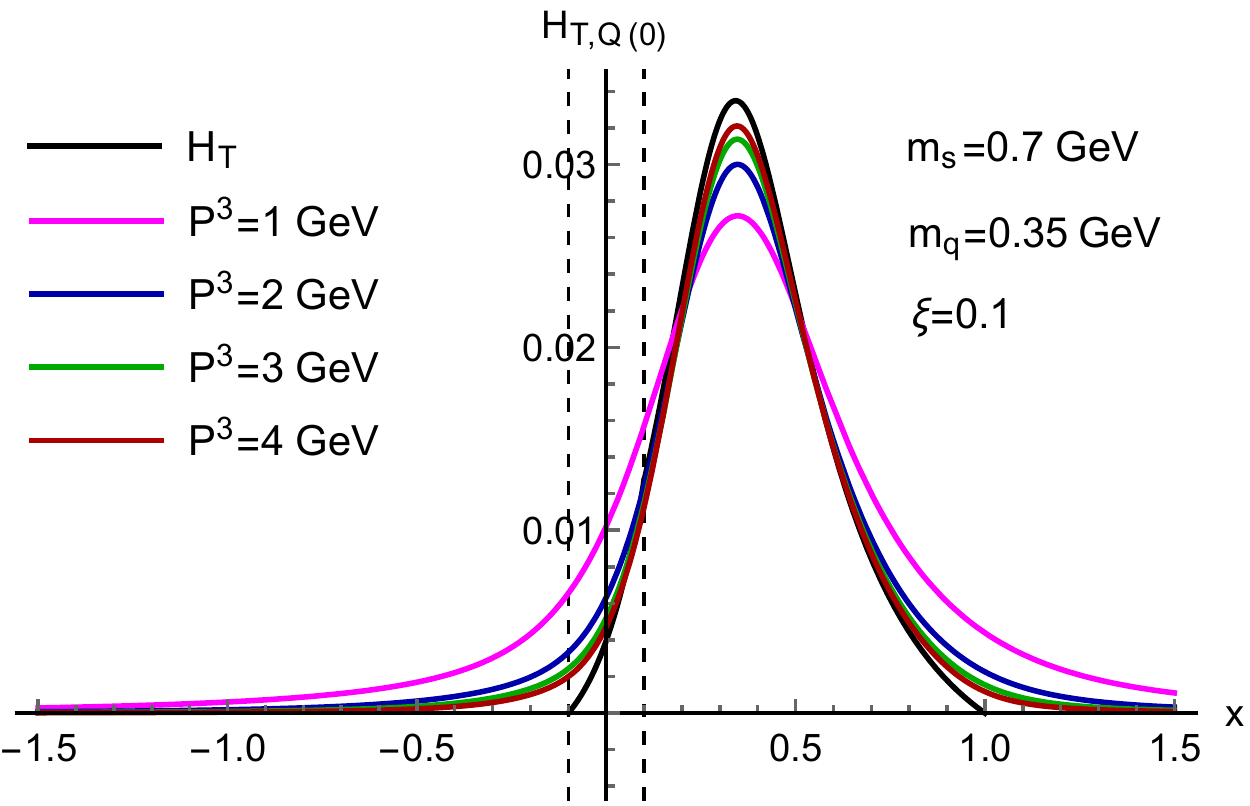}
\hspace{1.5cm}
\includegraphics[width=6.5cm]{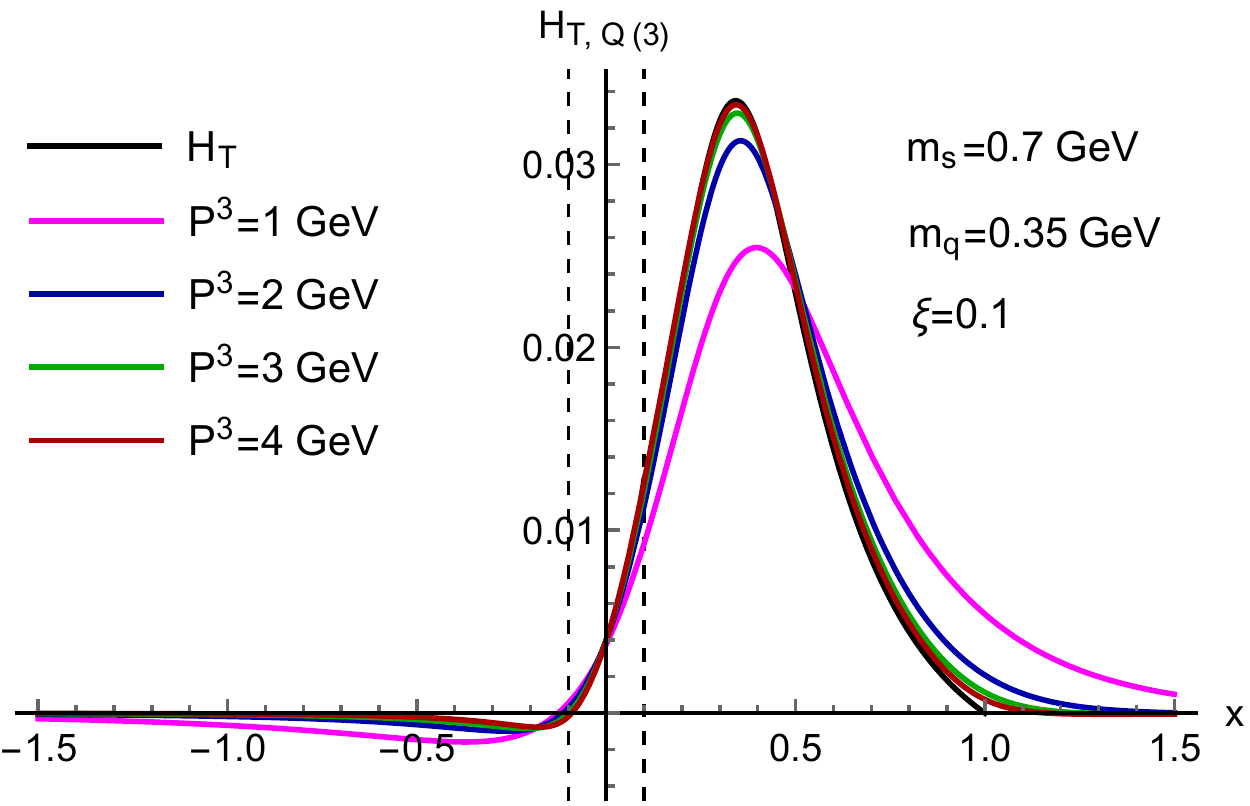}
\caption{Quasi-GPDs $H_{T, {\rm Q(0)}}$ and $H_{T, {\rm Q(3)}}$ as a function of $x$ for $\xi=0.1$ and different values of $P^3$.
The standard GPD $H_T$ is shown for comparison.
The limits of the ERBL region are indicated by vertical dashed lines.}
\label{f:HT}
\end{figure}

\begin{figure}[!]
\includegraphics[width=6.5cm]{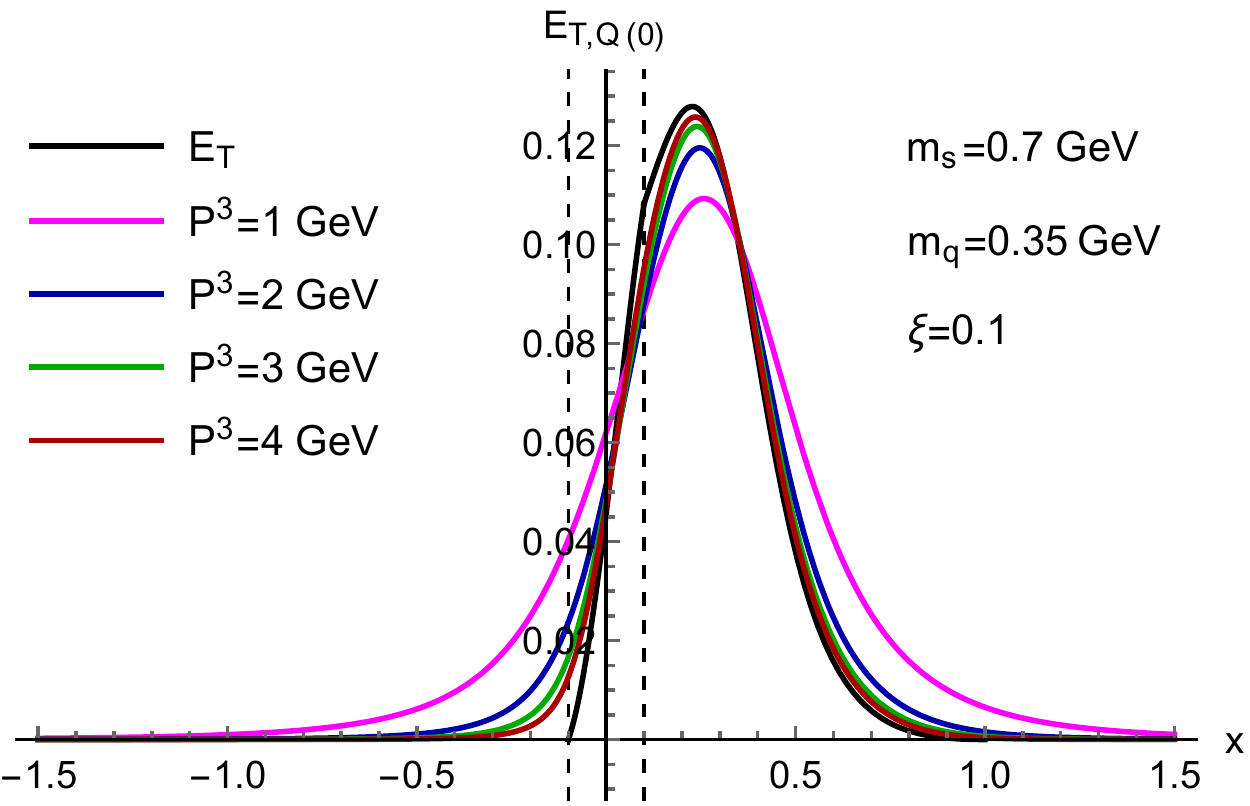}
\hspace{1.5cm}
\includegraphics[width=6.5cm]{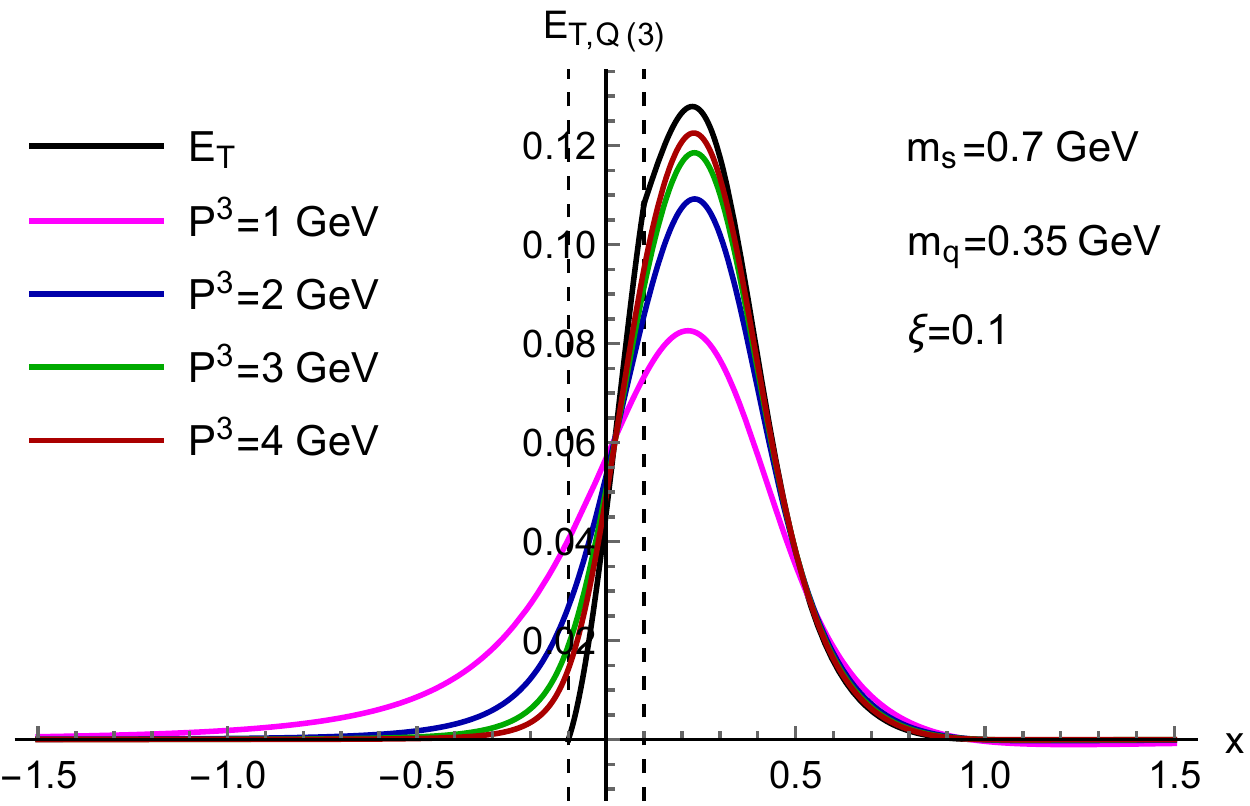}
\caption{Quasi-GPDs $E_{T, {\rm Q(0)}}$ and $E_{T, {\rm Q(3)}}$ as a function of $x$ for $\xi=0.1$ and different values of $P^3$.
The standard GPD $E_T$ is shown for comparison.
The limits of the ERBL region are indicated by vertical dashed lines.}
\label{f:ET}
\end{figure}

\begin{figure}[!]
\includegraphics[width=6.5cm]{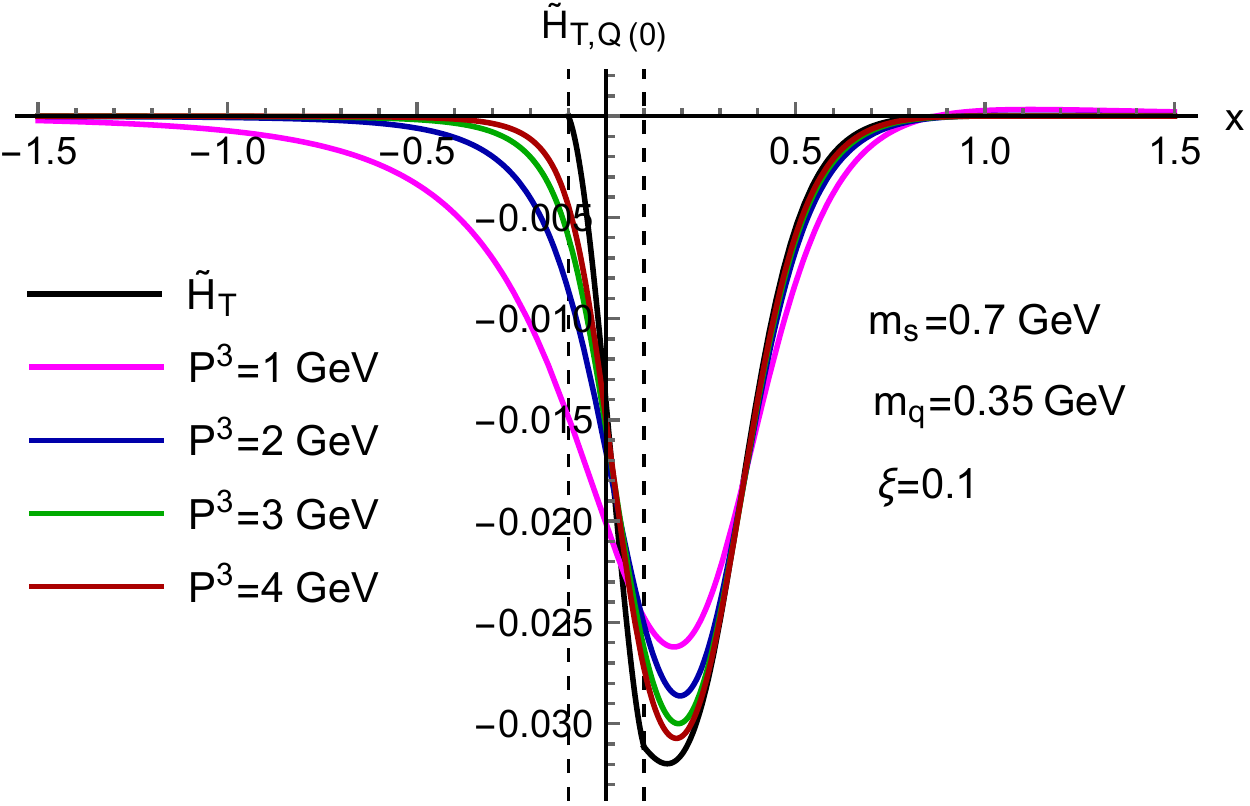}
\hspace{1.5cm}
\includegraphics[width=6.5cm]{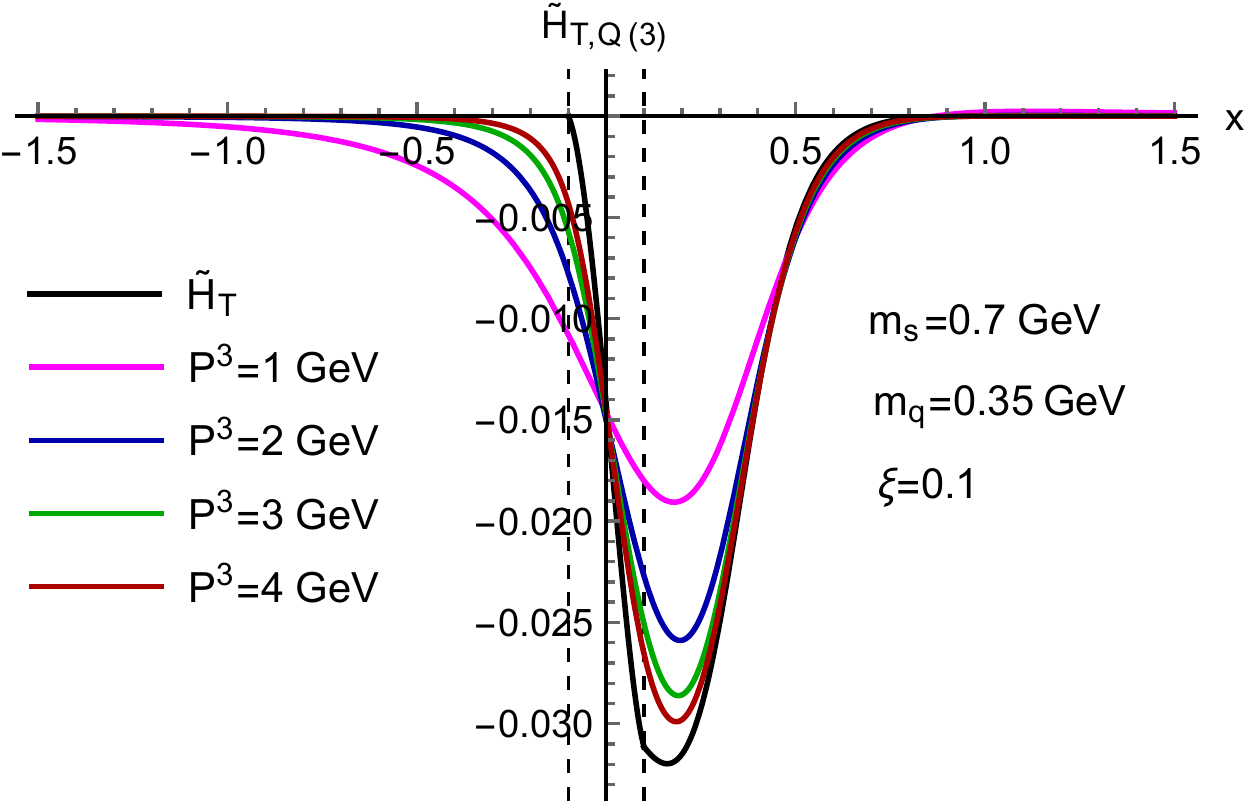}
\caption{Quasi-GPDs $\widetilde{H}_{T, {\rm Q(0)}}$ and $\widetilde{H}_{T, {\rm Q(3)}}$ as a function of $x$ for $\xi=0.1$ and different values of $P^3$.
The standard GPD $\widetilde{H}_T$ is shown for comparison.
The limits of the ERBL region are indicated by vertical dashed lines.}
\label{f:HT_tilde}
\end{figure}

\begin{figure}[!]
\includegraphics[width=6.5cm]{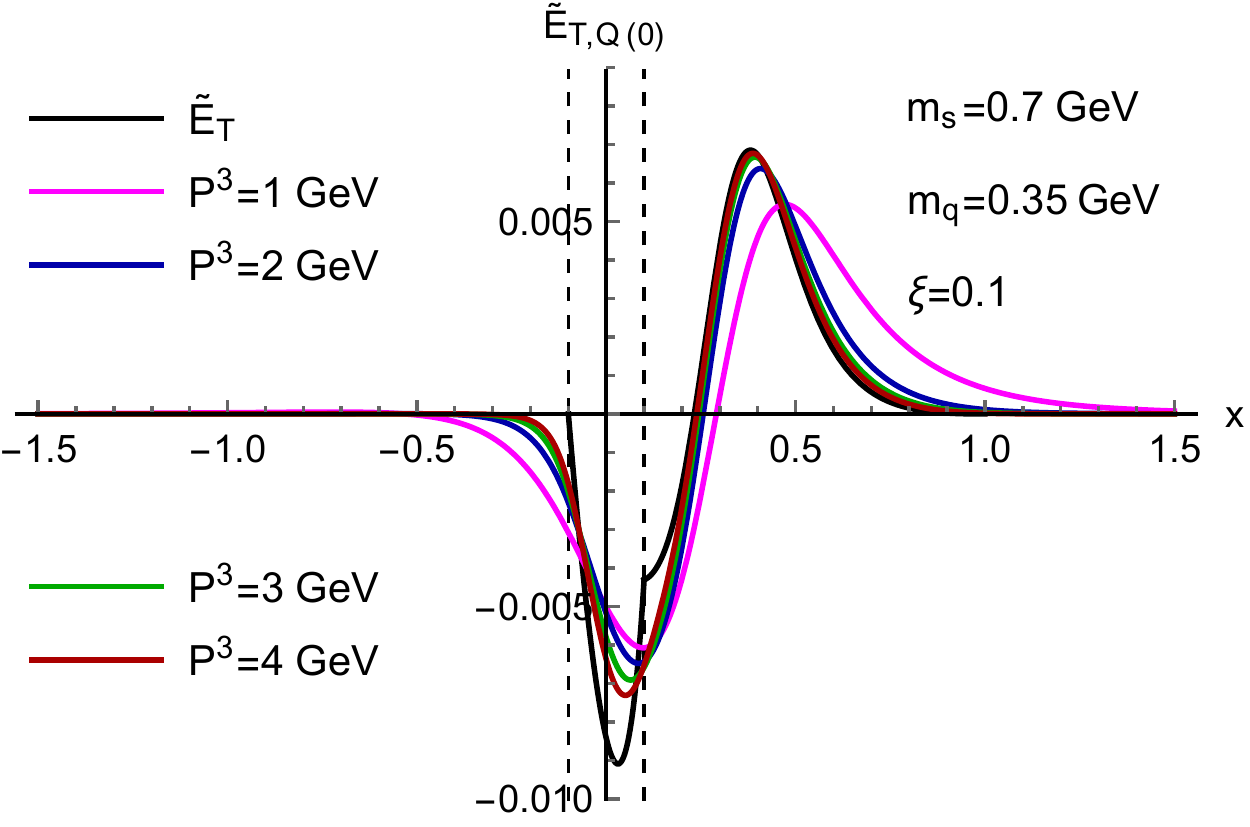}
\hspace{1.5cm}
\includegraphics[width=6.5cm]{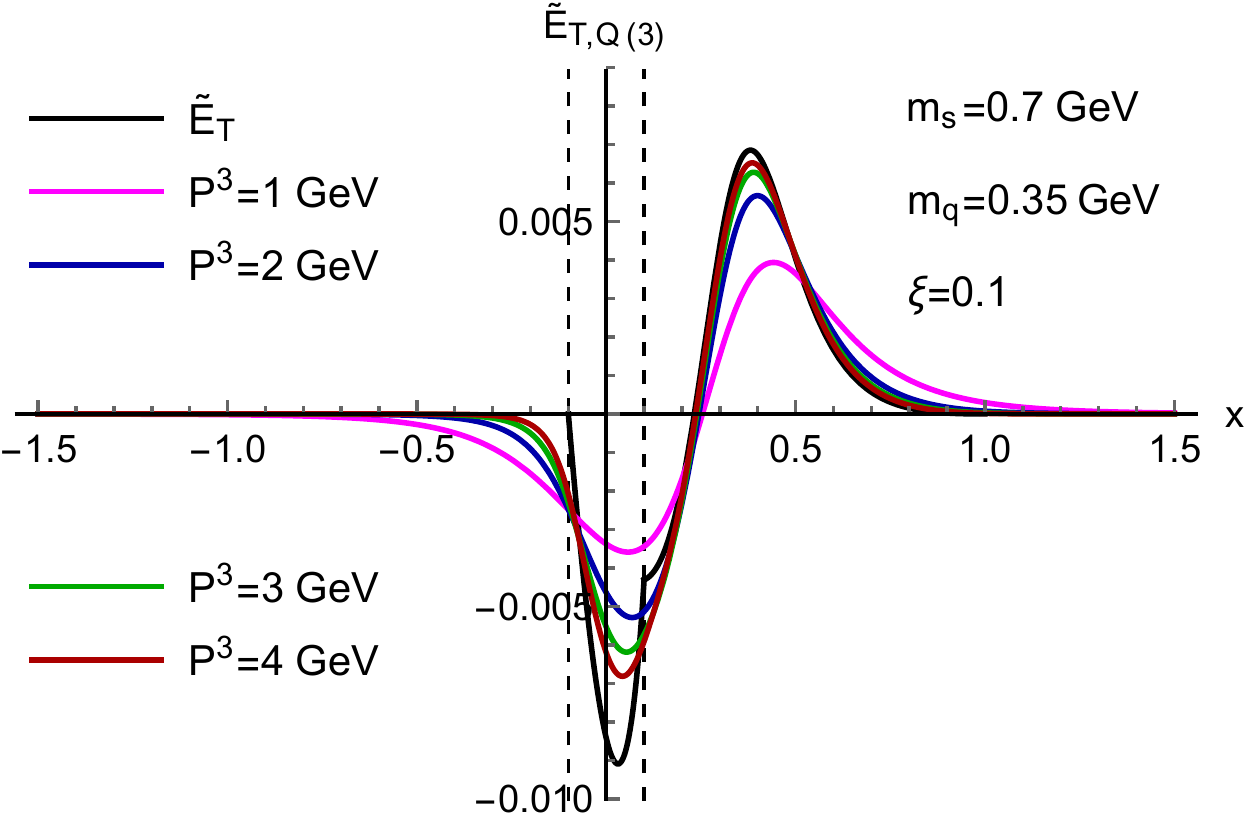}
\caption{Quasi-GPDs $\widetilde{E}_{T, {\rm Q(0)}}$ and $\widetilde{E}_{T, {\rm Q(3)}}$ as a function of $x$ for $\xi=0.1$ and different values of $P^3$.
The standard GPD $\widetilde{E}_T$ is shown for comparison.
The limits of the ERBL region are indicated by vertical dashed lines.}
\label{f:ET_tilde}
\end{figure}

\begin{figure}[!]
\includegraphics[width=6.5cm]{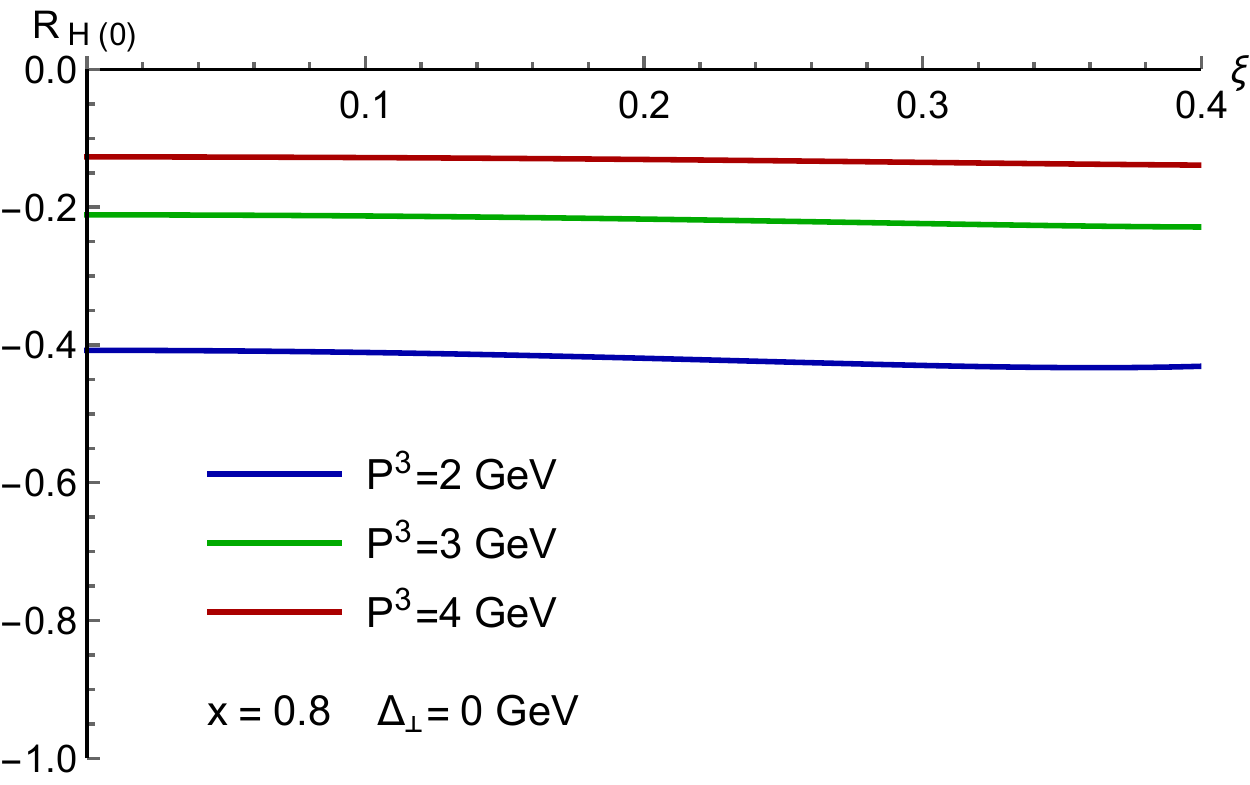}
\hspace{1.5cm}
\includegraphics[width=6.5cm]{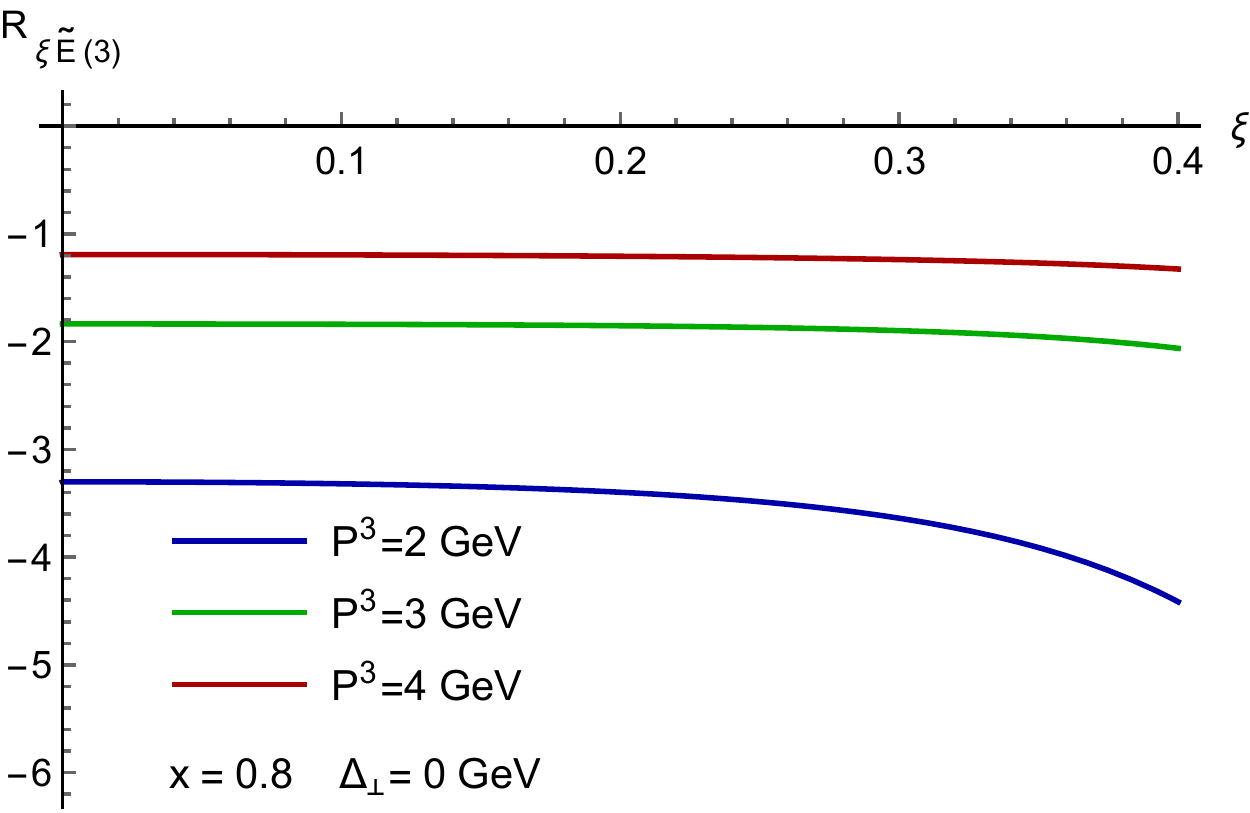}
\caption{Left panel: Relative difference between quasi-GPD $H_{\rm Q(0)}$ and $H$ as a function of $\xi$ for different values of $P^{3}$. Right panel: Relative difference between quasi-GPD $\xi \widetilde{E}_{\rm Q(3)}$ and $\xi \widetilde{E}$ as a function of $\xi$ for different values of $P^{3}$. The relative
difference with $P^{3} = 1$ GeV is much larger than what we observe here for $P^{3} \geq 2$ GeV (see for instance Fig.~\ref{f:h1Q_rel}).}
\label{f:H_tE_RD_Skew}
\end{figure}

\begin{figure}[h]
\includegraphics[width=6.5cm]{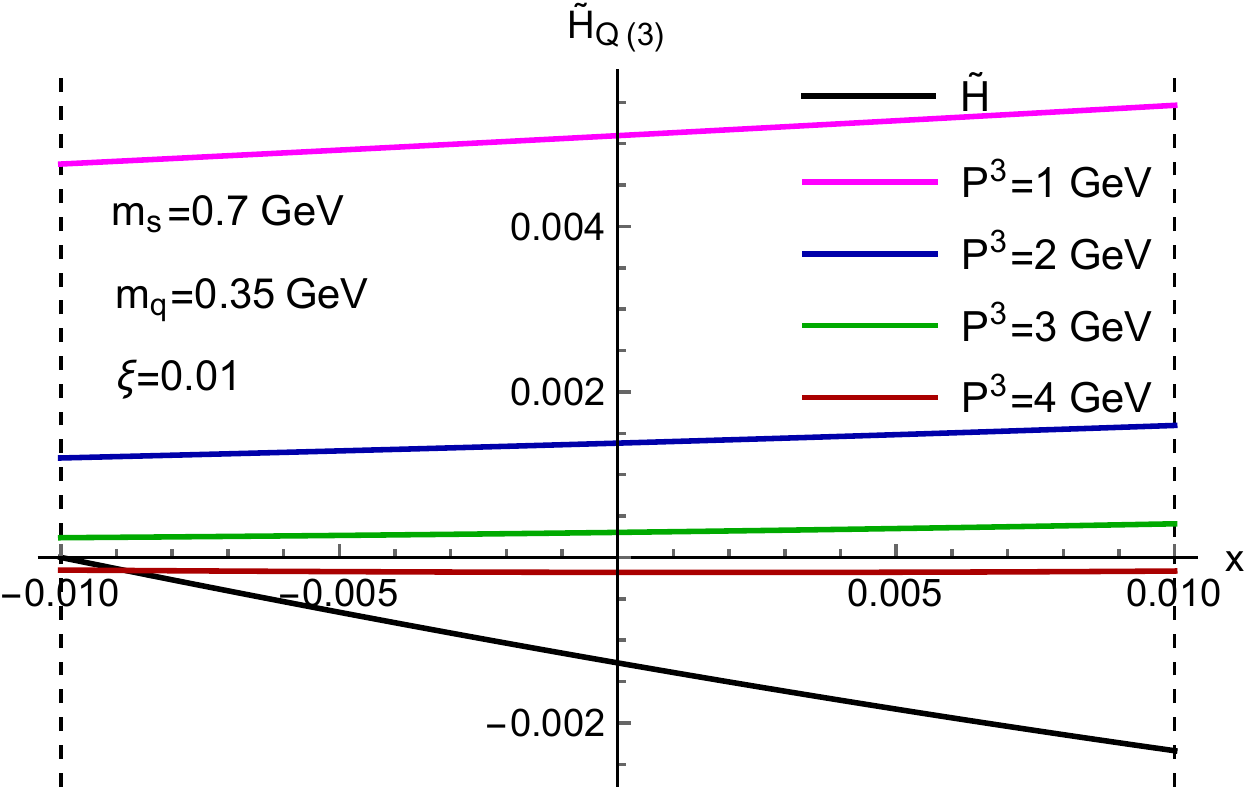}
\hspace{1.5cm}
\includegraphics[width=6.5cm]{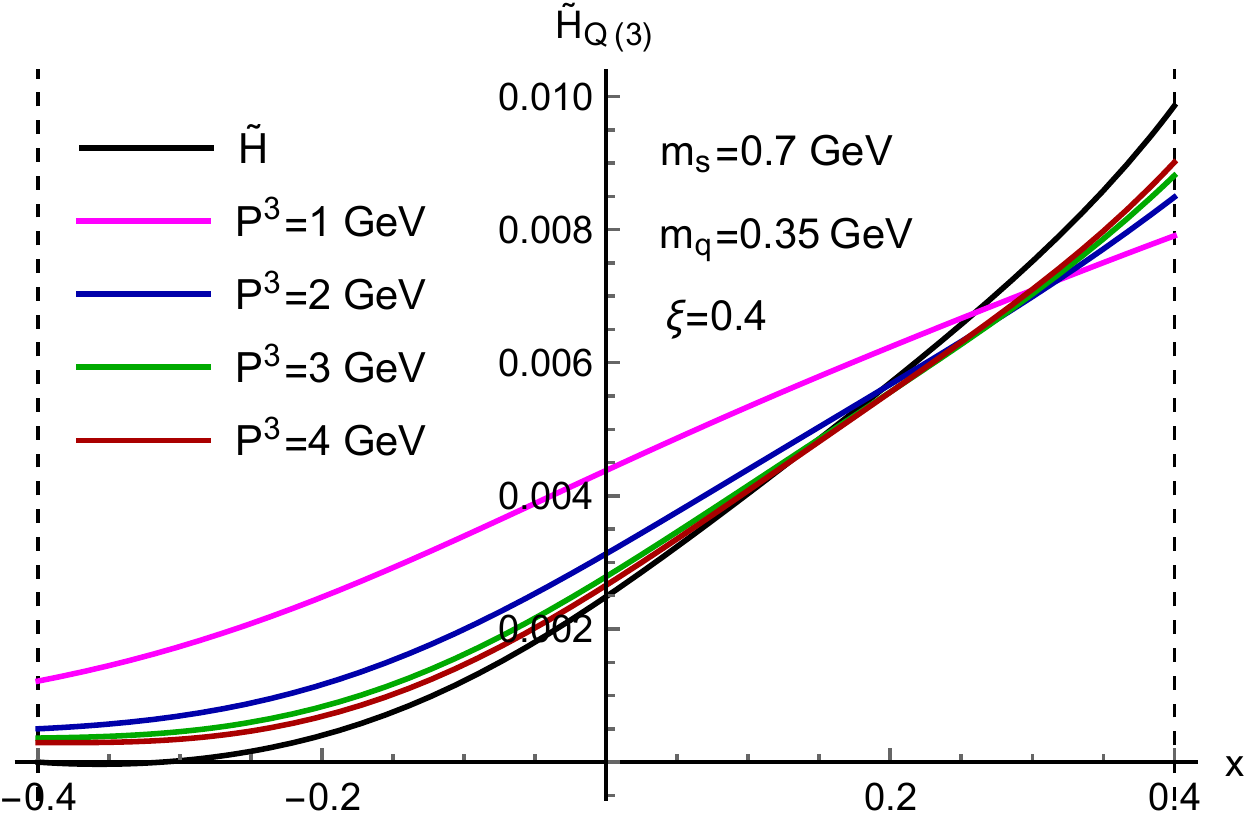}
\caption{Quasi-GPD $\widetilde{H}_{\rm Q(3)}$ as a function of $x$ in the ERBL region for different values of $P^3$. 
Left panel: results for $\xi = 0.01$.
Right panel: results for $\xi = 0.4$.
The standard GPD $\widetilde{H}$ is shown for comparison.}
\label{f:H_tilde_ERBL}
\end{figure}

\begin{figure}[h]
\includegraphics[width=6.5cm]{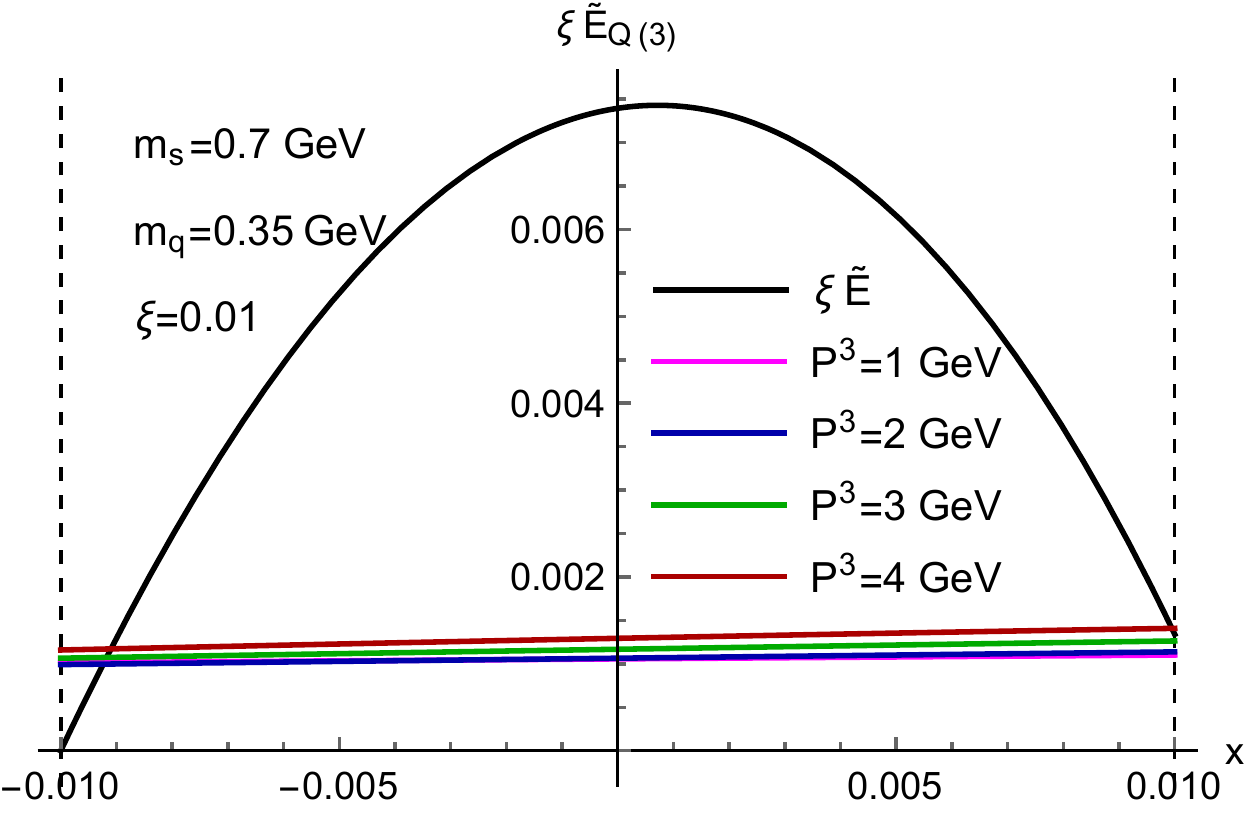}
\hspace{1.5cm}
\includegraphics[width=6.5cm]{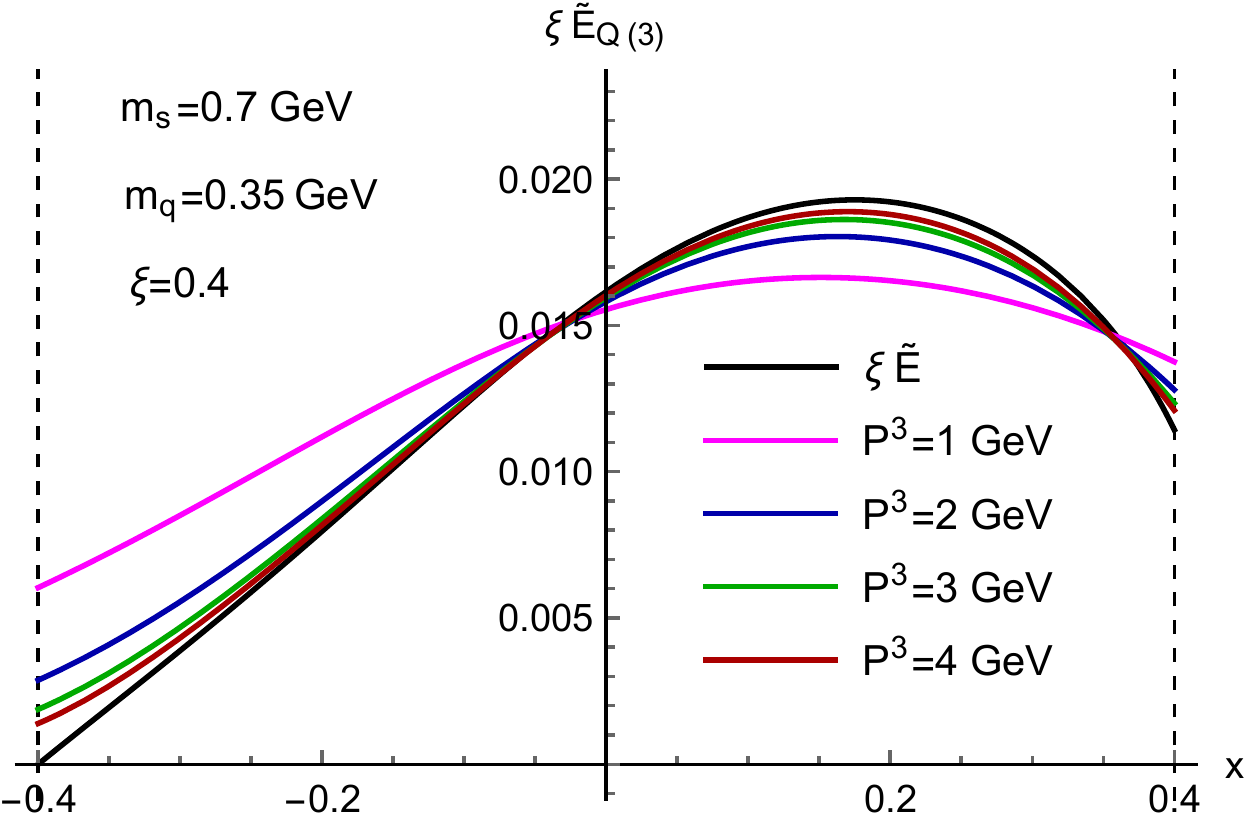}
\caption{Quasi-GPD $\xi \widetilde{E}_{\rm Q(3)}$ as a function of $x$ in the ERBL region for different values of $P^3$. 
Left panel: results for $\xi = 0.01$.
Right panel: results for $\xi = 0.4$.
The standard GPD $\xi \widetilde{E}$ is shown for comparison.}
\label{f:E_tilde_ERBL}
\end{figure}

\begin{figure}[!]
\includegraphics[width=6.5cm]{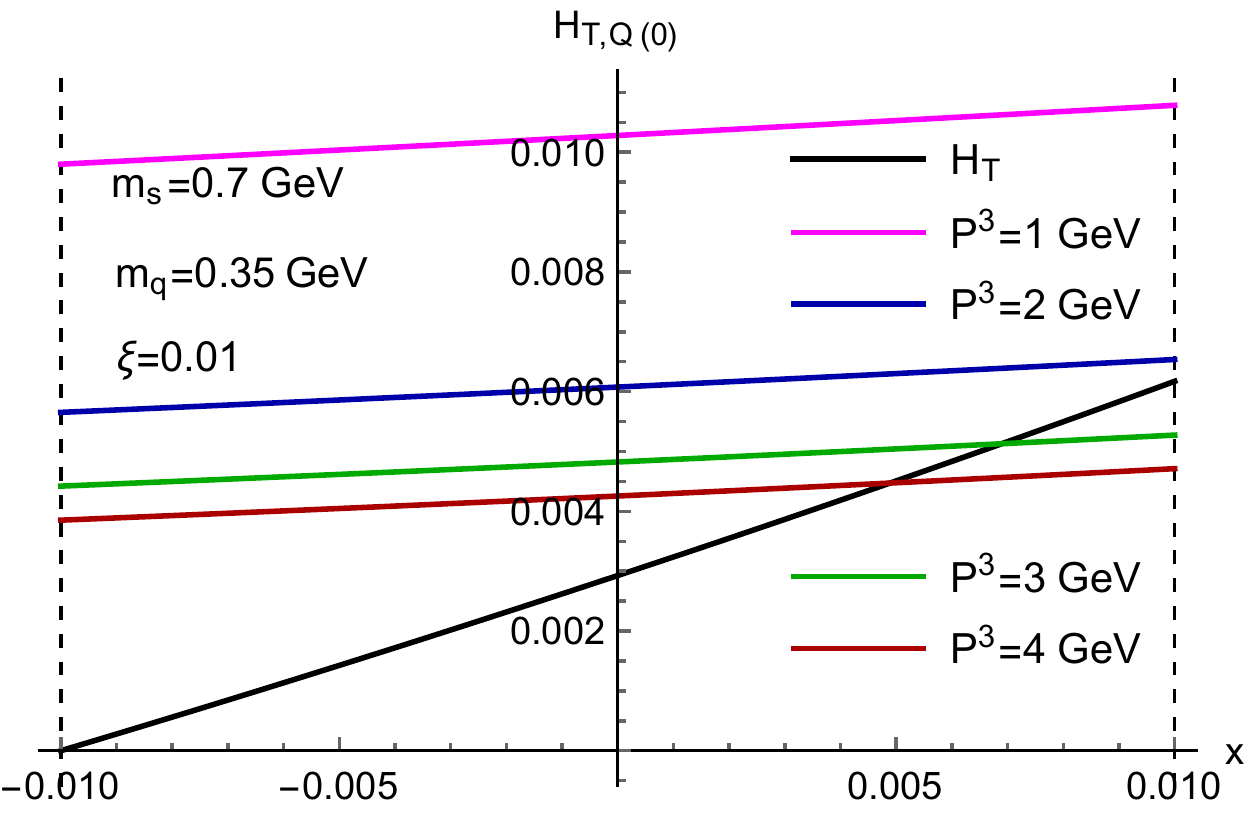}
\hspace{1.5cm}
\includegraphics[width=6.5cm]{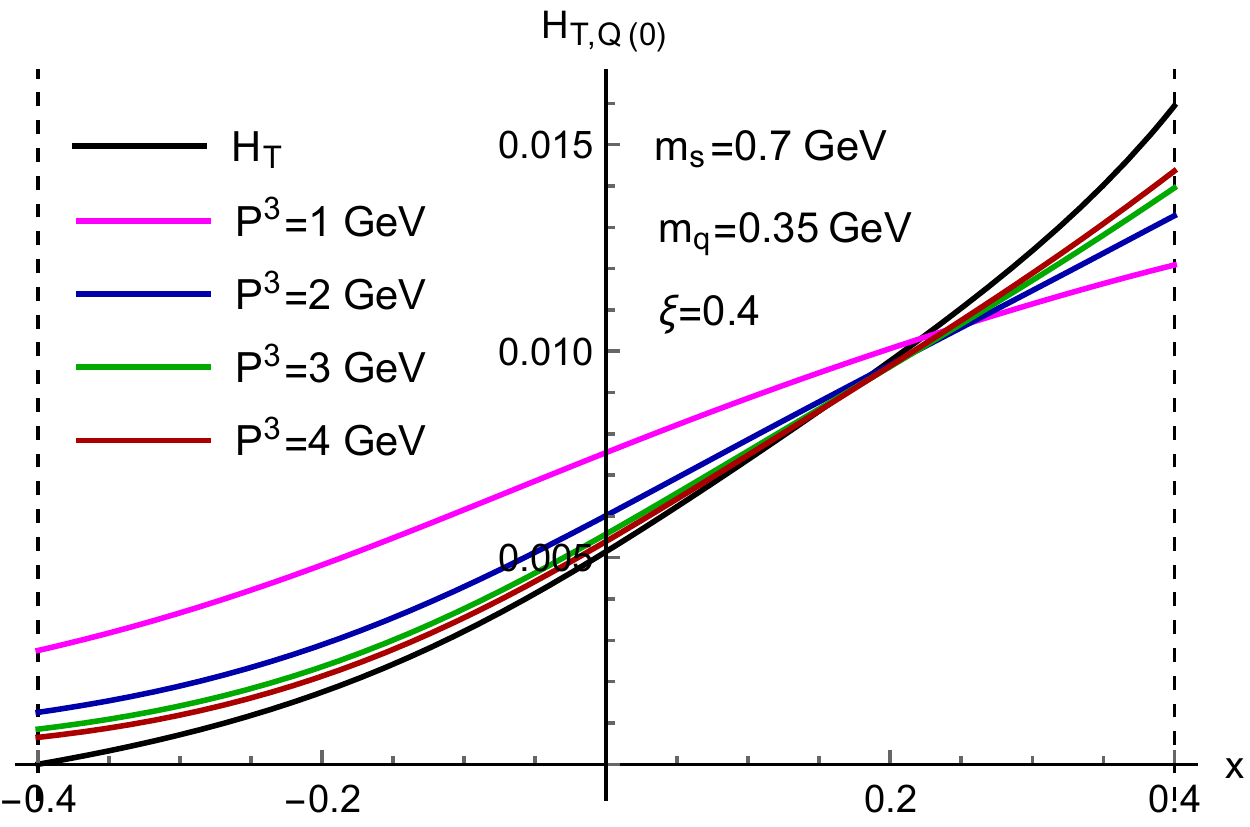}
\caption{Quasi-GPD $H_{T, {\rm Q(0)}}$ as a function of $x$ in the ERBL region for different values of $P^3$. 
Left panel: results for $\xi = 0.01$.
Right panel: results for $\xi = 0.4$.
The standard GPD $H_T$ is shown for comparison.}
\label{f:HT_ERBL}
\end{figure}

\begin{figure}[!]
\includegraphics[width=6.5cm]{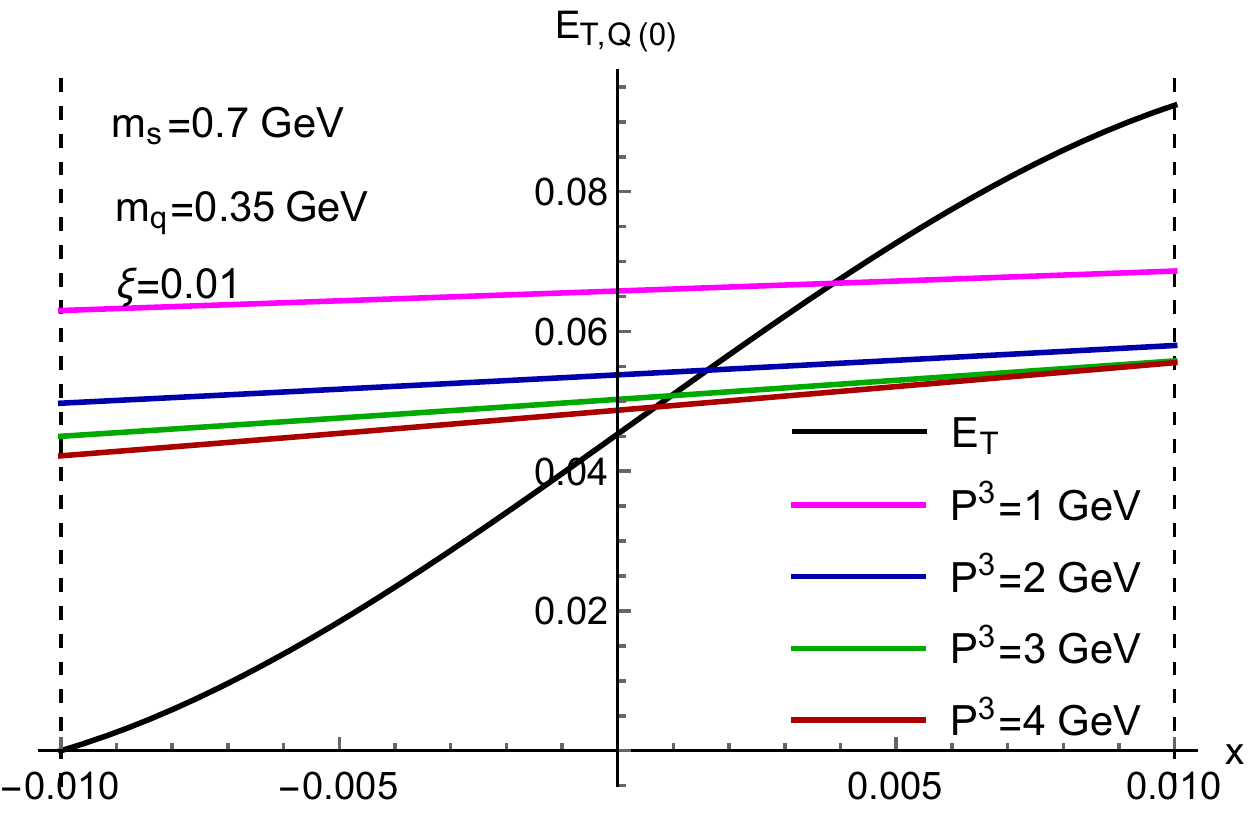}
\hspace{1.5cm}
\includegraphics[width=6.5cm]{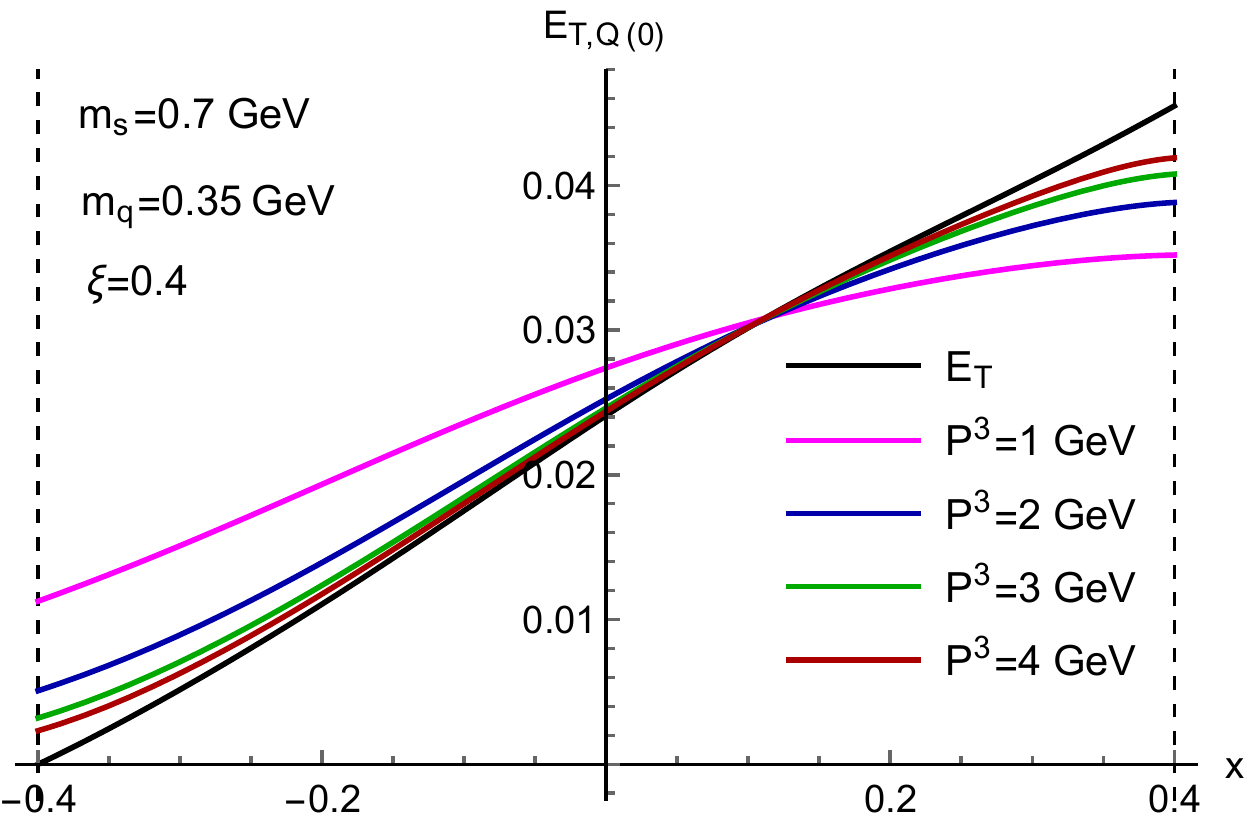}
\caption{Quasi-GPD $E_{T, {\rm Q(0)}}$ as a function of $x$ in the ERBL region for different values of $P^3$. 
Left panel: results for $\xi = 0.01$.
Right panel: results for $\xi = 0.4$.
The standard GPD $E_T$ is shown for comparison.}
\label{f:ET_ERBL}
\end{figure}

\begin{figure}[!]
\includegraphics[width=6.5cm]{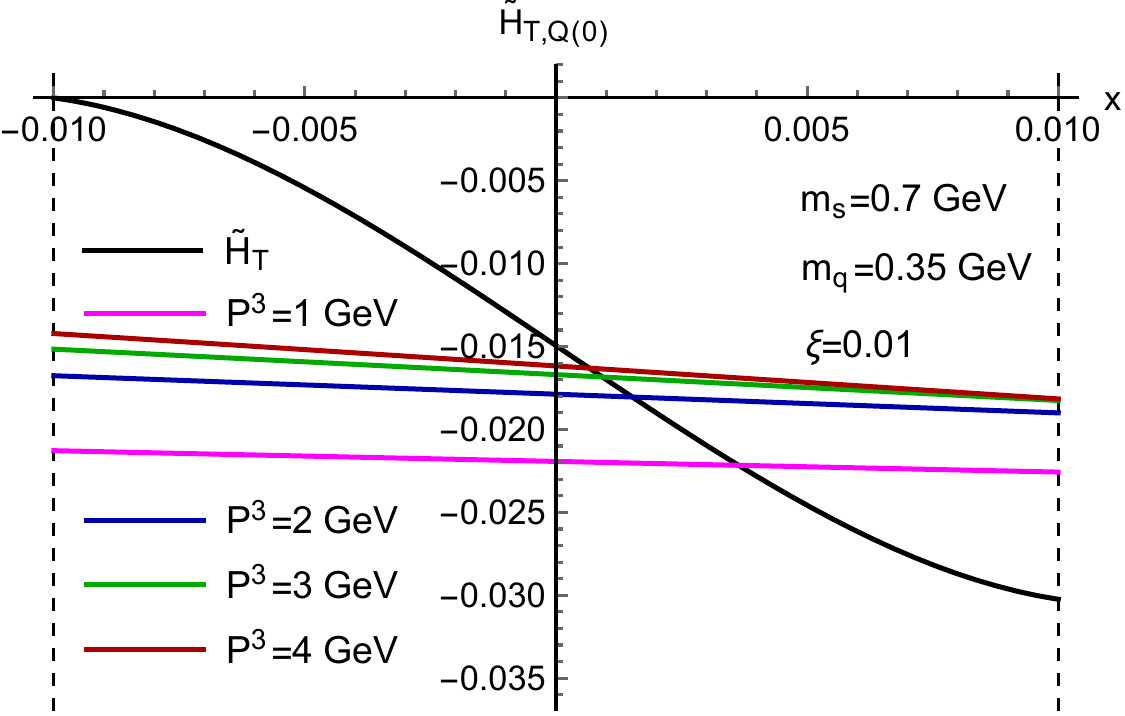}
\hspace{1.5cm}
\includegraphics[width=6.5cm]{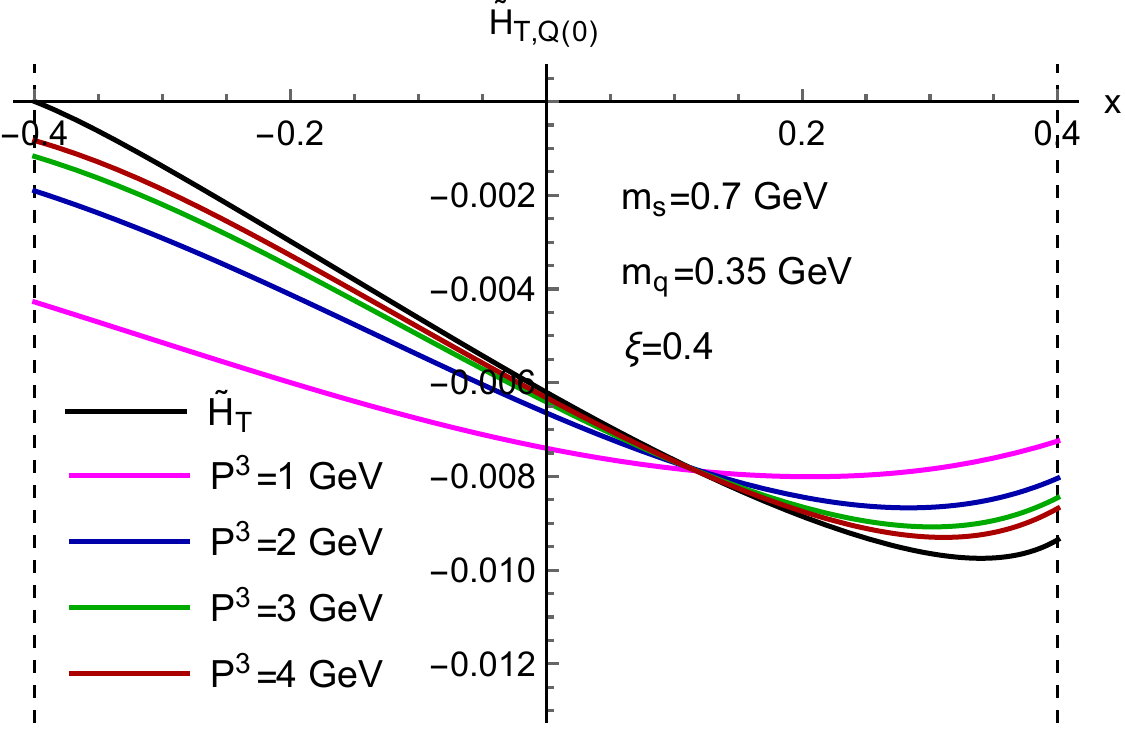}
\caption{Quasi-GPD $\widetilde{H}_{T, {\rm Q(0)}}$ as a function of $x$ in the ERBL region for different values of $P^3$. 
Left panel: results for $\xi = 0.01$.
Right panel: results for $\xi = 0.4$.
The standard GPD $\widetilde{H}_T$ is shown for comparison.}
\label{f:HT_tilde_ERBL}
\end{figure}

\begin{figure}[!]
\includegraphics[width=6.5cm]{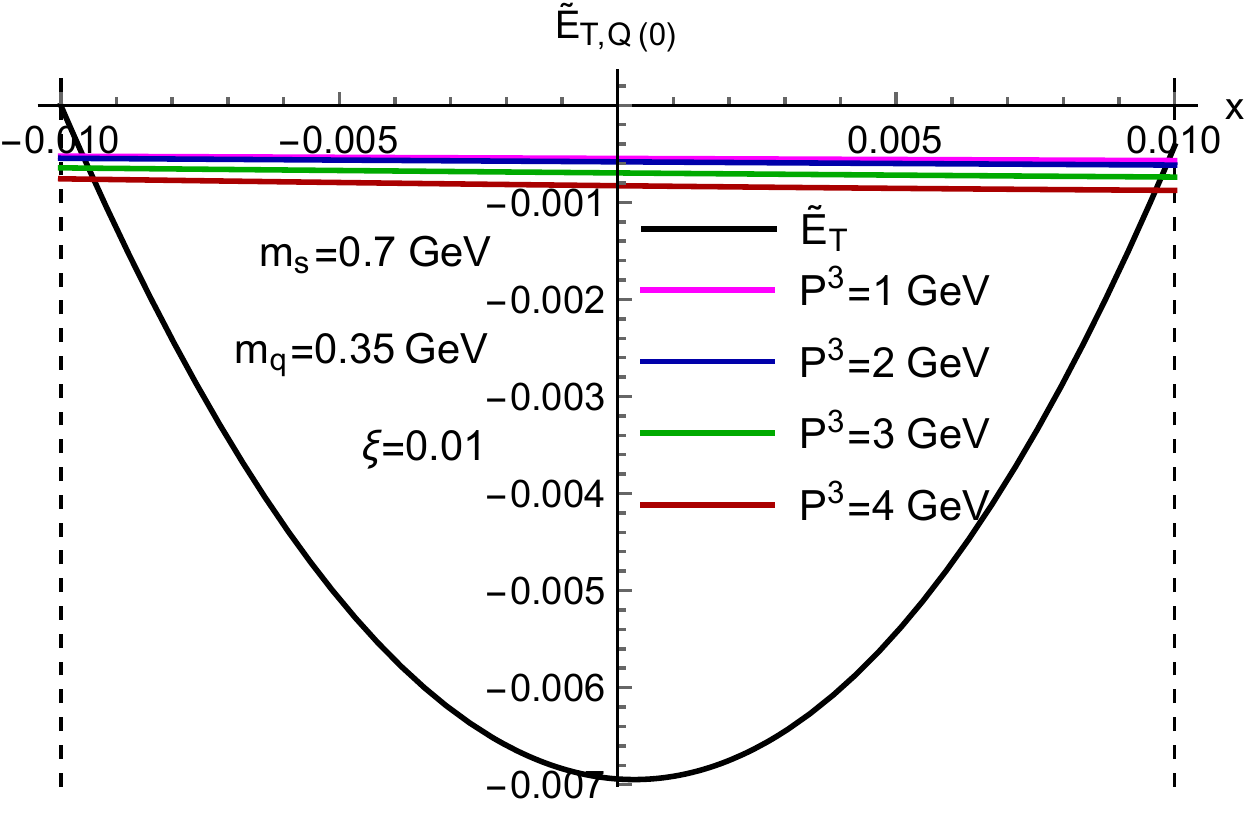}
\hspace{1.5cm}
\includegraphics[width=6.5cm]{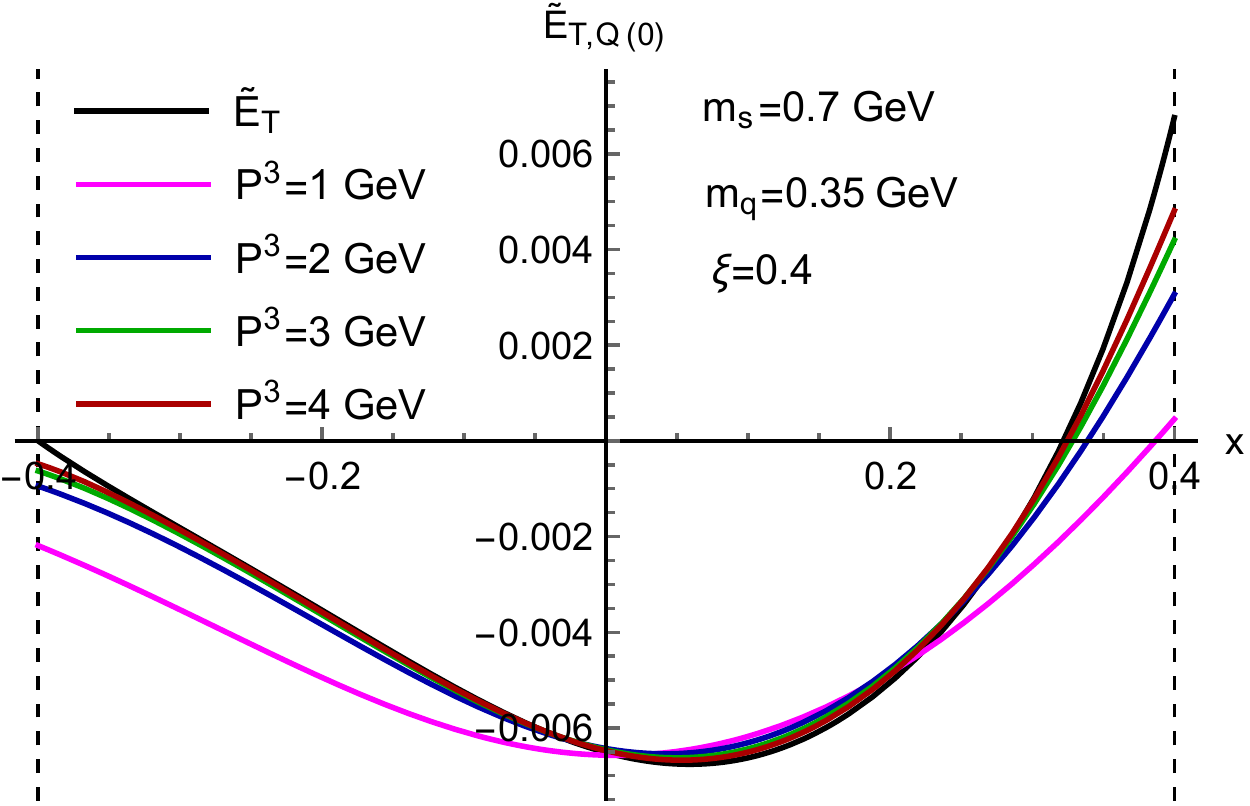}
\caption{Quasi-GPD $\widetilde{E}_{T, {\rm Q(0)}}$ as a function of $x$ in the ERBL region for different values of $P^3$. 
Left panel: results for $\xi = 0.01$.
Right panel: results for $\xi = 0.4$.
The standard GPD $\widetilde{E}_T$ is shown for comparison.}
\label{f:ET_tilde_ERBL}
\end{figure}

\begin{figure}[t]
\includegraphics[width=6.5cm]{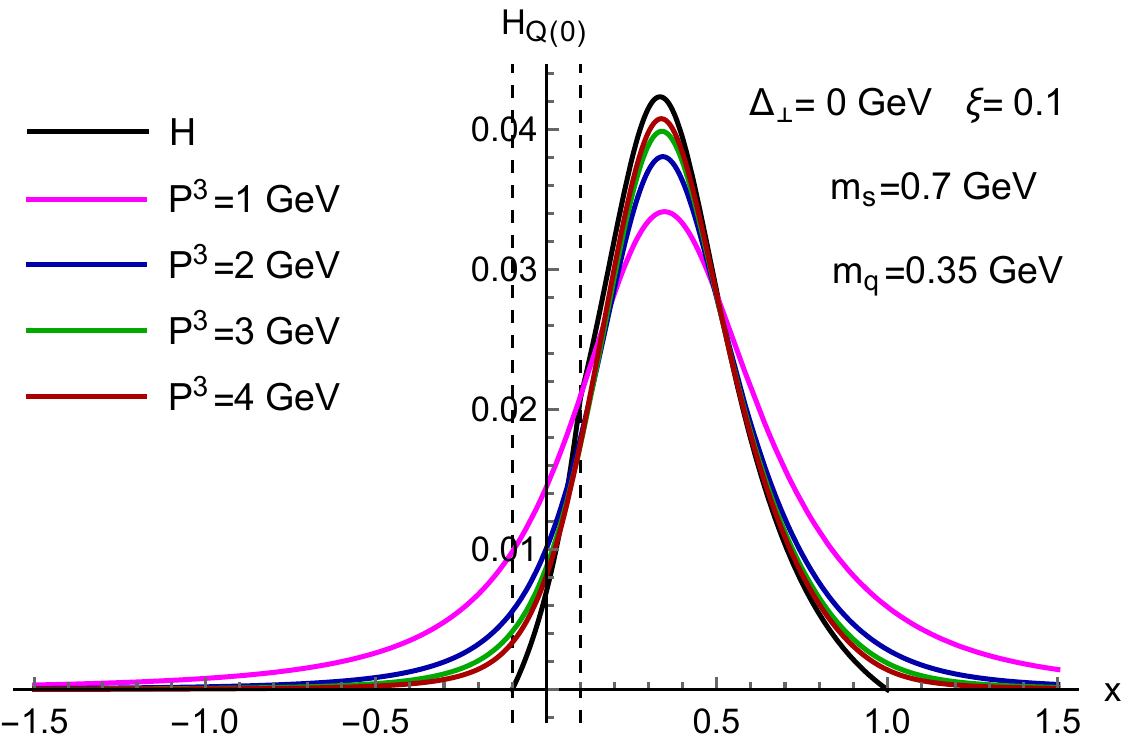}
\hspace{1.5cm}
\includegraphics[width=6.5cm]{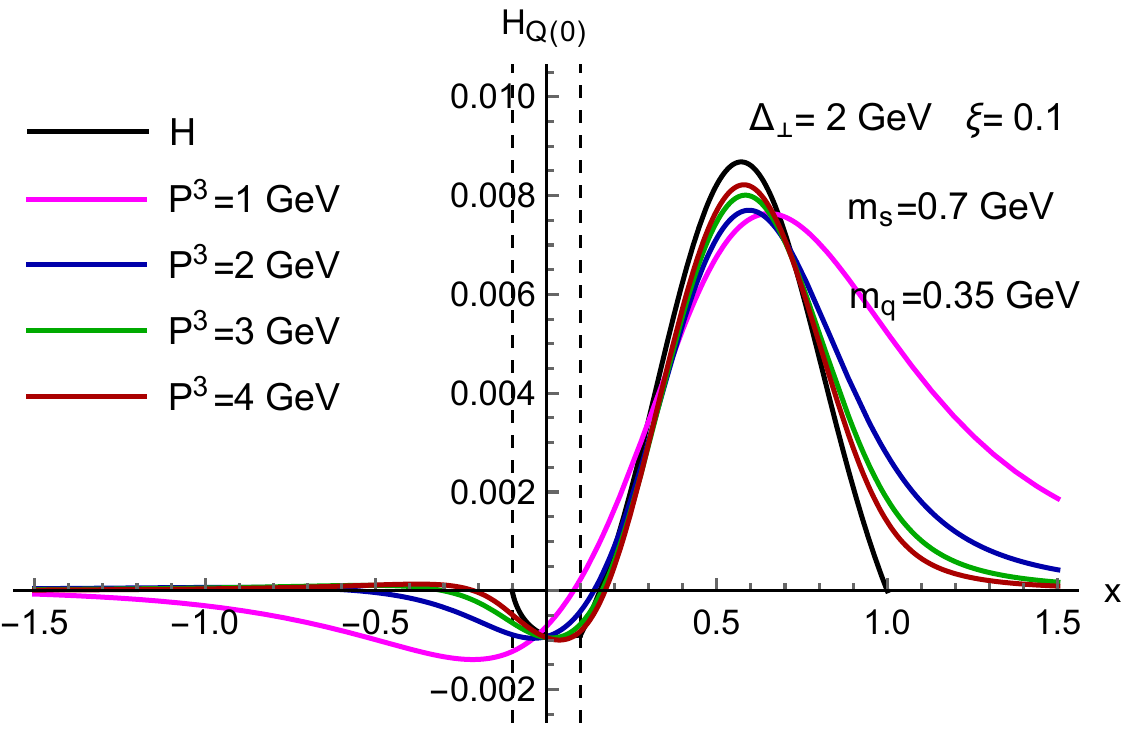}
\caption{Quasi-GPD $H_{\rm Q(0)}$ as a function of $x$ for different values of $P^3$ and two values of $\Delta_\perp = |\vec{\Delta}_\perp|$.
Left panel: results for $\Delta_\perp = 0 \, \textrm{GeV}$.
Right panel: results for $\Delta_\perp = 2 \, \textrm{GeV}$.}
\label{f:H_delta}
\end{figure}

\begin{figure}[!]
\includegraphics[width=6.5cm]{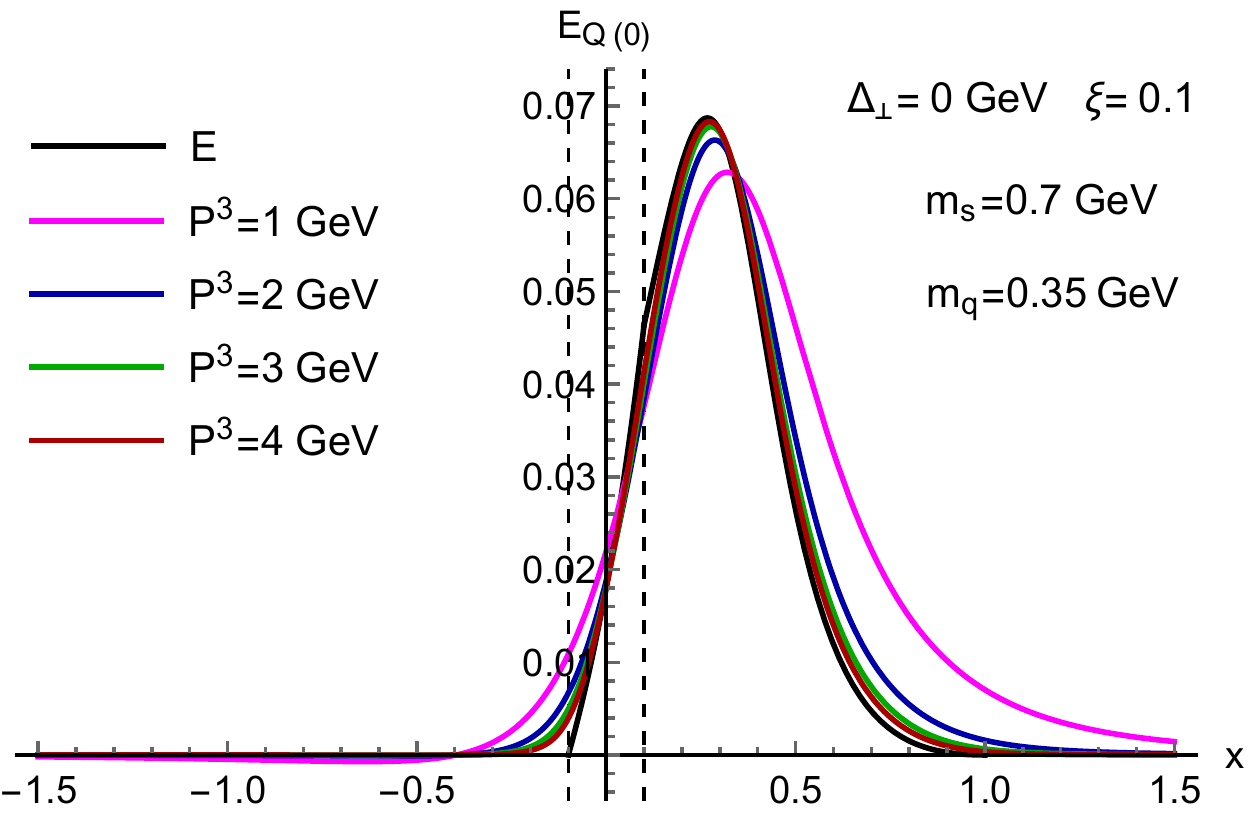}
\hspace{1.5cm}
\includegraphics[width=6.5cm]{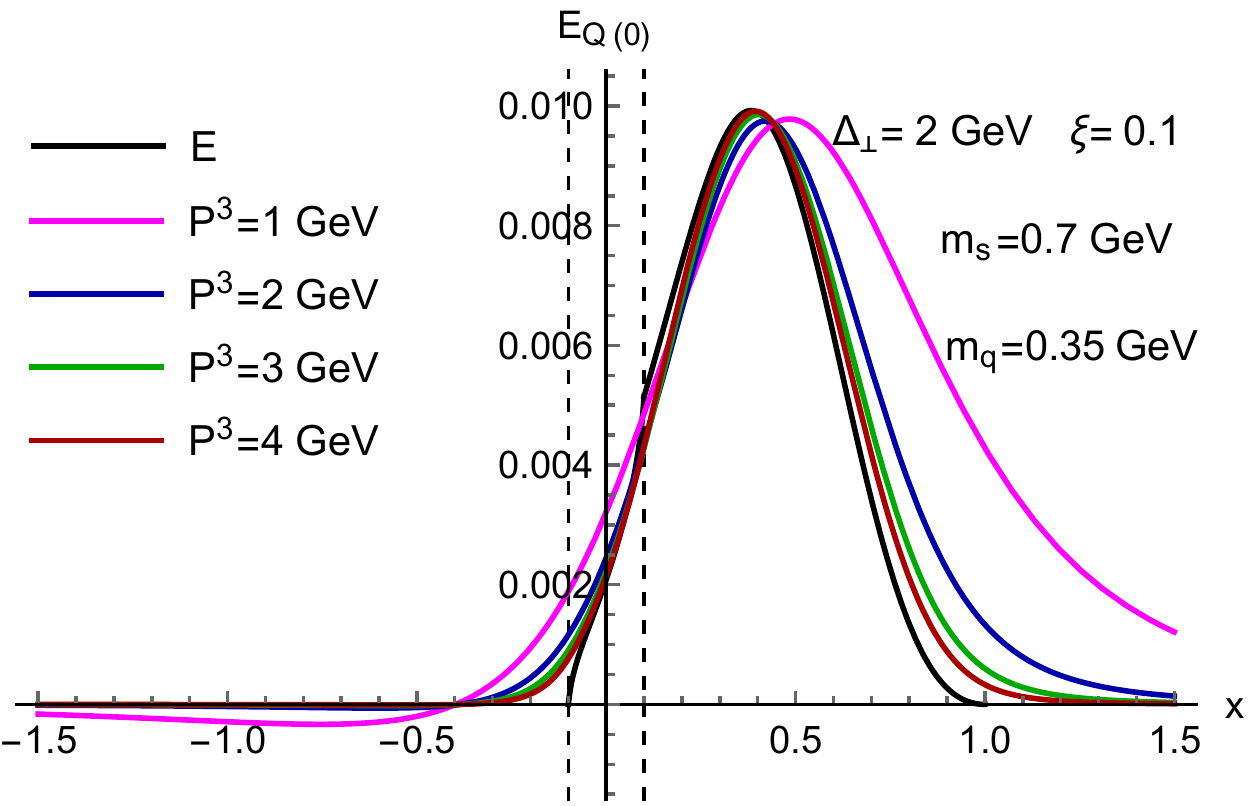}
\caption{Quasi-GPD $E_{\rm Q(0)}$ as a function of $x$ for different values of $P^3$ and two values of $\Delta_\perp = |\vec{\Delta}_\perp|$.
Left panel: results for $\Delta_\perp = 0 \, \textrm{GeV}$.
Right panel: results for $\Delta_\perp = 2 \, \textrm{GeV}$.}
\label{f:E_delta}
\end{figure}
Details about the numerics for the quasi-GPDs $H_{\rm Q}$ and $E_{\rm Q}$ can be found in Ref.~\cite{Bhattacharya:2018zxi}.
Here we therefore mostly focus on the remaining six quasi-GPDs, which are shown in Figs.~\ref{f:H_tilde} -- \ref{f:ET_tilde} for $\xi=0.1$.
For the skewness variable we have explored the range $0.01 \leq \xi \leq 0.4$ and below briefly comment on the $\xi$-dependence.
Like in the case of quasi-PDFs, for $P^3 \gtrsim 2 \, \textrm{GeV}$ there is no clear indication as to which of the two definitions (for each quasi-GPD)  one should prefer.
The convergence problem at large $x$ persists for the quasi-GPDs $H_{\rm Q}$ and $E_{\rm Q}$~\cite{Bhattacharya:2018zxi} and the other six quasi-GPDs.
We emphasize that this outcome is a robust feature of our model calculation.
In lattice calculations, the matching procedure could potentially improve the situation at large $x$, as was observed for the quasi-PDFs~\cite{{Alexandrou:2018pbm,Chen:2018xof}}. 
It has been shown that, at one-loop order, a nontrivial matching exists only for the quasi-GPDs $H_{\rm Q}$, $\widetilde{H}_{\rm Q}$ and $H_{T, {\rm Q}}$ --- the ones that survive in the forward limit~\cite{Ji:2015qla, Xiong:2015nua, Liu:2019urm}.
It remains to be seen whether in lattice studies it is more difficult to obtain good results at large $x$ for the quasi-GPDs that do not require a non-trivial matching. 
We also note that, in general, there is a tendency of the discrepancies at large $x$ to increase when $\xi$ gets larger.
The significance of this feature depends on the GPD under consideration, and it is most pronounced for the quasi-GPDs $\widetilde{E}_{\rm Q}$ and $\widetilde{E}_{T, {\rm Q}}$. This aspect is illustrated via Fig. \ref{f:H_tE_RD_Skew} which clearly shows an increase in the relative difference at large $x$ for larger values of $\xi$ for the GPD $\widetilde{E}$ for instance, compared to the GPD $H$.

We make a brief passing comment regarding the origin of sign changes in the $x < 0$ region depending upon the use of $\gamma^0$ vs. $\gamma^3$ operator in the quasi-GPD matrix element.  The reason for this are the terms proportional to $x \, M$.  If a quasi-distribution contains such a term, as $\widetilde{H}_{\rm Q(0)}$, $H_{T , \rm Q(3)}$, and $h_{1, \rm Q(3)}$ do (see Eqs. (32), (37), and (70)), then the distribution will change sign in the region $x < 0$ (see for instance Figs. \ref{f:h1Q} or \ref{f:H_tilde} or \ref{f:HT}). As this is the first model calculation of quasi-GPDs, it is difficult to speculate whether this behavior is model-independent. The purely mathematical origin of this behavior in our model gives no deeper insight.

The plots in the Figs.~\ref{f:H_tilde_ERBL} -- \ref{f:ET_tilde_ERBL} show the quasi-GPDs in the ERBL region for $\xi=0.01$ and $\xi=0.4$, while corresponding plots for $H_{\rm Q}$ and $E_{\rm Q}$ can be found in our previous work~\cite{Bhattacharya:2018zxi}.
Generally, for small $\xi$ one finds significant deviations between the quasi-GPDs and the corresponding standard GPDs.  
This situation is the GPD counterpart of the problem for quasi-PDFs around $x = 0$.
For small $\xi$, the standard GPDs rapidly approach zero at $x = - \, \xi$ in a very narrow $x$-range, whereas the quasi-GPDs are much smoother in that range. 
Once $\xi$ is increased, we observe a (much) better agreement between quasi-GPDs and the standard GPDs for a large fraction of the ERBL region.
To be more quantitative, we look at  $\widetilde{H}_{\rm Q(3)}$ as an example, with $\xi = 0.01$ at the point $x = 0.01$.  From Fig. 14 we can see that the agreement is extremely poor for $P^{3} = 4 \, \rm GeV$.  For decent agreement (relative difference less than $20\, \%$), one must go to $P^{3}$ values as high as $18 \, \rm GeV$, which is well beyond the present reach of lattice QCD.  On the other hand, if we instead choose $\xi = 0.4$ and focus on the point $x = 0.4$, one finds decent agreement (as defined above) with $P^{3}$ values as low as $1 \, \rm GeV$ which are currently accessible in lattice QCD. This outcome suggests that lattice calculations could provide very valuable information in the ERBL region, provided that the skewness is not too small.

We have also studied the dependence of our results on the transverse momentum transfer to the hadron $|\vec{\Delta}_{\perp}|$ or $t$, where Fig.~\ref{f:H_delta} and Fig.~\ref{f:E_delta} show results for $H_{\rm Q(0)}$ and $E_{\rm Q(0)}$, respectively. 
Apparently, at least at large $x$, the discrepancies get somewhat larger as $|\vec{\Delta}_{\perp}|$ is increased.
However, we also found that the relative difference as defined in Eq.~(\ref{e:rel_difference}) is hardly affected at all when $|\vec{\Delta}_{\perp}|$ gets larger.
This statement holds true for all the other quasi-GPDs. 
In fact none of the general conclusions discussed above are affected if $|\vec{\Delta}_{\perp}|$ is varied, where we have mostly explored the range $0 \, \textrm{GeV} \leq |\vec{\Delta}_{\perp}| \leq 2 \, \textrm{GeV}$ or $0 \, \textrm{(GeV})^{2} \leq |t| \leq 4 \, \textrm{(GeV})^{2}$. 

\subsection{Exploring different skewness variables}
\label{sec: skewnesses}
\begin{figure}[!]
\includegraphics[width=6.5cm]{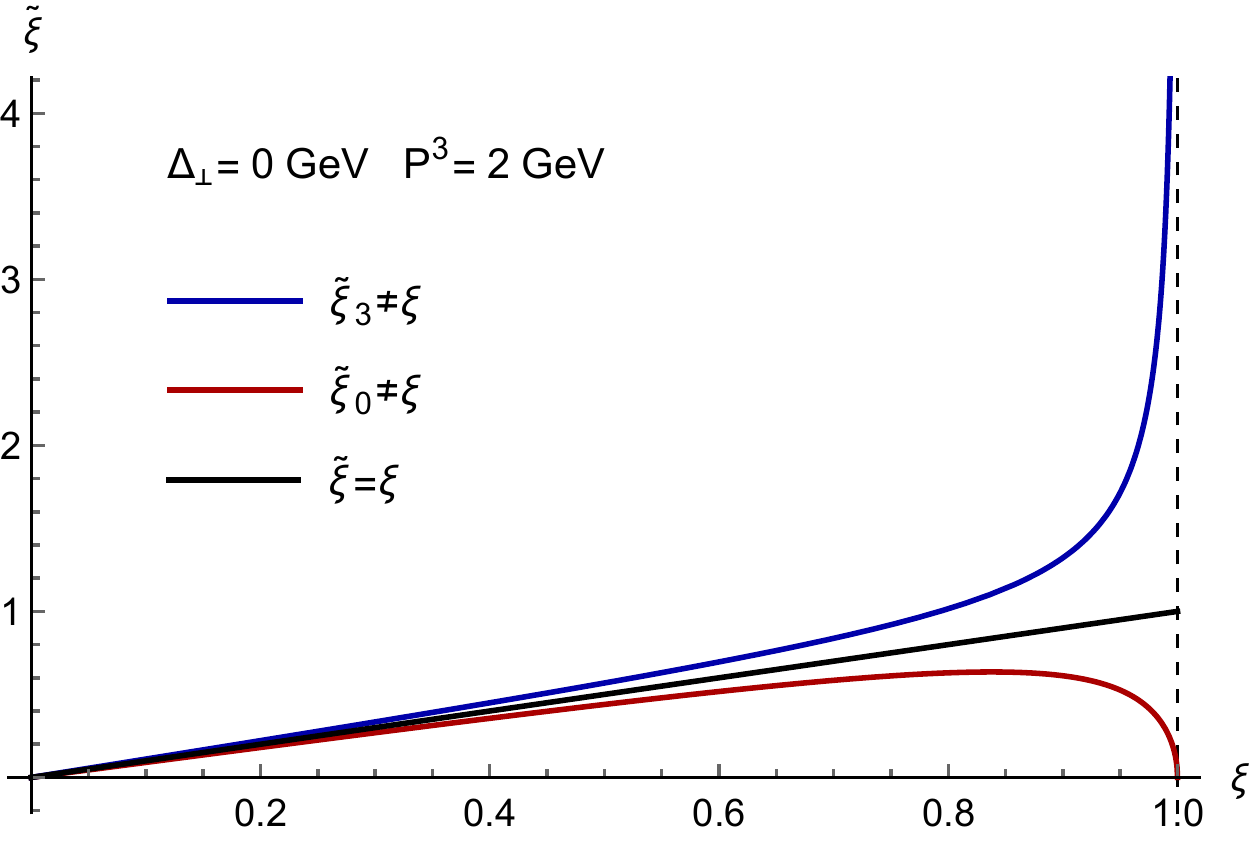}
\caption{Comparison of the skewness variables $\xi$, $\tilde{\xi}_3$ and $\tilde{\xi}_0$ for $P^3 = 2 \, \textrm{GeV}$ and $\Delta_\perp = |\vec{\Delta}_\perp| = 0 \, \textrm{GeV}$.}
\label{f:skewness}
\end{figure}

\begin{figure}[h]
\includegraphics[width=6.5cm]{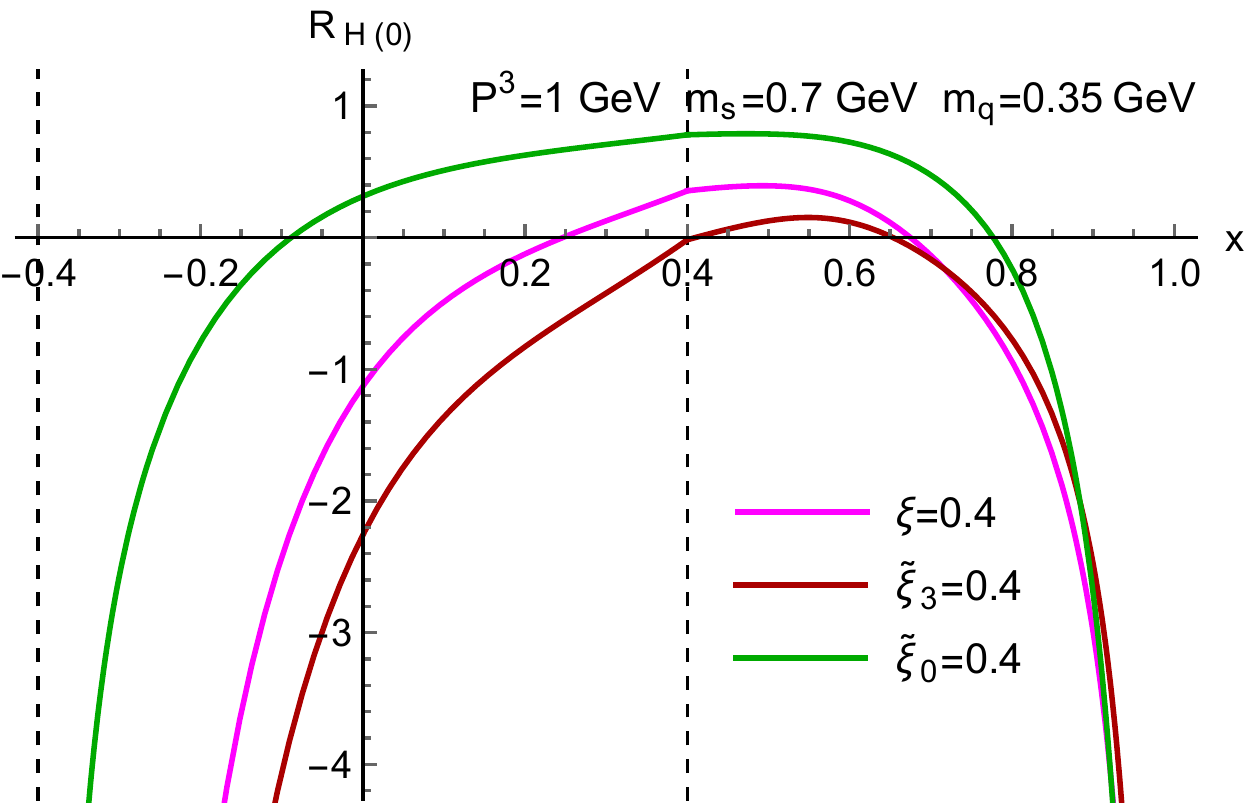}
\hspace{1.5cm}
\includegraphics[width=6.5cm]{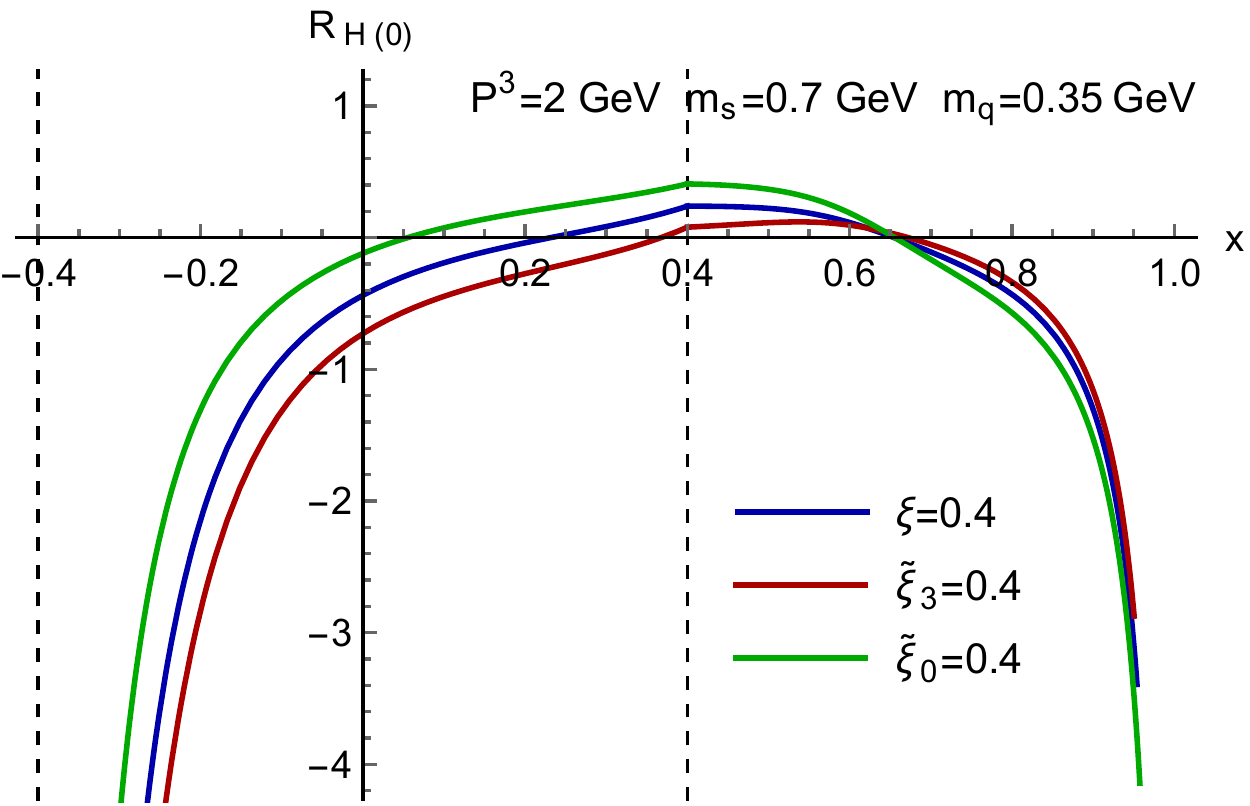}
\caption{Relative difference between $H_{\rm Q(0)}$ and $H$ as a function of $x$.
The quasi-GPD is evaluated for three different definitions of the skewness variable. 
(See text for more details.) 
Left panel: results for $P^3 = 1 \, \textrm{GeV}$.
Right panel: results for $P^3 = 2 \, \textrm{GeV}$.} 
\label{f:H_skewness}
\end{figure}

\begin{figure}[!]
\includegraphics[width=6.5cm]{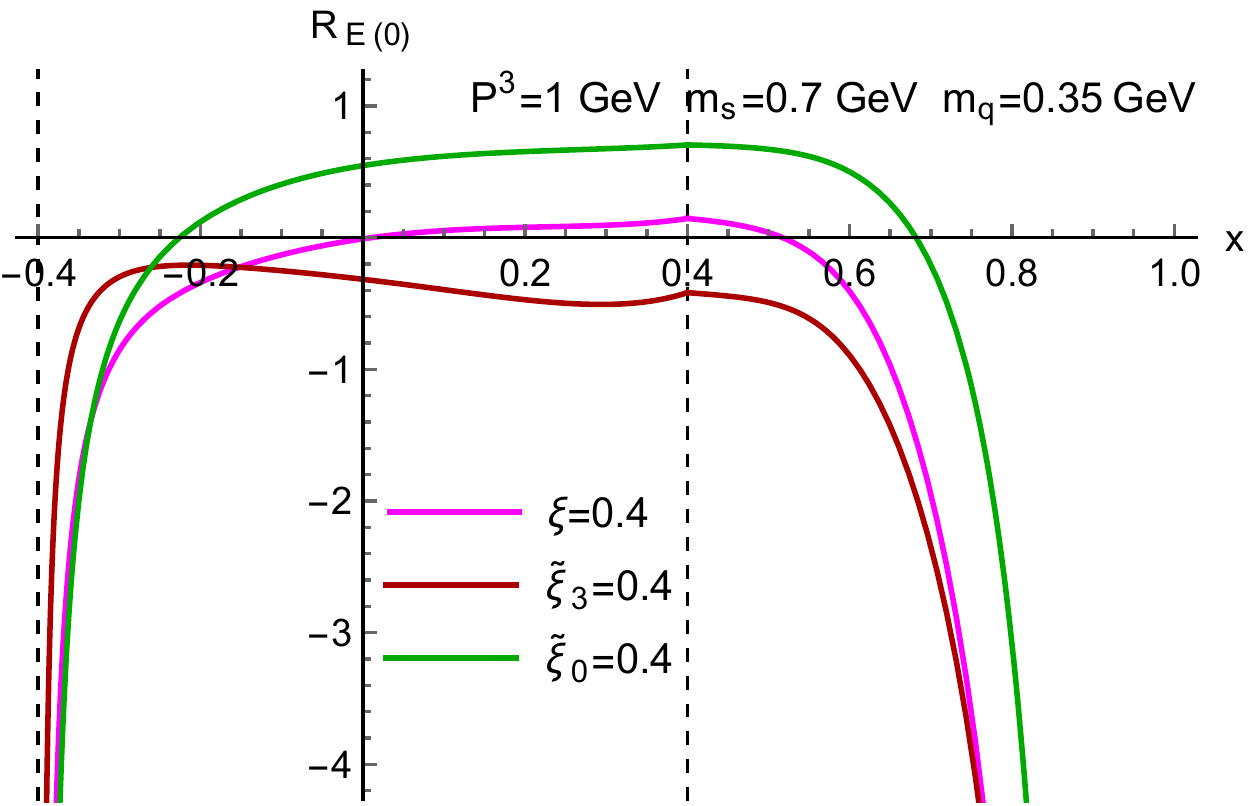}
\hspace{1.5cm}
\includegraphics[width=6.5cm]{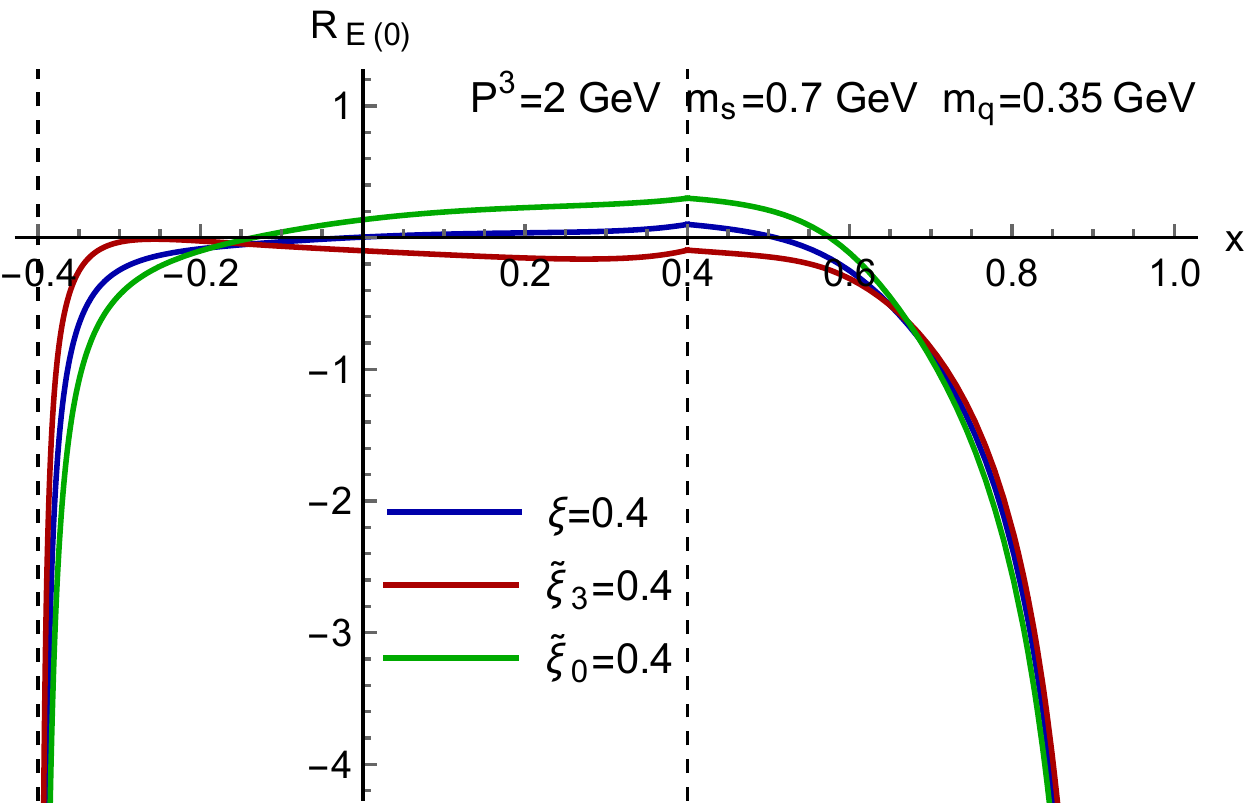}
\caption{Relative difference between $E_{\rm Q(0)}$ and $E$ as a function of $x$.
The quasi-GPD is evaluated for three different definitions of the skewness variable. 
(See text for more details.) 
Left panel: results for $P^3 = 1 \, \textrm{GeV}$.
Right panel: results for $P^3 = 2 \, \textrm{GeV}$.} 
\label{f:E_skewness}
\end{figure}

So far we have used the same skewness variable $\xi$ for both the standard GPDs and the quasi-GPDs.
However, in the case of quasi-GPDs one could in principle consider different variables to describe the longitudinal momentum transfer to the hadron.
Actually, in the matching calculations for quasi-GPDs the quantity $\tilde{\xi}_{3} = - \frac{\Delta^{3}}{2P^{3}}$ was used~\cite{Ji:2015qla, Xiong:2015nua, Liu:2019urm}.
The two variables are related via $\tilde{\xi}_{3} = \delta \xi$, with $\delta$ from Eq.~(\ref{e:delta}).
We emphasize that this relation is model-independent, which is in contrast to the situation for the parton momentum fractions $\frac{k^+}{P^+}$ and $\frac{k^3}{P^3}$ for which no model-independent relation exists.
Another possible skewness variable is $\tilde{\xi}_{0} = -\tfrac{\Delta^{0}}{2P^{0}} = \frac{\xi}{\delta}$, though admittedly $\tilde{\xi}_{0}$ is somewhat less natural than $\tilde{\xi}_{3}$ due to the dependence of quasi-GPDs on $\frac{k^3}{P^3}$.
In any case, the difference between the three variables is a higher-twist effect that vanishes for $P^3 \to \infty$.
For finite $P^3$, however, the differences can be substantial as illustrated in Fig.~\ref{f:skewness}, and they are largest for large $\xi$.
Note that as $\xi \to 1$ one has $|t| \to \infty$, and therefore also $\delta \to \infty$. 
Here we want to explore the impact of the difference between $\xi$, $\tilde{\xi}_3$, and $\tilde{\xi}_0$ on the quasi-GPDs.

In order to calculate quasi-GPDs using $\tilde{\xi}_{0/3}$ one can then no longer use Eq.~(\ref{e:t_s}), but rather needs 
\begin{eqnarray}
t(\tilde{\xi}_{0},\, \vec{\Delta}_{\perp};P^{3}) & = & -\, \frac{2}{\tilde{\xi}^{2}_{0}} \bigg [ (1-\tilde{\xi}^{2}_{0}) (P^{3})^{2} - 2 \tilde{\xi}^{2}_{0} M^2 - \sqrt {(1-\tilde{\xi}^{2}_{0})^{2} (P^{3})^{4} - \tilde{\xi}^{2}_{0}(4M^{2}+\vec{\Delta}^{2}_{\perp}) (P^{3})^{2} } \, \bigg] \,,
\label{e:t0} 
\\[0.1cm]
t(\tilde{\xi}_{3},\, \vec{\Delta}_{\perp};P^{3}) & = & 2 \bigg[ (1- \tilde{\xi}^{2}_{3}) (P^{3})^{2} + M^{2} -\frac{\vec{\Delta}^{2}_{\perp}}{4} \bigg ] 
\nonumber \\[0.1cm]
& & - \, 2 \, \sqrt {(1-\tilde{\xi}^{2}_{3})^{2} (P^{3})^{4} + 2 (1+\tilde{\xi}^{2}_{3}) \bigg (M^{2}+\dfrac{\vec{\Delta}^{2}_{\perp}}{4}\bigg ) (P^{3})^{2} + \bigg ( M^{2} +\dfrac{\vec{\Delta}^{2}_{\perp}}{4}\bigg )^{2}} \,,
\label{e:t3}
\end{eqnarray}
to compute the Mandelstam variable $t$.
For $P^3 \to \infty$, both~(\ref{e:t0}) and~(\ref{e:t3}) reduce to Eq.~(\ref{e:t_s}), while non-negligible numerical differences exist when $P^{3}$ is finite. 
From Fig.~\ref{f:skewness} one finds that the allowed range for $\tilde{\xi}_0$ is smaller than $[0,1]$.
As a consequence, $t$ would become imaginary if in Eq.~(\ref{e:t0}) one plugs in a value for $\tilde{\xi}_0$ that is too large.

In Fig.~\ref{f:H_skewness} and Fig.~\ref{f:E_skewness} we show the following comparisons.
The standard GPDs in these figures, which enter the relative difference $R$ in Eq.~(\ref{e:rel_difference}),  are all evaluated for $\xi = 0.4$, while the quasi GPDs are calculated using the three different skewness variables $\xi$, $\tilde{\xi}_0$ and $\tilde{\xi}_3$ and choosing for them in each case again the value $0.4$.
One observes considerable differences between the three cases, especially once $P^3$ is relatively low.
Interestingly, in the case of $H_{\rm Q(0)}$ the relative difference is smaller for most of the DGLAP region (in particular, in the range where the GPDs have their maximum) if one uses $\tilde{\xi}_3$ instead of $\xi$.
We find this pattern for most of the quasi-GPDs, while in the ERBL region no general pattern exists.
The only outliers in that regard are $E_{\rm Q(0)}$, $\widetilde{E}_{\rm Q(0/3)}$ and $E_{T, {\rm Q(0)}}$, where $E_{\rm Q(0)}$ is shown in Fig.~\ref{f:H_skewness} as a representative case.
We also observe that using the variable $\tilde{\xi}_{0}$ typically gives poorer convergence for the quasi-GPDs.
This feature is again most pronounced in the range where the GPDs are largest.
Our conclusions also hold for even larger values of $\xi$, where the numerical discrepancy between the three skewness variables increases further --- see Fig.~\ref{f:skewness}.

We take a moment to briefly discuss the nature of two distinct higher-twist effects that we encounter in our model study. The higher-twist effect associated with the longitudinal momentum transfer to the target (relating $\tilde{\xi}_{0/3}$ to $\xi$) is kinematical. Such an effect, expressed through the parameter $\delta$, is model-independent and simply describes the relationship between the variables $P^{0}$ and $P^{3}$. The impact of this effect, however, on the individual GPDs is model-dependent. On the other hand, the higher-twist effect associated with the longitudinal parton momenta (relating $\tilde{x}$ to $x$) is a dynamical one which stems from modeling of the spectator as an onshell diquark. We note that these effects are entirely separate from the higher-twist effects associated with QCD higher-twist operators.

\section{Axial-vector Diquark Results}
\label{sec:axial-vector}
It is well known that both a scalar diquark and an axial-vector diquark are needed to describe the phenomenology of up quarks and down quarks in the nucleon --- see, for instance, Refs.~\cite{Mezrag:2017znp, Kumar:2017dbf, Bednar:2018htv}.
In this section, we therefore explore contributions from the axial-vector diquark. 
We repeat that we do not aim at a fine-tuned quantitative description of the standard GPDs, which would be beyond the scope of the present work, but rather just focus on how well the quasi-GPDs compare with their corresponding standard GPDs.

In order to study the impact of the axial-vector diquark, we examine in detail the effects on the GPD $H$. For the scalar diquark, the vertex factor is given by $i g_{s}\gamma^{\mu}$, where $g_{s}$ is the scalar coupling constant, and the propagator is given by $\frac{i}{(P-k)^{2}-m_{s}^{2}+i\varepsilon}$. In contrast, for the axial-vector diquark the vertex factor is given by $\frac{i g_{a}}{\sqrt{2}}\gamma^{\mu}\gamma_{5}$, where $g_{a}$ is the axial vector coupling constant, and the propagator by $\frac{i d^{\mu\nu}}{(P-k)^{2}-m_{a}^{2}+i\varepsilon}$, where $d^{\mu\nu}$ and $m_{a}$ are, respectively, the polarization tensor and mass of the axial-vector diquark. There are several possible choices for the polarization tensor (see Ref.~\cite{Bacchetta:2008af}), but we choose the definition
\begin{equation}
d^{\mu\nu}=-g^{\mu\nu}+\frac{P^{\mu}P^{\nu}}{m_{a}^{2}} \, ,
\label{pol_ten}
\end{equation}
which was analyzed in Ref.~\cite{Jakob:1997wg}. The other choices for the polarization tensor are explored in Refs.~\cite{Bacchetta:2008af,Brodsky:2000ii,Gamberg:2007wm,Bacchetta:2003rz}. We now replace the scalar diquark vertex and propagator in the light-cone correlation function with the axial vector diquark vertex and propagator. Note that with this replacement the light-cone correlation function no longer follows from a Lagrange density, and thus this direct replacement is also a part of our model for the axial vector diquark. The result is
\begin{equation}
F^{a[\Gamma]}(x,\Delta)=\frac{i g_{a}^{2}}{4(2\pi)^{4}}\int dk^{-}d^{2}\vec{k}_{\perp} d_{\mu\nu} \frac{\bar{u}(p',\lambda')\gamma^{\mu}\gamma_{5} (\slashed{k}+\frac{\slashed{\Delta}}{2}+m_{q}) \Gamma(\slashed{k}+\frac{\slashed{\Delta}}{2}-m_{q}) \gamma^{\nu}\gamma_{5} u(p,\lambda)}{D_{\rm GPD}^{a}} \, ,
\end{equation}
where $D_{\rm GPD}^{a}$ is the same as $D_{\rm GPD}$ with the replacement $m_{s}\rightarrow m_{a}$. For the figures below we always choose $g_{a}=g_{s}=1$ and $\vec{\Delta}_{\perp}=0 $ GeV. We also use $m_{a}=1$ GeV as our standard value, as when quarks couple to a higher spin-state, the resulting state tends to have a larger mass~\cite{Bacchetta:2008af}.

For the standard GPD $H$, one again uses $\Gamma=\gamma^{+}$. The
result for the axial-vector diquark is given by Eq. (16) in our paper~\cite{Bhattacharya:2018zxi} with the replacement $N_{H}\rightarrow N_{H}^{a}$, where
\begin{eqnarray}
2N_{H}^{a} & = & \bigg ( 2+\dfrac{M^{2}}{m_{a}^{2}}+\dfrac{t}{4 m_{a}^{2}} \bigg ) \big ( \vec{k}_{\perp}^{2}+ m_{q}^{2} + x^{2} M^{2}\big )+2\bigg [4x-x \dfrac{M^{2}}{m_{a}^{2}}+(2+x)\dfrac{t}{4m_{a}^{2}}\bigg ]m_{q} M \nonumber \\
&& +(1+x)\bigg [2(1+x)+(1-3x)\dfrac{M^{2}}{m_{a}^{2}}+(1+x)\dfrac{t}{4m_{a}^{2}}\bigg] \dfrac{t}{4}+(1+x)\bigg [2-\dfrac{M^{2}}{m_{a}^{2}}+\dfrac{t}{4m_{a}^{2}}\bigg ]\xi t\dfrac{\vec{k}_{\perp}\cdot\vec{\Delta}_{\perp}}{\vec{\Delta}_{\perp}^{2}} \, .
\end{eqnarray}
The quasi--GPD correlator for the axial-vector diquark is given by
\begin{equation}
F^{a[\Gamma]}(x,\Delta; P^{3})=\frac{i g_{a}^{2}}{4(2\pi)^{4}}\int dk^{0}d^{2}\vec{k}_{\perp} d_{\mu\nu} \dfrac{\bar{u}(p',\lambda')\gamma^{\mu}\gamma_{5} (\slashed{k}+ \dfrac{\slashed{\Delta}}{2}+m_{q})\Gamma(\slashed{k}+\dfrac{\slashed{\Delta}}{2}-m_{q})\gamma^{\nu}\gamma_{5}u(p,\lambda)}{D_{\rm GPD}^{a}} \, ,
\end{equation}
and the results for the axial-vector diquark are given by Eq. (27) but with the replacement $N_{H(0/3)}\rightarrow N_{H(0/3)}^{a}$ where
\begin{eqnarray}
\dfrac{2}{\delta} N^{a}_{H(0)} &=& \bigg ( 2 + \dfrac{M^{2}}{m^{2}_{a}} + \dfrac{t}{4 m^{2}_{a}} \bigg ) (k^{0})^{2} \nonumber \\
&& + \dfrac{2}{\delta P^{3}} \bigg [ -x \bigg ( 2 + \dfrac{M^{2}}{m^{2}_{a}} + \dfrac{t}{4 m^{2}_{a}} \bigg ) (P^{3})^{2} + \bigg ( 4 - \dfrac{M^{2}}{m^{2}_{a}} + \dfrac{t}{4 m^{2}_{a}} \bigg ) m_{q} M + \bigg ( 2 - \dfrac{M^{2}}{m^{2}_{a}} + \dfrac{t}{4 m^{2}_{a}} \bigg ) \bigg ( x \dfrac{t}{4} + \dfrac{\delta}{2} \xi t \dfrac{\vec{k}_{\perp}\cdot \vec{\Delta}_{\perp}}{\vec{\Delta}^{2}_{\perp}} \bigg ) \bigg ] k^{0} \nonumber \\
&& + \bigg ( 2 + \dfrac{M^{2}}{m^{2}_{a}} +\dfrac{t}{4 m^{2}_{a}} \bigg ) \bigg ( \vec{k}^{2}_{\perp}+\dfrac{t}{4} + m^{2}_{q} + x^{2} (P^{3})^{2} \bigg ) +\dfrac{m_{q} M}{m^{2}_{a}} t + 2 \bigg ( 2 - \dfrac{M^{2}}{m^{2}_{a}} +\dfrac{t}{4 m^{2}_{a}} \bigg ) \bigg ( x \dfrac{t}{4} + \dfrac{\delta}{2} \xi t \dfrac{\vec{k}_{\perp}\cdot \vec{\Delta}_{\perp}}{\vec{\Delta}^{2}_{\perp}} \bigg ) \, ,\\ [0.3cm]
2N^{a}_{H (3)} &=& - \bigg ( 2 + \dfrac{M^{2}}{m^{2}_{a}} +\dfrac{t}{4 m^{2}_{a}} \bigg ) (k^{0})^{2} + 2 \delta P^{3} \bigg [\bigg ( 2 + \dfrac{M^{2}}{m^{2}_{a}} +\dfrac{t}{4 m^{2}_{a}} \bigg ) x + \bigg ( 2 - \dfrac{M^{2}}{m^{2}_{a}} +\dfrac{t}{4 m^{2}_{a}} \bigg ) \dfrac{1+x}{\delta^{2} (P^{3})^{2}} \dfrac{t}{4} \bigg ] k^{0} \nonumber \\
&& + \bigg ( 2 + \dfrac{M^{2}}{m^{2}_{a}} +\dfrac{t}{4 m^{2}_{a}} \bigg )\bigg ( \vec{k}^{2}_{\perp}+\dfrac{t}{4} + m^{2}_{q} - x^{2} (P^{3})^{2} \bigg ) + 2 \bigg [ 4x - x \dfrac{M^{2}}{m^{2}_{a}} +(2+x)\dfrac{t}{4 m^{2}_{a}} \bigg ] m_{q} M \nonumber \\
&& + \bigg ( 2 - \dfrac{M^{2}}{m^{2}_{a}} +\dfrac{t}{4 m^{2}_{a}} \bigg ) \dfrac{1+x}{\delta} \xi t \dfrac{\vec{k}_{\perp}\cdot \vec{\Delta}_{\perp}}{\vec{\Delta}^{2}_{\perp}} \, .
\end{eqnarray}

\begin{figure}[t]
\begin{center}
\includegraphics[width=6.5cm]{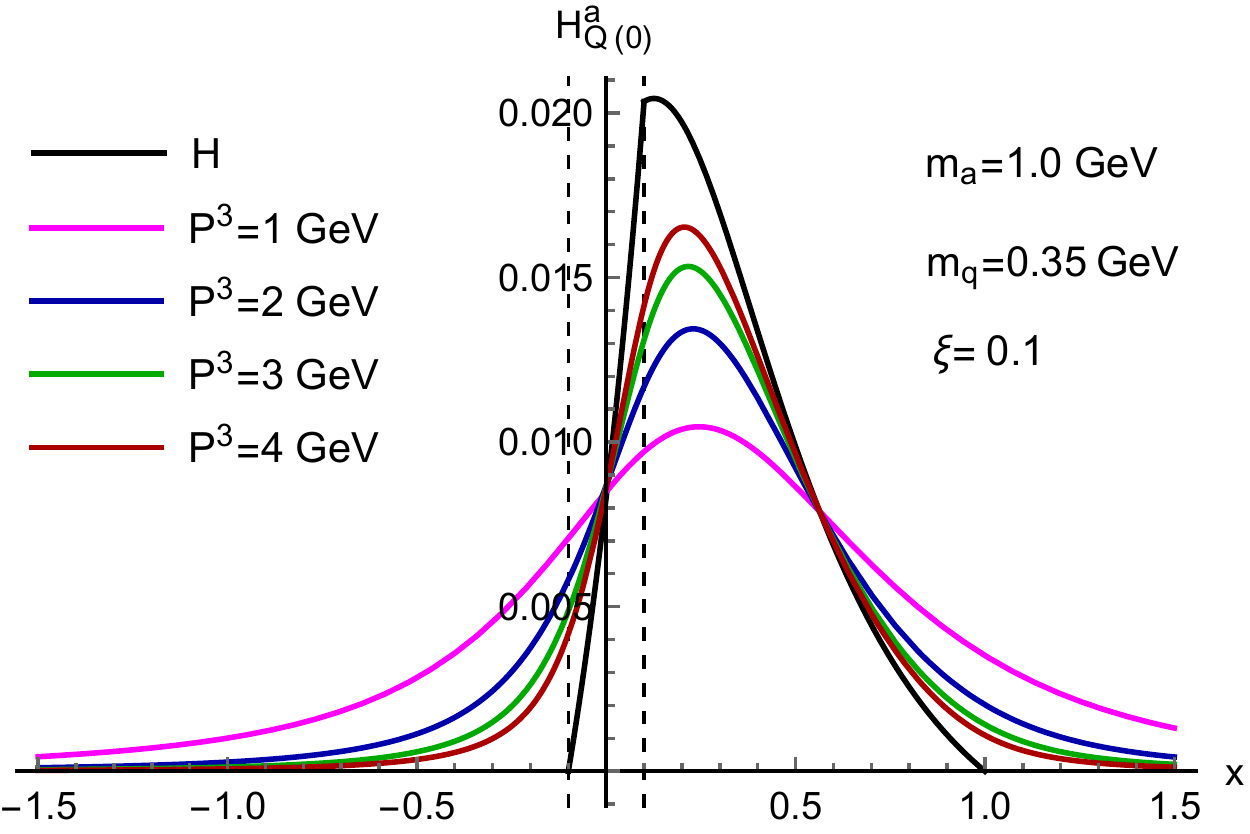}
\hspace{1.5cm}
\includegraphics[width=6.5cm]{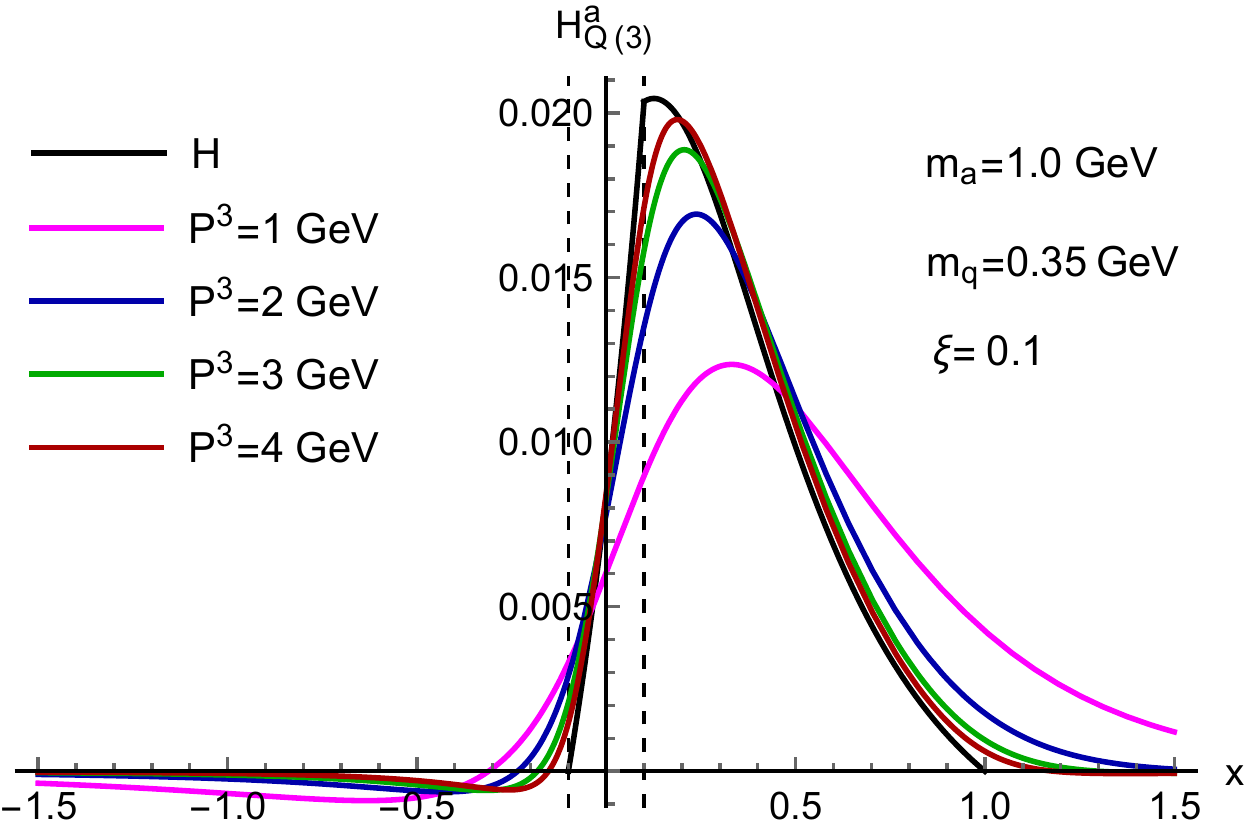}
\end{center}
\caption{Quasi-GPD $H_{\rm {Q}}^{a}$ as a function of $x$ for $\xi=0.1$
and different values of $P^{3}$. Left panel: results for $H_{\rm Q(0)}^{a}$.
Right panel: results for $H_{\rm Q(3)}^{a}$. The standard GPD $H^{a}$
is shown for comparison.}
\label{H_axial_vector}
\end{figure}

\begin{figure}[t]
\begin{center}
\includegraphics[width=6.5cm]{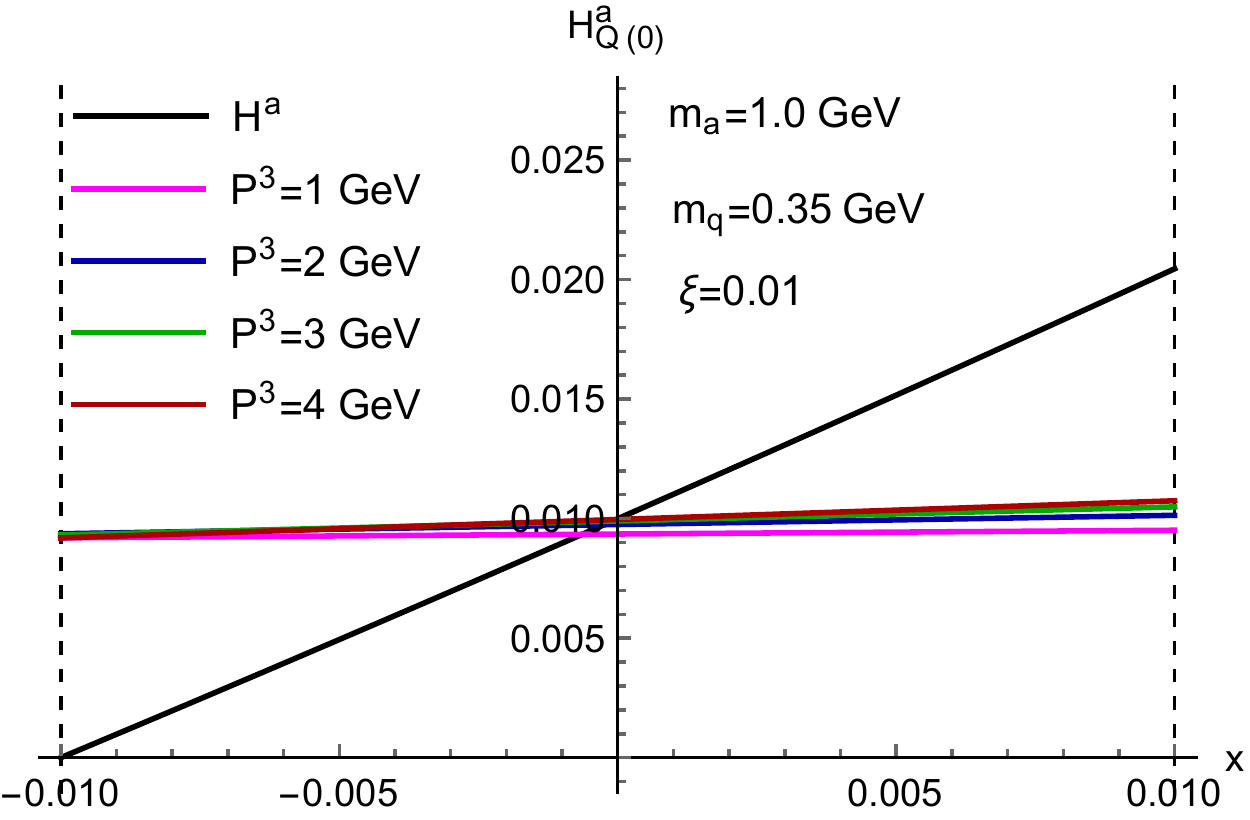}
\hspace{1.5cm}
\includegraphics[width=6.5cm]{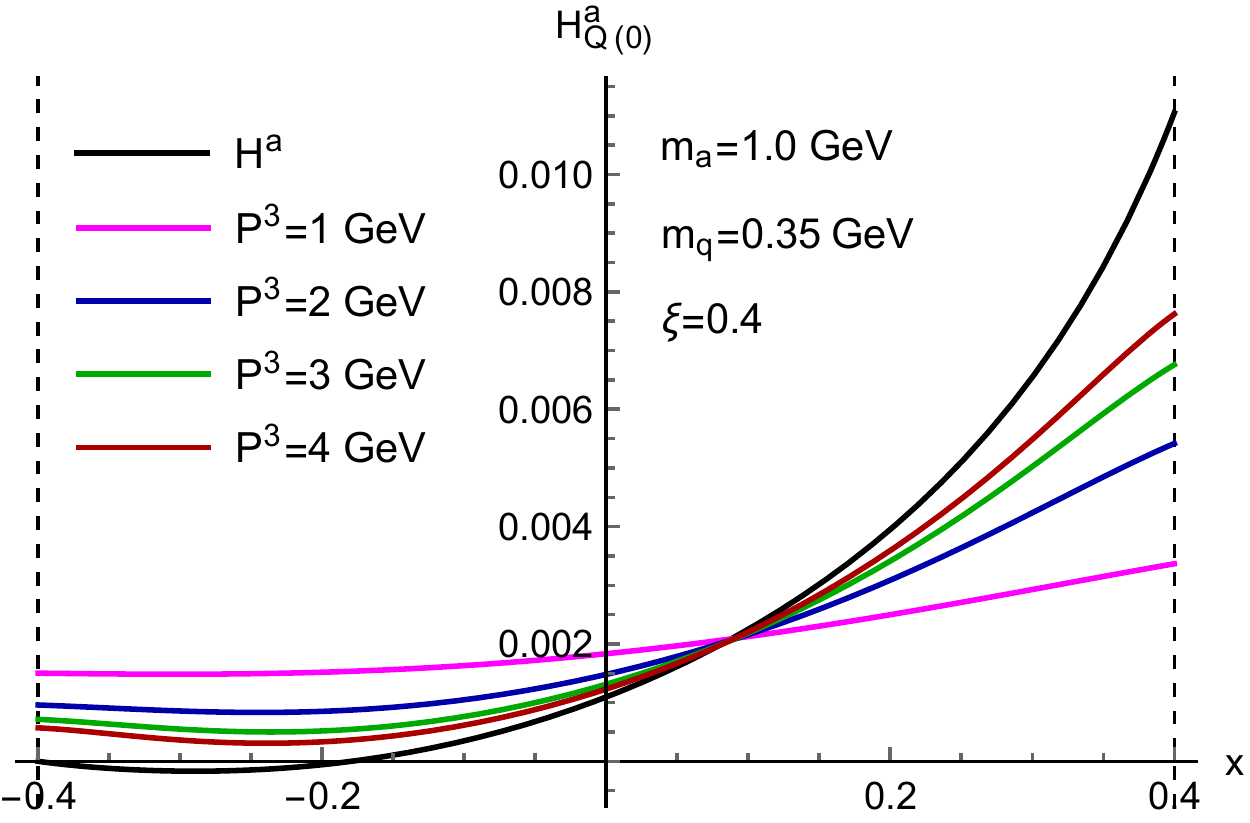}
\end{center}
\caption{Quasi-GPD $H_{\rm Q(0)}^{a}$ in the ERBL region for
different values of $P^{3}$. Left panel: results for $\xi=0.01$.
Right panel: results for $\xi=0.4$. The standard GPD $H^{a}$ is
shown for comparison.}
\label{H_axial_vector_ERBL}
\end{figure}

In Figs. \ref{H_axial_vector} and \ref{H_axial_vector_ERBL} we show the results obtained for the quasi-GPD $H_{\rm Q}$. Comparison with Figs. 6 and 9 in Ref.~\cite{Bhattacharya:2018zxi} shows that the qualitative features of the GPD $H$ remain the same regardless of the type of diquark. Fig. \ref{H_axial_vector_ERBL} shows that while convergence in the ERBL region is poor at extremely small values of $\xi$, it is reasonable at larger values.
Based on this, we conclude that with the polarization tensor chosen as in Eq. (\ref{pol_ten}), there are non-negligible contributions to the GPDs and PDFs from the axial vector diquark. However, these contributions have the exact same qualitative features as those of the scalar diquark contributions, and thus our conclusions based on the scalar diquark contributions alone are robust.
Our general findings therefore also apply for faithful GPDs for up and down quarks in a spectator model.

\section{Moments of quasi distributions}
\label{sec:moments}
Recently, moments of quasi-PDFs have attracted some attention~\cite{Rossi:2017muf, Rossi:2018zkn, Radyushkin:2018nbf, Karpie:2018zaz}.
Specifically, in Refs~\cite{Rossi:2017muf,Rossi:2018zkn} concerns have been raised over divergences of moments of quasi-PDFs, while Ref.~\cite{Radyushkin:2018nbf} argues that the two lowest moments are well defined. 
While the whole point of exploring quasi-PDFs is to go beyond the calculation of moments, it can still be instructive to look at them.

We first consider the lowest moments of quasi-GPDs and recall also the well-known results for the lowest moments of the corresponding standard GPDs.
Including a flavor index `$q$' one finds the model-independent relations
\begin{eqnarray}
\int\limits_{-1}^{1} dx \, H^{q}(x,\xi,t) &=& \int\limits_{-\infty}^{\infty} dx \, \frac{1}{\delta} \, H^{q}_{\rm Q(0)}(x,\xi,t;P^3) = \int\limits_{-\infty}^{\infty} dx \, H^{q}_{\rm Q(3)}(x,\xi,t;P^3) = F^{q}_{1}(t) \,,
\label{e:FF_1}  
\\[0.1cm] 
\int\limits_{-1}^{1} dx \, E^{q}(x,\xi,t) &=& \int\limits_{-\infty}^{\infty} dx \, \frac{1}{\delta} \, E^{q}_{\rm Q(0)}(x,\xi,t;P^3) = \int\limits_{-\infty}^{\infty} dx \, E^{q}_{\rm Q(3)}(x,\xi,t;P^3) = F^{q}_{2}(t) \,,
\label{e:FF_2}  
\\[0.1cm]
\int\limits_{-1}^{1} dx \, \widetilde{H}^{q}(x,\xi,t) &=& \int\limits_{-\infty}^{\infty} dx \, \widetilde{H}^{q}_{\rm Q(0)}(x,\xi,t;P^3) = \int\limits_{-\infty}^{\infty} dx \, \frac{1}{\delta} \, \widetilde{H}^{q}_{\rm Q(3)}(x,\xi,t;P^3) = G^{q}_{A}(t) \,,
\label{e:FF_3}  
\\[0.1cm]
\int\limits_{-1}^{1} dx \, \widetilde{E}^{q}(x,\xi,t) &=& \int\limits_{-\infty}^{\infty} dx \, \widetilde{E}^{q}_{\rm Q(0)}(x,\xi,t;P^3) = \int\limits_{-\infty}^{\infty} dx \, \frac{1}{\delta} \, \widetilde{E}^{q}_{\rm Q(3)}(x,\xi,t;P^3) = G^{q}_{P}(t) \,, 
\label{e:FF_4}  
\\[0.1cm]
\int\limits_{-1}^{1} dx \, H^{q}_{T}(x,\xi,t) &=& \int\limits_{-\infty}^{\infty} dx \, \frac{1}{\delta} \, H^{q}_{T, {\rm Q(0)}}(x,\xi,t;P^3) = \int\limits_{-\infty}^{\infty} dx \, H^{q}_{T, {\rm Q(3)}}(x,\xi,t;P^3) = F^{q}_{1, T}(t) \,, 
\label{e:FF_5} 
\\[0.1cm]
\int\limits_{-1}^{1} dx \, E^{q}_{T}(x,\xi,t) &=& \int\limits_{-\infty}^{\infty} dx \, \frac{1}{\delta} \, E^{q}_{T, {\rm Q(0)}}(x,\xi,t;P^3) = \int\limits_{-\infty}^{\infty} dx \, E^{q}_{T, {\rm Q(3)}}(x,\xi,t;P^3) = 2F^{q}_{2, T}(t) \,, 
\label{e:FF_6}  
\\[0.1cm]
\int\limits_{-1}^{1} dx \, \widetilde{H}^{q}_{T}(x,\xi,t) &=& \int\limits_{-\infty}^{\infty} dx \, \dfrac{1}{\delta} \, \widetilde{H}^{q}_{T, {\rm Q(0)}}(x,\xi,t;P^3) = \int\limits_{-\infty}^{\infty} dx \, \widetilde{H}^{q}_{T, {\rm Q(3)}}(x,\xi,t;P^3) = F^{q}_{3, T}(t) \,.
\label{e:FF_7} 
\end{eqnarray}
In the above equations, $F_{1}$ is the Dirac form factor, $F_{2}$ the Pauli form factor, $G_{A}$ the axial form factor, $G_{P}$ the pseudo-scalar form factor, and $F_{1, T}$, $F_{2, T}$ and $F_{3, T}$ are the form factors of the local tensor current~\cite{Adler:1975he}. 
Note that time-reversal invariance leads to a vanishing first moment for $\widetilde{E}_{T}$~\cite{Diehl:2003ny}. 
The results for the moments of the forward PDFs $f_1$, $g_1$ and $h_1$ can be extracted from Eq.~(\ref{e:FF_1}),~(\ref{e:FF_3}) and~(\ref{e:FF_5}), respectively.
The lowest moment of standard GPDs depends on $t$, but does not depend on $\xi$.
The quasi-GPDs depend in addition on $P^3$, but it is remarkable that also that dependence drops out in the lowest moment. 
(A corresponding discussion for $f_{1, {\rm Q}}$ can be found in Ref.~\cite{Radyushkin:2018nbf}.)
However, according to Eqs.~(\ref{e:FF_1})--(\ref{e:FF_7}) one must divide half of the quasi-GPDs by the (simple) kinematical factor $\delta$ in order to arrive at this result. 
(Our numerical results in the SDM comply with Eqs.~(\ref{e:FF_1})--(\ref{e:FF_7}).)
For $P^3 \lesssim 2 \, \textrm{GeV}$ inclusion of this factor leads to a visible difference for the numerics.
Of course $\delta$ describes a higher-twist effect, and therefore including this factor is in principle a matter of taste. 
But the moment analysis suggests that taking into account $\delta$ like in Eqs.~(\ref{e:FF_1})--(\ref{e:FF_7}) appears natural.
(This suggestion is in line with the definition of quasi-GPDs used in the very recent matching study in Ref.~\cite{Liu:2019urm}.)
In the case of quasi-PDFs, such a definition implies that $f_{1, {\rm Q(0)}}$, $g_{1, {\rm Q(3)}}$ and $h_{1, {\rm Q(0)}}$ are to be divided by $\delta_0$ in comparison to what so far has been used mostly in the literature.

We now turn our attention to the second moment of quasi-distributions, but consider the vector operator $\bar{\psi}^q \gamma^\mu \psi^q$ only.
It is well known that the corresponding local operators are related to the form factors of the energy momentum tensor (EMT) $T^{\mu\nu}$.
The EMT, when evaluated between different hadron states, has five independent structures~\cite{Lorce:2018egm},
\begin{eqnarray}
\langle p',\lambda '|T^{\mu \nu , q}(0) |p,\lambda \rangle &=& \bar{u}(p',\lambda ') \bigg [ \frac{P^{\mu} P^{\nu}}{M} A^{q}(t) + \frac{\Delta^{\mu}\Delta^{\nu}-g^{\mu \nu}\Delta^{2}}{M} C^{q}(t) +M g^{\mu \nu} \bar{C}^{q}(t) 
\nonumber \\[0.1cm]
&& + \frac{P^{\{\mu}i\sigma^{\nu \}\alpha}\Delta_{\alpha}}{4M}\big (A^{q}(t)+B^{q}(t)\big )+ \frac{P^{[\mu}i\sigma^{\nu ]\alpha}\Delta_{\alpha}}{4M}D^{q}(t) \bigg ] u(p,\lambda) \,,
\end{eqnarray}
where $T^{\mu \nu , q}(0) = \bar{\psi}^{q}(0)\gamma^{\mu} \frac{i}{2} \overleftrightarrow{D^{\nu}} \psi^{q}(0)$ with $D^\mu$ the covariant derivative, $a^{\{\mu}b^{\nu \}}=a^{\mu}b^{\nu}+a^{\nu}b^{\mu}$ and $a^{[\mu}b^{\nu ]}=a^{\mu}b^{\nu}-a^{\nu}b^{\mu}$. 
The connection between the quasi-GPDs and the form factors of the EMT is established through the equation
\begin{equation}
(P^{3})^{2} \int\limits_{-\infty}^{\infty} dx \, x \, \int\limits_{-\infty}^{\infty} \frac{dz^{3}}{2\pi} e^{ixP^{3}z^{3}} \langle p',\lambda ' | \bar{\psi}^{q}(-\tfrac{z^{3}}{2}) \gamma^{\mu} \, {\cal W}_{\rm Q}(-\tfrac{z}{2}, \tfrac{z}{2}) \, \psi^q (\tfrac{z^{3}}{2}) | p, \lambda \rangle \bigg |_{z^{0}=0, \vec{z}_{\perp}=\vec{0}_{\perp}} 
= \langle p',\lambda '|T^{\mu 3 , q}(0) |p,\lambda \rangle \,,
\end{equation}
where the index $\mu$ can be $0$ or $3$. 
In close analogy to the celebrated expression for the second moment of $H + E$, namely $\int\limits_{-1}^{1} dx \, x \, \big(H^{q}(x,\xi ,t)+E^{q}(x,\xi, t)\big) = A^{q}(t)+B^{q}(t)$ where $A^{q}(0)+B^{q}(0) = J^q$ is the total angular momentum for the quark flavor `$q$'~\cite{Ji:1996ek}, one then finds for the quasi-GPDs
\begin{eqnarray}
\int\limits_{-\infty}^{\infty} dx \, x \, \frac{1}{\delta} \, \big(H^{q}_{\rm Q(0)}(x,\xi ,t;P^3)+E^{q}_{\rm Q(0)}(x,\xi, t; P^3)\big) &=& \frac{1}{2}  (\delta^{2} + 1) \big(A^{q}(t)+B^{q}(t)\big)  + \frac{1}{2} (\delta^{2} - 1)D^{q}(t) \,, 
\label{e:smoment_1}
\\[0.1cm]
\int\limits_{-\infty}^{\infty} dx \, x \, \big(H^{q}_{\rm Q(3)}(x,\xi ,t; P^3)+E^{q}_{\rm Q(3)}(x,\xi, t; P^3)\big) &=& A^{q}(t)+B^{q}(t) \,.
\label{e:smoment_2} 
\end{eqnarray}
Note that in Eq.~(\ref{e:smoment_1}) the form factor $D^q$ of the anti-symmetric part of the EMT enters.
One can conclude that the second moment of $H_{\rm Q(3)} + E_{\rm Q(3)}$ is directly related to the angular momentum of quarks, while for $H_{\rm Q(0)} + E_{\rm Q(0)}$ this relation contains a higher-twist ``contamination.''
Our numerics are in accord with these two equations in the sense that the l.h.s.~of~(\ref{e:smoment_2}) is independent of $P^3$ and agrees with what we find from the second moment of $H + E$, while the l.h.s.~of~(\ref{e:smoment_1}) does depend on $P^3$ and converges to $A^{q}(t)+B^{q}(t)$ for large $P^3$.
In table~\ref{table}, we show that the first moments of $H_{\rm Q(3)}$ and $E_{\rm Q(3)}$ match with the first moments of $H$ and $E$, respectively, and are independent of $P^3$. We also show that Ji's spin-sum rule holds for the $\gamma^{3}$ projection regardless of the value of $P^3$. The conclusions remain the same if one uses $\gamma^{0}$ as the projection.
\begin{table}[t]
\centering
\begin{tabular}{|| c c c c ||}
\hline
$P^{3}$ (\rm GeV) & $\int dx H_{\rm Q(3)}$ & \quad $\int dx E_{\rm Q(3)}$ & \quad $\int dx \, x \, (H_{\rm Q(3)}+ E_{\rm Q(3)} )$\\
\hline \hline 
1 & 0.0105746 & \qquad 0.0136164 & 0.00904580 \\
2 & 0.0105743 & \qquad 0.0136165 & 0.00904583 \\
3 & 0.0105744 & \qquad 0.0136164 & 0.00904580 \\
4 & 0.0105743 & \qquad 0.0136163 & 0.00904584 \\
\hline
\end{tabular}
\caption{Left column: First moment of $H_{\rm Q(3)}$ for various values of $P^3$. Note that $\int dx \, H = 0.0105741$. Middle column: First moment of $E_{\rm Q(3)}$ for various values of $P^3$. Note that $\int dx \, E = 0.0136164$. Right column: Ji's spin-sum rule for the $\gamma^{3}$ projection. Note that $\int dx x \, (H + E) = 0.00904572$. All the numerical values have been obtained for $\xi = 0.1$ and $t = - 1 \, \rm (GeV)^{2}$.}
\label{table}
\end{table}


We now briefly take up the case of the second moment for $f_1$. 
In that case one has
\begin{equation}
\int\limits_{-1}^{1} dx \, x \, f_{1}(x) = A^{q}(0) \,,
\end{equation} 
and the corresponding equations for the quasi-PDFs read
\begin{eqnarray}
\int\limits_{-\infty}^{\infty} dx \, x \, \frac{1}{\delta_0} \, f_{1, {\rm Q(0)}}(x; P^3) &=& A^{q}(0) \,, 
\\[0.1cm]
\int\limits_{-\infty}^{\infty} dx \, x \, f_{1, {\rm Q(3)}}(x; P^3) &=& A^{q}(0) - (\delta_0^2 - 1) \bar{C}^{q}(0) \,.
\end{eqnarray}
The second moment of the quasi-PDF $f_{1, {\rm Q(0)}}$ is independent of $P^3$ only if the function is divided by $\delta_0$, which agrees with the situation for the lowest moment.
On the other hand, the second moment of $f_{1, {\rm Q(3)}}$ does depend on $P^3$.
Once again, our numerical results align with these analytical results. 
We also find that the third moments of the quasi-PDFs $f_{1, \rm Q}$, $g_{1, \rm Q}$ and $h_{1, \rm Q}$ and their corresponding quasi-GPDs ($H_{\rm Q}$, $\widetilde{H}_{\rm Q}$, and $H_{T, \rm Q}$) diverge. 
On the other hand, the third moments of the quasi-GPDs without forward counterparts do not diverge. We emphasize again that these moments relations are model-independent. For the regulated results, the moments are finite for both the standard and quasi-distributions. Of course, renormalization of the quasi-distributions needs to be considered as well. However, this point is equally relevant for the moments of the standard GPDs.

The model-independent expressions for the moments of the quasi-distributions are potentially significant as they may be useful for studying the systematic uncertainties of results from lattice QCD, especially due to the fact that the $P^3$-dependence of the moments is either computable or nonexistent.

\section{Summary}
\label{sec:summary}
We have presented results for all the quasi-GPDs corresponding to the eight leading-twist standard GPDs in the SDM. 
While the results for the vector quasi-GPDs $H_{\rm Q}$ and $E_{\rm Q}$ were already included in our previous work~\cite{Bhattacharya:2018zxi}, all the other ones are new.
For each standard GPD we have studied two quasi-GPDs.
Taking the forward limit, we have also obtained the quasi-PDFs $f_{\rm 1, Q}$, $g_{\rm 1, Q}$ and $h_{\rm 1, Q}$ as byproducts. 
In the limit $P^3 \to \infty$, all quasi-GPDs analytically reduce to the respective standard GPDs.
This outcome further supports the idea of computing quasi-GPDs in lattice QCD to get information on standard GPDs.
Though the forward PDFs (in the SDM) are discontinuous at $x = 0$, for $P^{3} \rightarrow \infty$ they are exactly reproduced by the corresponding quasi-PDFs.
Numerically, in the case of PDFs we have found significant discrepancies between the quasi-distributions and the standard distributions around $x = 0$ and $x = 1$.
We have also elaborated on the underlying cause of these discrepancies.
For instance, the disparities near $x = 1$, which also exist for quasi-GPDs, are due to higher-twist effects that grow as $x \to 1$.
For GPDs these disparities tend to increase with an increase of the skewness $\xi$.
On the other hand, for large $\xi$ we have found quite good agreement between quasi-GPDs and standard GPDs for a significant part of the ERBL region.
Furthermore, at least in the DGLAP region we have observed for most quasi-GPDs a better agreement with the standard GPDs if $\xi$ is replaced by $\tilde{\xi}_3 = - \frac{\Delta^3}{2 P^3}$.
The difference between $\xi$ and $\tilde{\xi}_3$ is a higher-twist effect. 
Generally, we have tried to identify robust results in the SDM.
We have therefore studied the dependence of the result on the free parameters --- the quark mass $m_q$, the spectator mass $m_s$, the cut-off for the integration upon the transverse quark momentum $\Lambda$, and the momentum transfer to the hadron.
We have also studied the contributions from the axial diquark and found no qualitative differences compared to the scalar diquark.
We have also clarified the behavior of quasi-GPDs under the transformation $\xi \to - \xi$.
Moreover, we have derived model-independent results for the first and second moments of quasi-distributions. 
It is remarkable that these moments either do not depend on $P^3$, or their $P^3$-dependence can be computed.
A particularly interesting case is the second moment of $H_{\rm Q} + E_{\rm Q}$, which is related to the total angular momentum of quarks.
The results for the moments suggest a preferred definition of several quasi-distributions.
Moments of quasi-distributions might allow one to explore systematic uncertainties of results in lattice QCD.
In conclusion, we believe it is worthwhile to further study quasi-GPDs from a conceptual point of view as well as numerically in lattice QCD and in other models.

\begin{acknowledgments}
We thank Martha Constantinou and Jianhui Zhang for discussions.
This work has been supported by the National Science Foundation under grant No. PHY-1812359, and by the U.S. Department of Energy, Office of Science, Office of Nuclear Physics, within the framework of the TMD Topical Collaboration.
\end{acknowledgments}


\end{document}